\documentclass[a4paper,11pt]{article}
\pdfoutput=1 % if your are submitting a pdflatex (i.e. if you have
\usepackage{jheppub} % for details on the use of the package, please
\usepackage{caption} 
\usepackage{subcaption}
\usepackage{breakurl}

\usepackage[latin9]{inputenc}
\usepackage{float}
\usepackage{amsmath}
\usepackage{amssymb}
\usepackage{graphicx}
\usepackage{esint}
\usepackage{hyperref}
\usepackage{comment}
\usepackage{color}
\usepackage{microtype}
\usepackage{cleveref}
\usepackage{breakurl}
\usepackage{bbm}
\newcommand{\be}{\begin{equation}}
\newcommand{\ee}{\end{equation}}
\newcommand{\ben}{\begin{displaymath}}
\newcommand{\een}{\end{displaymath}}
\newcommand{\bea}{\begin{eqnarray}}
\newcommand{\eea}{\end{eqnarray}}
\def\K{K{\"a}hler}
   \newcommand{\rf}[1]{(\ref{#1})}

\def\be{\begin{equation}}
\def\ee{\end{equation}}
\def\bea{\begin{eqnarray}}
\def\eea{\end{eqnarray}}
\def\ba{\begin{array}}
\def\ea{\end{array}}
\def\bit{\begin{itemize}}
\def\eit{\end{itemize}}

\def\a{\alpha}

\def\rmi{{\rm i}}

\newcommand{\cN}{\mathcal{N}}

 \makeatletter

\allowdisplaybreaks

\makeatother

\makeatletter
\DeclareRobustCommand{\rcite}[1]{%
  \rcite@aux#1,\@nil{#1}%
}
\def\rcite@aux#1,#2\@nil#3{%
  \if\relax#2\relax
    Ref.~\cite{#3}%
  \else
    Refs.~\cite{#3}%
  \fi
}
\makeatother

\hypersetup{
    colorlinks = true,
    citecolor = {blue},
    linkcolor = {blue},
    urlcolor = {blue},
}

 \title{\rm {\bf \huge    Mass Production of IIA and IIB dS Vacua}}

\author[a]{Niccol\`o Cribiori,}  
\author[b]{Renata Kallosh,}
\author[b]{Andrei Linde,}
\author[a]{Christoph Roupec}
\affiliation[a]{Institute for Theoretical Physics, TU Wien,\\ Wiedner Hauptstrasse 8-10/136, A-1040 Vienna, Austria}
\affiliation[b]{Stanford Institute for Theoretical Physics and Department of Physics,\\ Stanford University, Stanford, CA 94305, USA}
\emailAdd{niccolo.cribiori@tuwien.ac.at}
\emailAdd{kallosh@stanford.edu}
\emailAdd{alinde@stanford.edu}
\emailAdd{christoph.roupec@tuwien.ac.at}
\notoc

\abstract{We describe several applications of the mass production procedure proposed in \cite{Kallosh:2019zgd} to stabilize multiple moduli in a dS vacuum, in supergravity models  inspired by string theory. The construction involves a small downshift of an initial supersymmetric Minkowski minimum to a supersymmetric AdS minimum, and a consequent small uplift to a dS minimum. Our  type IIA examples include dS stabilization in a 7-moduli  model with $[SL(2, \mathbb{R})]^7$ tree level symmetry, and its simplified version, a 3-moduli STU model. In these models, we use uplifting anti-D6 branes. In type IIB models, we present  2- and 3-moduli examples of stable dS vacua in CY three-folds, with an uplifting anti-D3 brane. These include K3 fibration models, a CICY model and a multi-hole Swiss cheese model. We also address the issue whether this procedure is limited to a very small parameter range or if large deviations from the progenitor Minkowski vacuum are possible.}

\begin{document}

\maketitle

   \newpage

 \tableofcontents{}

  \parskip 8pt

 \section{Introduction}\label{intro}
 
In this paper, we continue the investigation of the mechanism of mass production of stable dS vacua in type IIA string theory, originally proposed in  \cite{Kallosh:2019zgd}, and we generalize it to dS stabilization in type IIB string theory. 
 
The main results of \cite{Kallosh:2019zgd} can be summarized as follows. At the level of $d=4$, $\cN=1$ supergravity it was established that there is a simple, systematic procedure to construct stable de Sitter minima in  theories with many moduli and many different \K\ potentials and superpotentials. The necessary conditions for this are the following:
 \begin{enumerate}
  \item The model requires a progenitor:  a supersymmetric Minkowski minimum without flat directions. This happens if $\partial_i W=0$ and $W=0$ at a finite point (or a series of disconnected points) in the moduli space, $W$ being the superpotential of the model. 
  \item The progenitor model needs a deformation: a downshift to a supersymmetric AdS minimum, via a parametrically small deformation of $W$. This can be achieved by adding to $W$ a small constant (or a small function) $\Delta W$. 
  \item For the next stage of uplifting to a dS minimum, a nilpotent  chiral multiplet has to be added to the original model. If the supersymmetry breaking scale after the uplift is parametrically small (we will be more precise about this during our presentation), then the resulting potential inherits the shape of the original potential in the vicinity of the Minkowski minimum, i.e.  the final dS state is also a minimum.
\end{enumerate}

Note that, if our only goal was to construct stable dS vacua, we could directly uplift a supersymmetric Minkowski vacuum to a stable dS vacuum. However, if one produces a dS vacuum with an extremely small cosmological constant, such as $V_{dS} \sim 10^{{-120}}$, in this way, one would arrive at an almost exactly supersymmetric state, which is very different from the one we observe.  The role of the downshift stage is thus to disentangle the strength of the supersymmetry breaking and the smallness of the cosmological constant. 

A technically separate but very important issue is the relation of this new $d=4$, $\cN=1$ supergravity multi-moduli construction of dS minima to string theory. In particular, the relation of the new construction to $d=10$ supergravity including local sources, branes, orientifolds, anti-branes as well as non-perturbative corrections via gaugino condensates and/or instanton corrections needs a better understanding. In general, string theory inspired $d=4$, $\cN=1$ supergravity multi-moduli models have to satisfy a number of additional conditions, besides the ones specified above, which are necessary and sufficient for the existence of dS minima in supergravity.

In the present work, our examples of  mass production of dS vacua will be based on a choice of \K\ potential $K$ and  superpotential $W$  motivated by string theory.  Moreover, the introduction of a nilpotent chiral superfield, an ingredient that we add to the standard supergravity description at the third step, will be motivated by the presence of uplifting anti-D3  \cite{Ferrara:2014kva,Kallosh:2014wsa} and anti-D6 branes \cite{Kallosh:2018nrk,Cribiori:2019bfx}  in type IIB and type IIA models, respectively.
In \cite{Kallosh:2019zgd}, the general procedure of a mass production of dS vacua in $d=4$ supergravity was designed in a way that can be applied both to type IIA as well as to type IIB string theory inspired models, since $d=4$, $\cN=1$ supergravity does not differentiate between them. However, examples presented in  \cite{Kallosh:2019zgd}  involved only the choices of $K, \,W$ and uplifting related to the type IIA case, which explains why the title of \cite{Kallosh:2019zgd} had `type IIA' in it. Here, we will show explicitly that the general results which can be found in section 5 of \cite{Kallosh:2019zgd} remain valid also in application to type IIB examples.

If we were interested in stable dS minima in supergravity, without relation to string theory, we could begin with a simple superpotential  of the type $W(\Phi) =(\Phi-\Phi_{0})^{2}$, which produces a potential with a supersymmetric Minkowski minimum at $\Phi = \Phi_{0}$, where the superpotential and its derivative vanish. More generally, one could consider any superpotential $W(\Phi^{i})$ depending on $n$ different moduli. Independently of the \K\ potential, the universal conditions for the existence of a supersymmetric Minkowski state at $\Phi^{i} = \Phi_{0}^{i}$ are 
\be
 W = 0 \ , \qquad  D_i W=0 \ ,
\label{cond0}\ee
which imply $\partial_i W= {\partial W\over \partial \Phi^{i}} {\big |_{\Phi^{i} = \Phi_{0}^{i}}}=0$. In essence, any superpotential which has a local extremum (a minimum, a maximum, or a saddle point, but not a flat direction) at some point $\Phi_{0}^{i}$, namely $\partial_i W=0$, will satisfy both conditions \eqref{cond0}, if one subtracts from $W(\Phi^{i})$ its value at $\Phi_{0}^{i}$. The important remaining issue is then to find string theory motivated superpotentials of this type. 

The simplest non-perturbative superpotential 
\be
W_{{\rm KKLT}} = W_{0} + A e^{\rmi a T},
\label{cond}\ee
used in the KKLT string theory construction \cite{Kachru:2003aw}, does not satisfy the required condition $\partial_T W=0$ at any finite value of $T$. Nevertheless, one can successfully construct stable dS vacua in type IIB models using such a  superpotential, together with an uplift contribution, from a supersymmetric AdS state \cite{Kachru:2003aw}. Recently, it was shown that one can use this class of superpotentials to stabilize type IIA theory as well  \cite{Cribiori:2019bfx}. 
However, in order to use the three-stage mass production mechanism, which starts from a supersymmetric Minkowski vacuum, one should first generalize this class of superpotentials, along the lines of \cite{Kallosh:2004yh,BlancoPillado:2005fn,Kallosh:2011qk}.

The characteristic feature of non-perturbative superpotentials used in \cite{Kallosh:2004yh,BlancoPillado:2005fn,Kallosh:2011qk} and in  \cite{Kallosh:2019zgd}, which we will also use in this paper,  is that they are of the racetrack type. The general feature of racetrack superpotentials is that they can be used for all directions in the moduli space, based on U-duality, as suggested in  \cite{Cribiori:2019bfx}.  Starting with a superpotential with two such exponents for each modulus,
\be
W=W_0 + \sum_i \bigl(A_i\, e^{\rmi a_i \Phi^i}- B_i\, e^{\rmi b_i \Phi^i}\bigr) \ ,
\ee
 one can find a progenitor model with a supersymmetric Minkowski minimum with the superpotentials satisfying \eqref{cond0}, as shown in \cite{Kallosh:2004yh,Kallosh:2019zgd}. If the the remaining conditions 1-3 are satisfied, the existence of a metastable dS minimum is  guaranteed.

An advantage of the three-stage mechanism based on racetrack superpotentials is that one can strongly stabilize supersymmetric Minkowski states, making the stabilizing barriers in the moduli potentials extremely high, even in the absence of supersymmetry breaking  \cite{Kallosh:2004yh,Kallosh:2011qk}. In this case, the subsequent  small deformations of the potential during the downshift and uplift do not affect the stability of the slightly downshifted and uplifted supersymmetric Minkowski minimum. The corresponding stability criterion is that the gravitino mass related to the downshift and uplift should remain parametrically smaller than the masses of the string theory moduli in the Minkowski vacuum \cite{Kallosh:2004yh,BlancoPillado:2005fn,Kallosh:2011qk,Kallosh:2019zgd}. See section 5 of \cite{Kallosh:2019zgd} for a detailed discussion.

It is interesting that the original racetrack superpotentials were studied in the context of moduli stabilization in string theory a decade before the discovery of dark energy, see for example  \cite{Krasnikov:1987jj,Taylor:1990wr,Casas:1990qi,deCarlos:1992kox,Kaplunovsky:1997cy}. During that time, the focus was on Minkowski vacua with a vanishing cosmological constant and broken supersymmetry, such that $V=0$, $DW\neq 0$ and $W\neq 0$. We would like to stress here that the authors of   \cite{Krasnikov:1987jj,Taylor:1990wr,Casas:1990qi,deCarlos:1992kox,Kaplunovsky:1997cy} made the following key observation: in the context of moduli stabilization in string theory   {\it the racetrack superpotentials, in general, are expected to have  many exponents}. This was explained in \cite{Kaplunovsky:1997cy} as follows.

A semi-realistic compactified string theory model is expected to have four spacetime dimensions, $\cN=1$ supersymmetry and a large gauge symmetry $G = \prod _a G_a$,  including the $SU(3) \times SU(2) \times U(1)$ of the Standard Model, as well as additional `hidden' factors. In particular, the non-perturbative string theory or F-theory description should allow for a large number of such hidden sectors. Each of these sectors might produce an exponential in the superpotential, depending on some moduli related to gauge-group couplings. Specific examples of superpotentials with several gaugino condensates in string theory were studied in \cite{Taylor:1990wr,Casas:1990qi,deCarlos:1992kox}, for the product of non-abelian gauge groups like $G_1\otimes G_2\otimes \cdots \otimes G_p$, in general, and for $SU(M)\otimes SU(N) \otimes \cdots \otimes SU(Q)$, in particular. For example, the gauge group $SU(N) \otimes SU(M)$ leads to a sum of exponentials, where  both pre-exponentials $A$ and $B$ are nonzero, while $a ={2\pi\over N}$  and $b = {2\pi\over M}$.

Note also that the five types of superstring theories: Type I, Type IIA, Type IIB, Heterotic SO(32) and Heterotic $E_8\times E_8$ are believed to be all linked by non-perturbative dualities,  in this way representing a unifying M-theory.  See e.g. Fig. 26 in \cite{Palti:2019pca} showing the  parameter space of M-theory. In the context of the non-perturbative string theory, very large gauge groups are possible, like $E_8\times E_8$ in the heterotic case. In \cite{Kaplunovsky:1997cy}, in the context of non-perturbative superpotentials, it is found that F-theory compactifications have 251 simple gauge group factors of total non-abelian rank of $302896$.  Therefore, many gaugino condensates and many exponents for each of the moduli in the superpotential are expected at the non-perturbative level in string theory.

The reason why we would like to stress this key feature of racetrack potentials, namely the fact that they are expected to have many exponents, is that we can easily find stable dS minima starting with the Minkowski progenitor model with at least two exponents for each of the moduli, and it is even easier to find it with many exponents.

In addition to many possible gaugino condensates, the existence of many world-sheet instantons and Euclidean brane instantons suggest significantly more possibilities for the racetrack potentials. 
Another reason to point out that many exponential contributions to $W$ are possible, is to stress that there are many possible choices of parameters, in these models, allowing stable dS states. This, in turn, gives yet another motivation  why we call the construction a `mass production of dS vacua'.

The original, two-exponents KL scenario  \cite{Kallosh:2004yh} was initiated due to   cosmological considerations. Indeed, in the simplest versions of the KKLT scenario  the height of the barrier stabilizing the dS vacuum is proportional to the square of the gravitino mass. This could lead to vacuum destabilization during the cosmological evolution in the early universe with a large Hubble constant $H \gtrsim m_{3/2}$    \cite{Kallosh:2004yh}. A similar problem appears in the large volume stabilization (LVS) models for $H \gtrsim m^{3/2}_{3/2}$ \cite{Conlon:2008cj}. This problem was addressed in the KL model, where the vacuum can be strongly stabilized even for very small gravitino mass \cite{Kallosh:2004yh,BlancoPillado:2005fn,Kallosh:2011qk}. It is therefore interesting that  generalized versions of the KL scenario  lead  to  mass production of stable dS vacua in type IIA theory \cite{Kallosh:2019zgd}.

This paper is organized as follows:
In section \ref{KL}, we review the original KL model and we present its generalization to the multi-moduli and multi-exponentials cases.
In section \ref{stability}, we present some details on the properties of the mass matrix in our scenario, which will be consequently confirmed in numerical examples.
We will further develop this scenario  in section \ref{sec:IIAexamples} of this paper by describing stabilization in a seven-moduli $ \Phi^i=   \{S,  T_1, T_2, T_3, U_1, U_2, U_3\}$ model and in its  simplified version, a three-moduli $ \Phi^i=  \{S, T, U\}$ model, with an uplifting anti-D6 brane. Then, in section \ref{IIB},  we will show that this mechanism can be applied to the stabilization of moduli in type IIB theory  with more  complex  \K\ potentials. In section \ref{big}, we describe models with dS minimum which are valid even for strong deviation from the progenitor models. In section \ref{concl}, we summarize our results. More technical details can be found in the appendix \ref{appA}.

\section{The KL model  and its generalizations} \label{KL}

In this section we first review the KL model in its original formulation \cite{Kallosh:2004yh}, together with the addition of a nilpotent chiral multiplet \cite{Ferrara:2014kva}. Then, we provide some generalizations, which can be directly employed in the mass production mechanism.

\subsection{The basic KL model}\label{KLBASIC}

The simplest string theory related realization of the general three-stage mechanism described 	in the introduction is given by a \K\ potential of the type
\be 
K = -3 \ln (-\rmi(\Phi -\bar \Phi)) + X\bar X, 
\ee
where $X$ is a nilpotent chiral multiplet,  which does not contain scalars \cite{Rocek:1978nb}, 
and a racetrack superpotential  with two exponents 
\be
W(\Phi ,X)  =W_\text{KL}(\Phi ,X) + \Delta W + \mu^2 X\ ,
\label{adssup}
\ee
with
\be
W_\text{KL}(\Phi ,X)  =W_{0} +Ae^{\rmi a\Phi}- Be^{\rmi b\Phi} \,. \label{adssupX}
\ee
We have also already introduced the down shift  as $W_0 \to W_0 + \Delta W$ and the uplift via the term $W_{\rm uplift} = \mu^2 X$. To find a supersymmetric Minkowski minimum in this model, with $\Delta W = 0$ and $\mu^2= 0$, one should find a point $\Phi_{0} =  \phi_{0} + i \theta_{0}$ in moduli space such that (we  also set $X=0$ as long as $\mu^2=0$)
\be\label{susy} 
W(\Phi_{0})=0 \ , \qquad D_\Phi W|_{\Phi=\Phi_{0}}=0 \ .
\ee
Considered together, these conditions imply
\be\label{susy1} 
\partial_\Phi W|_{\Phi=\Phi_{0}}=\rmi aAe^{\rmi a\Phi _{0}} - \rmi bBe^{\rmi b\Phi _{0}}= 0 \ 
\ee
and are satisfied, in particular, for $\theta_0=0$ and
\be\label{fullcompl}
\Phi_{0} = \phi _{0} = {\rmi \over a-b} \ln  {a\,A\over b\,B}\ , \qquad W_0=  -A \left({a\,A\over
b\,B}\right)^{a\over b-a} +B \left ({a\,A\over b\,B}\right) ^{b\over b-a}  \ .
\ee
The supersymmetric vacuum \eqref{fullcompl} exists for any value of $a, A, b, B$ satisfying the conditions $a > b$ and $aA > bB$ (or $a < b$ and $aA <bB$). 

The mass squared of both real and imaginary components of the field $\Phi$ at the supersymmetric Minkowski minimum is given by
\be\label{klmod}
m^{2}_{\Phi} =  {2\phi_{0}\over 9} \bigl(W''(\phi _{0})\bigr)^{2} = {2\over 9} a A b B  (a -b )\left(a A \over b B  \right)^{-{a +b \over a -b }}  \ln {a\, A\over b\, B}\ .
\ee
 Adding a small correction $\Delta W$ to $W_{0}$ makes this minimum a supersymmetric AdS. In particular, in order to preserve supersymmetry, the position of the minimum has to shift at leading order by an amount of
 \be
 \label{DeltaPhi}
 \Delta \Phi =- \frac{K_\Phi \Delta W}{\partial_\Phi^2W},
 \ee
 and the value of the potential at its minimum is downshifted, from $V = 0$, to
 \be
V_{\rm AdS}  = -{3(\Delta W)^2\over 8 \phi_{0}^{3}}= - {3\over 8} \left({a - b \over  \ln \left({a A\over b B}\right)}\right)^{3}\, (\Delta W)^2 \ .
\label{vads}
\ee 
Eventually, one finds that the uplift of the AdS state with scalar potential \eqref{vads} with small $\Delta W$, to a dS state with a nearly vanishing cosmological constant, $V_{dS} \sim 10^{{-120}}$,   can be obtained with an uplifting parameter 
\be
\mu^{4}\approx  3\, (\Delta W)^2 \ .
\ee
In particular, the gravitino mass after uplift is given by 
\be
m^{2}_{3/2} =  e^{K}(\Delta W)^2 = { (\Delta W)^2\over 8 \phi_{0}^{3}} ={|V_{AdS}|\over 3}= { \mu^4\over 24 \phi_{0}^{3}} \ .
\ee
%Once again, for sufficiently small values of $\Delta W$, the required uplifting does not affect the strong stabilization of the original Minkowski vacuum state. An approximate criterion  of applicability of these conclusions and the estimates is the requirement that the gravitino mass should be much smaller than other mass parameters describing the potential, see a detailed discussion of this issue in Sect. \ref{mass}. 

\subsection{A single-field generalization with many exponents}\label{manyexp}
In the spirit of the earlier work in  \cite{Casas:1990qi,Kaplunovsky:1997cy}, we consider now a more general multi-exponent racetrack superpotential: 
\be
W(\Phi ,X)  =W_{0} +\sum\limits_{k=1}^n A^{k}e^{\rmi a^{k}\Phi}  +   \Delta W + \mu^2 X\ .
\label{adssupn}
\ee
The  condition  \eqref{susy1} for a supersymmetric Minkowski vacuum becomes then
\be
\partial_\Phi W|_{\Phi=\Phi _{0}}  =\rmi \sum\limits_{k=1}^n a^{k }A^{k}e^{\rmi a^{k}\Phi _{0}} = 0\ .
\label{susy2}
\ee

One may solve this equation, together with $W(\Phi _{0}) = 0$, to find the position of the supersymmetric Minkowski minimum  $\Phi=\Phi _{0}$ as a function of the parameters. Alternatively, it is possible to find a set of parameters which ensure that a given fixed point  $\Phi_{0}$ corresponds to a supersymmetric Minkowski vacuum state. This second method is often much simpler, and it clearly demonstrates the existence of supersymmetric Minkowski vacua.

Indeed, one can, for example, take arbitrary values for $n$ coefficients $a^{k}$ and for the first $n-1$ coefficients $A^{k}$. The condition \eqref{susy2} is then satisfied, if one does not take the $n$-th coefficient $A^n$ to be independent, but rather given by
\be
A^{n} = \rmi \,  {\sum\limits_{k=1}^{n-1} a^{k }A^{k}e^{\rmi a^{k}\Phi_{0}}\over a^{n }e^{ \rmi a^{n}\Phi_{0}}} \ .
\ee
This fixes all parameters in the sum $\sum\limits_{k=1}^n A^{k}e^{\rmi a^{k}\Phi}$. As  the remaining step, one chooses 
\be
W_{0} = -\sum\limits_{k=1}^n A^{k}e^{\rmi a^k \Phi _{0}} \ ,  
\ee
which ensures that $W(\Phi _{0}) = 0$ for $ \Delta W = \mu^2 = 0$, i.e. before the downshift and uplift. Thus, for any set of $2n$ free parameters, namely $\Phi _{0}=\phi_0$, $n$ coefficients $a^{k}$ and  the first $n-1$ coefficients $A^{k}$, one can always find two remaining parameters $A^{n}$ and $W_{0}$ such that the theory has a supersymmetric vacuum state at $\Phi = \Phi _{0}$.

There are also other ways to reach a similar goal. For example, instead of calculating $W_{0}$, one can fix it from the very beginning (e.g. take $W_{0} = 0$), and  calculate the value of $a^{n}$, or $A^{n-1}$, required for satisfying the conditions $W(\Phi _{0}) = \partial_\Phi W|_{\Phi=\Phi _{0}} = 0$.

\subsection{A generalization with many moduli and many exponents }\label{manymoduli}

The previous model with a multi-exponent racetrack superpotential for one field $\Phi$ can be directly generalized to the multi-field case. Indeed, consider the set of models with $m$ moduli $\Phi^i$, and assume, as suggested in \cite{Casas:1990qi,Kaplunovsky:1997cy}, that each of them gives many exponential contributions  to the superpotential. Concretely, let us suppose that the superpotential is given by
\be\label{racen}
W=W_0 + \sum_{i=1} ^{m} \sum_{k=1} ^{n} A_i^k e^{\rmi a_i^k \Phi^i} + \Delta W + \mu^2 X \ .
\ee
Then, the equation  $\partial_i W\equiv {\partial W\over \partial \Phi^{i}}  =0$  is satisfied for each modulus $\Phi^i$, for any given value of the parameters $\Phi _{0 i }$,  $a^{k}_i $, and for the first $(n-1)$ coefficients $A^{k}_i$, if one  takes 
\be
A^{n}_i = \rmi \, {\sum\limits_{k=1}^{n-1} a^{k }_i A^{k}_i e^{\rmi a^{k}_i \Phi _{0}^i}\over a^{n }_ie^{\rmi a^{n}_i \Phi _0^i}} \ .
\ee
To satisfy $W(\Phi _{0i})=0$, we also need
\be
W_{0} = -  \sum_{i=1} ^{m} \sum\limits_{k=1}^n A^{k}_i e^{\rmi a^k_i \Phi _0^i } \ .
\ee

Thus, for any set of $2mn$ free parameters (for each $i = 1,...,m$ one can take various values of $\Phi _0^i=\phi_{0}^i$, $n$ coefficients $a^{k}_i $, and  the first $n-1$ coefficients $A^{k}_i$), one can always find $m+1$ remaining parameters $A^{n}_i$ and $W_{0}$ such  that the theory has a supersymmetric vacuum state at $\Phi^{i}  = \Phi_0^i$. Alternatively, instead of calculating $W_{0}$, one can fix it from the very beginning (e.g. take $W_{0} = 0$ or any other desired value), and  calculate the value of $a^{n}_i$, or $A^{n-1}_i$, required for satisfying the conditions $W = 0, \, \partial_i W=0$ at  some point $\Phi _{0 i}$.

The general conclusion is that in a theory with any number $m$ of fields $\Phi _i$ with racetrack potentials containing $n$ exponents, with $n\geq 2$  for each of the fields, one can always find a large number of supersymmetric Minkowski vacua depending on $2m n$ free parameters. Since the values of the parameters $A_{i}^k$ depend on the properties of compactification, one may expect that the total number of different supersymmetric Minkowski vacua can be extremely large. 

Each of these vacua can be downshifted to a supersymmetric AdS and uplifted to a stable dS vacuum in many different ways. In particular, in most of the examples to be considered in this paper we will make a downshift by adding a small constant $\Delta W$ to the superpotential, but the main results will not be qualitatively affected if instead of that we add any sufficiently small function $\Delta W(\Phi^i)$.

\section{Properties of the mass matrix at the dS minimum }\label{stability}

In this section, we review  some general properties of the mass matrix that are important in order to study the stability of vacua configurations. We refer the reader to Sec. 5 of \cite{Kallosh:2019zgd} for a more detailed discussion. Then, we focus our analysis to specific examples of dS vacua in type IIA and type IIB theory. We inspect the corresponding mass matrices and we compare their properties.  

\subsection{General properties of the mass matrix at the dS minimum }\label{generalstability}

In the class of models we are going to consider, the moduli space is made up of complex scalar fields $\{\Phi^i, \bar \Phi^{\bar \imath}\}$, living inside chiral and antichiral multiplets respectively. To this set up, we add then a nilpotent chiral multiplet $X$,  which can be conveniently used to describe the uplifting contribution to the scalar potential given by an anti-Dp-brane in string theory. Indeed, thanks to the nilpotent constraint on $X$, the total scalar potential, including the uplift, can be given by the standard supergravity formula
\be
V^{dS} = e^{K^{dS}} (|D_I W^{dS}|^2  - 3 |W^{dS}|^2) \ ,
\ee
where the index $I$ runs over all the chiral multiplets $\Phi^I=\{\Phi^i,X\}$ and where we define
\be
\label{genmodel}
K^{dS}= K(\Phi, \bar \Phi)+ K_{X, \bar X}(\Phi, \bar \Phi)  X\bar X \ , \qquad W^{dS} = W(\Phi)+ \mu^2 X \,,
\ee
with $X^2=0$.

At an extremum, $\partial _i V=0$, the scalar mass matrix in supergravity takes the following form
\begin{equation}
\mathcal{M}^2=\, \left ( \begin{array}{cc} V_{i\bar \jmath} & V_{ij} \cr V_{\bar \imath \bar \jmath} & V_{\bar \imath j} \cr  \end{array} \right ) .
\label{massMn}
\end{equation}
In particular, the upper left corner gives the holomorphic-anti-holomorphic part of the second derivative of the potential at the minimum, $V_{i\bar \jmath}=\partial_i \partial_{\bar \jmath}V$, the upper right corner has the holomorphic-holomorphic part of the second derivative of the potential at the minimum, $V_{ij}=\partial_i \partial_{j}V$, and so on. The exact expressions for each entry as a function of $K$ and $W$ were derived in \cite{Denef:2004ze}.

To analyse the vacua in supergravity, we find it convenient to use the notation introduced in   \cite{Kallosh:2000ve} and used in 
\cite{Kallosh:2019zgd}. We define the covariantly holomorphic gravitino mass $m(\Phi,\bar\Phi)\equiv e^\frac{K}{2}W$, together with its \K\ covariant derivative $m_i\equiv D_i m = \partial_i m+\frac12 (\partial_i K)m=e^\frac K2 D_i W $. In a supersymmetric Minkowski vacuum, which is one of the cases of interest in our discussion, the explicit expression for the mass matrix can be considerably simplified. At a Minkowski minimum, the conditions $W=0$ and $D_i W=0$ imply respectively $m=0$ and $m_i=0$ and the quantities in \eqref{massMn} reduce to
 \bea \label{Min}
&V_{i\bar \jmath}^{Mink}&=  {\it m}_{ik}g^{k\bar k} \,\bar {\it m}_{\bar k \bar \jmath}\ \bigg|_{m_i=m=0}\ , \\
\cr
& V_{ij}^{Mink} \bigg|_{m=0}&= 0  \ ,
\label{Min1}\eea
where we define the chiral fermion mass matrix $m_{ij}\equiv D_i D_j m$. This, in turn, can be read directly from the fermionic bilinear term in the action,
\be
e^{-1}\mathcal{L}\supset-{1\over 2} {\it m}_{ij}\bar \chi^i P_L \chi^j +h. c.,
\ee
see for example \cite{Freedman:2012zz}.

An important consequence of having unbroken supersymmetry in a Minkowski vacuum is that, within each chiral multiplet, there is a mass degeneracy between scalars and pseudoscalars.  Indeed, by switching to the real basis, $\Phi^i= \phi^i+i \theta^i$, it follows from \rf{Min1} that \footnote{ We recall that, when passing from the complex basis $\{\Phi^i, \bar \Phi^{\bar \imath}\}$ to the real basis $\{\phi^i, \theta^i\}$, the derivatives change as 
\be
\partial_i = \frac12 \frac{\partial}{\partial \phi^i} - \frac i2 \frac{\partial}{\partial \theta^i}, \qquad \partial_{\bar \imath} = \frac12 \frac{\partial}{\partial \phi^i} +\frac i2 \frac{\partial}{\partial \theta^i}.
\ee
}
\be
{\rm Re} \, V_{ij}^{Mink}= {\rm Re} \, \partial_i \partial_j V^{Mink}={1\over 4} \Big( {\partial \over\partial  \phi^i} {\partial \over \partial \phi^j}-  {\partial \over \partial \theta^i} {\partial \over \partial \theta^j} \Big ) V^{Mink}=0 \ .
\ee
This means that, at the supersymmetric Minkowski minimum, the masses of the scalars $\phi^i$ are equal to the masses of the pseudoscalars $\theta^i$.

Another configuration of interest in the present work is a supersymmetric AdS vacuum. It corresponds to the second step of the procedure outlined in the Introduction and is obtained by shifting $W$, as explained in the previous section. It is given by $m_i=0$ but $m\neq 0$ and therefore we now have: 
\begin{equation}
%V_{a\bar b}^{AdS} &=(m_{ac} g^{c\bar d}\bar m_{\bar b\bar d}-2 g_{a \bar b}m \bar m)\bigg |_{m_a=0},\\
\label{VabAdS}
V_{ij}^{AdS}= -m_{ij} \, \bar  m\bigg |_{m_i=0}\ .
\end{equation}
The expression for $V_{i\bar \jmath}^{AdS} $ is also different from the previous case and can be found in \cite{Kallosh:2019zgd}. The fact that \eqref{VabAdS} is now non-vanishing, due to a non-vanishing $W$ in the vacuum, results in a difference between the masses of scalars $\phi^i$ and masses of the pseudoscalars $\theta^i$
\be
{1\over 4} \Big( {\partial \over\partial  \phi^i} {\partial \over \partial \phi^j}-  {\partial \over \partial \theta^i} {\partial \over \partial \theta^j} \Big ) V^{AdS}=- {\rm Re} \Big ( m_{ij} \, \bar m\Big )\bigg |_{m_a=0}.
\ee

The third configuration we look at is a dS minimum, obtained by uplifting the supersymmetric AdS vacuum discussed previously. As discussed at the beginning of this section, at the third step of the procedure we enlarge the set of chiral multiplets $\Phi^i$ by adding a nilpotent chiral multiplet X. Since generically $m_I\neq 0$ in a dS vacuum\footnote{In particular, we have to require at least $m_X\neq 0$ in order for the description in terms of the nilpotent $X$ to be consistent.}, the mass matrix is now given by
\be
V_{ij}^{dS} =-    {\it m}_{ij}  \bar {\it m} + m_{ijI} \bar m^I  ,
\label{hh}
\ee
where we defined  $m_{ijI}\equiv D_i D_j D_I  m$ and we raised the indices with the \K\ metric, namely $\bar m^I = g^{I\bar J}\bar m_{\bar J}$. Again, the mass degeneracy between scalars and pseudoscalars present in Minkowski is broken and moreover, with respect to the AdS case, there are also additional contributions, due to supersymmetry breaking terms $m_I\neq 0$,
\be
{1\over 4} \Big( {\partial \over\partial  \phi^i} {\partial \over \partial \phi^j}-  {\partial \over \partial \theta^i} {\partial \over \partial \theta^j} \Big ) V^{dS}=- {\rm Re} \Big ( m_{ij} \, \bar m - m_{ijI} \bar m^I\Big ).
\ee

We now proceed to discuss the stability of the vacua in these three situations, by analysing the positiveness of the associated mass matrix. In the Minkowski case, it can be directly deduced from \eqref{Min} and \eqref{Min1} that the vacuum is stable and without flat directions (condition 1 in the Introduction), as long as all the fermions in the theory are massive \cite{Kallosh:2019zgd}. The AdS case, in general, is more involved. However, if we assume that the AdS vacuum is constructed by downshifting the previous Minkowski one, and if the perturbation $\Delta W$ of the superpotential is parametrically small (condition 2 in the Introduction), then the AdS configuration is again stable and there are no flat directions  \cite{Kallosh:2019zgd}. 

By uplifting the AdS vacuum to a dS vacuum with a parametrically small breaking of supersymmetry, namely $|m_I|^2 \approx 0$, (condition 3 in the Introduction), one finds that, up to parametrically small terms, the upper left part of the mass matrix is positive definite, being the square of the fermion mass matrix, which has no flat directions by assumption,
 \be
V_{i\bar \jmath}^{dS}\approx  {\it m}_{ij}g^{j\bar k} \,\bar {\it m}_{\bar k\, \bar \jmath} > 0 \ .
\label{dSmassN}\ee
See also the appendix \ref{appA} for more details.
Similarly, the holomorphic-holomorphic part of the second derivative of the potential, namely the right upper corner of \rf{massMn}, is approximately zero, if the conditions 1-3 in the Introduction are satisfied:
 \be
V_{ij}^{dS}\approx  0 \, \qquad  \Rightarrow   \qquad m^2_{\phi^i \phi^j} \approx  m^2_{\theta^i \theta^j}\ .
\label{dSmassCorner}
\ee
In other words, the mass degeneracy between scalars and pseudoscalars, if present at the beginning, is then approximately preserved through the entire three-steps procedure.

Before concluding, one comment is in order. It was explained in  \cite{Kallosh:2019zgd} that the uplifting stage preserves the stability of the AdS minimum,  such that the dS minimum remains stable, under the condition that the total supersymmetry breaking and the mass of   the  gravitino are  small 
\be
m_\chi^2 \gg |m_I^2|\, \qquad m_\chi^2 \gg  m_{3/2}^2\, ,  \qquad |m_I^2| > 3 m_{3/2}^2\, .
\ee
Here the masses of the physical fermions $m_\chi^2$ are approximately of the same order of those of the scalars. During the uplift from a supersymmetric AdS vacuum, the quantity $D_iW=0$ can change only due to the shift of the position of the minimum. As shown in  \cite{Kallosh:2019zgd}, this shift is parametrically small and, as a consequence, one finds the hierarchy  $ |m_X|^2 \gg |m_i^2|$. We stress that this hierarchy is predicted by the procedure and not assumed a priori.

To summarize, since the moduli space metric $g_{I\bar J}$ in \eqref{dSmassN} is positive definite, under the conditions 1-3 of the introduction the dS mass matrix can be (symbolically) expressed as
\begin{equation}
\mathcal{M}^2=\, \left ( \begin{array}{cc} V_{i\bar \jmath} & V_{ij} \cr V_{\bar \imath \bar \jmath} & V_{\bar \imath j} \cr  \end{array} \right )\approx m_f  m^\dagger_f  > 0 \, ,
\label{massMapprox}
\end{equation}
where $m_f$ is the fermionic mass matrix. The stability of the dS vacuum is thus manifest.

We conclude the discussion by stressing that this outcome, namely the existence of stable dS minima, is valid for a large class of  superpotentials and with no restrictions on the \K\, potential. The only restriction on the superpotential is shown in \eqref{cond0}: the progenitor model must have a local extremum of the superpotential, without flat directions. The rest follows, if the conditions 1-3 of the Introduction are satisfied.

\subsection{Special choices in type IIA}

The discussion so far was valid for a generic \K\ potential and under mild assumptions on the superpotential. By imposing additional restrictions on $K$ and $W$, we can derive more specific properties for the mass matrix in \eqref{massMapprox}. Therefore, we devote this subsection to the study of models with type IIA motivated 	\K\ potential and superpotential, while in the next subsection we will focus on type IIB.

A large class of type IIA inspired models was studied in \cite{Kallosh:2019zgd}. They have a superpotential with two exponents for each modulus $\Phi^{i}$ and a \K\ potential of the form
\be\label{KLsup}
W = W_{0}   +\sum\limits_{i=1}^m  A_{i} e^{\rmi a_{i}\Phi^{i}} -  \sum\limits_{i=1}^m  B_{i} e^{\rmi b_{i}\Phi^{i}}  \ , \qquad K= -\sum\limits_{i=1}^m  N_{i}\, \ln\left(-\rmi (\Phi^{i} - \bar{\Phi}^{\bar \imath})\right) \ ,
\ee
where $W_0$,  $A_i$, $a_i$, $B_i$, $b_i$ and $N_i$ are real parameters. In \cite{Kallosh:2019zgd}, section 3.1, it was shown that for this class of models the Minkowski mass matrix in the basis of real fields $\{\phi^i, \theta^i\}$ is diagonal 
\be
\begin{pmatrix}
V_{\phi^i \phi^j}      &  V_{\phi^i \theta^j}  \\
V_{\theta^i \phi^j}      &  V_{\theta^i \theta^j}  
\end{pmatrix} \Biggl |_{Mink }=   \begin{pmatrix}
m^2_{\phi^i \phi^i}      &   0  \\
0     &  m^2_{\theta^i \theta^i}  
\end{pmatrix} \ .
\ee
In particular, along each direction in moduli space, we observe the mass degeneracy between scalars and pseudoscalars, as explained above
\be
m^2_{\phi^i \phi^i} = m^2_{\theta^i \theta^i} \ .
\ee
For example, in the case with seven moduli
\be
V_{\phi^i \phi^j} ^{Mink} =  V_{\theta^i \theta^j} ^{Mink} =
\left(\begin{array}{ccccccc}m_1^{\;2}& 0 & 0 & 0 & 0 & 0 & 0 \\0 & m_2^{\;2} & 0 & 0 & 0 & 0 & 0 \\0 & 0 & m_3^{\;2} & 0 & 0 & 0 & 0 \\0 & 0 & 0 & m_4^{\;2} & 0 & 0 & 0 \\0 & 0 & 0 & 0 & m_5^{\;2} & 0 & 0 \\0 & 0 & 0 & 0 & 0 & m_6^{\;2} & 0 \\0 & 0 & 0 & 0 & 0 & 0 & m_7^{\;2}\end{array}\right) . \label{Mink}\ee

When perturbing the Minkowski solution with a parametrically small $\Delta W$, in order to find a supersymmetric AdS vacuum, the mass matrix in the real basis remains block diagonal, at least for the class of models given by \eqref{KLsup}. Therefore, we have
 \be
\begin{pmatrix}
V_{\phi^i \phi^j}      &  V_{\phi^i \theta^j}  \\
 V_{\theta^i \phi^j}      &  V_{\theta^i \theta^j}  
\end{pmatrix} \Biggl |_{AdS  }=   \begin{pmatrix}
V_{\phi^i \phi^j} ^{Mink} +  \Delta_{\phi^i \phi^j}     &   0  \\
0     &  V_{\theta^i \theta^j}^{Mink} +   \Delta_{\theta^i \theta^j}
\end{pmatrix}  .
\label{corrAdS}\ee
We notice the possible appearance of non-diagonal terms in $ \Delta_{\phi^i \phi^j} $ and $ \Delta_{\theta^i \theta^j} $ which, however, are parametrically smaller than the terms on the diagonal $i=j$.

Finally, when uplifting this supersymmetric AdS vacuum to dS, we introduce additional terms in the potential of the form 
\be
V^{uplift}=V^{uplift} \left( -\rmi (\Phi^i - \bar \Phi^i)\right) \ .
\ee
In particular, these terms do not depend on $\phi^i$ and therefore cannot generate off-diagonal blocks in the mass matrix. Eventually, the mass matrix takes the form
\be
\begin{pmatrix}
V_{\phi^i \phi^j}      &   V_{\phi^i \theta^j}  \\
 V_{\theta^i \phi^j}      &  V_{\theta^i \theta^j}  
\end{pmatrix} {\Biggl |_{dS } }=   \begin{pmatrix}
V_{\phi^i \phi^j} ^{Mink} +  \Delta_{\phi^i \phi^j}  + \tilde \Delta_{\phi^i \phi^j}   &   0  \\
0     &  V_{\theta^i \theta^j}^{Mink} +   \Delta_{\theta^i \theta^j}
 + \tilde\Delta_{\theta^i \theta^j}
\end{pmatrix}.
\label{corrdS}\ee
The new all diagonal and non-diagonal terms in in $ \tilde \Delta_{\phi^i \phi^j} $ and $ \tilde \Delta_{\theta^i \theta^j} $ are parametrically smaller than the terms on the diagonal $i=j$. Therefore, the mass eigenvalues in the dS vacuum are approximately equal to the original Minkowski masses of the progenitor model and the masses of the scalars and pseudoscalars are approximately equal to each other.

\subsection{Special choices in type IIB}
\label{subsec:IIBgen}

The type IIB models we are going to consider will have the same form of the superpotential as in the IIA case, but they differ generically in the \K\, potential. Indeed, in CY examples, the \K\ potentials are more complicated: they are still functions of  $-\rmi (\Phi^i-\bar \Phi^{\bar \imath} )$, but not simple products of functions as in the type IIA case \rf{KLsup}. When discussing type IIB models in general, we will therefore denote $K$ and $W$ as
\be\label{KLsupB}
K= K(-\rmi (\Phi^{i} - \bar \Phi^i)), \qquad W = W_{0}   +\sum\limits_{i=1}^m  A_{i} e^{\rmi a_{i}\Phi^{i}} -  \sum\limits_{i=1}^m  B_{i} e^{ \rmi b_{i}\Phi^{i}}   \ .
\ee
These complications immediately lead to some changes in the Minkowski mass matrix. In particular, one finds that for these models the matrix is block diagonal, but not diagonal 
\be
\mathcal{M}^2=
\begin{pmatrix}
 V_{\phi^i \phi^j}      &   0  \\
0     &  V_{\theta^i \theta^j}  
\end{pmatrix} {\Biggl |_{Mink} }.
\label{nond1}\ee
For example, in the 3-moduli case we have a matrix where all entries can be non-zero:
\be
V_{\phi^i \phi^j}^{Mink} = V_{\theta^i \theta^j}^{Mink} = \left(\begin{array}{ccc}V_{11}  & V_{12}  & V_{13}  \\V_{21}  & V_{22}  & V_{23}  \\V_{31}  & V_{32}  & V_{33} \end{array}\right).
\label{nond}\ee
Nevertheless, it follows from \eqref{Min} that all of the eigenvalues of $V_{\phi^i \phi^j}= V_{\theta^i \theta^j}$ must be positive definite. In other words, the fact that in the models \eqref{KLsup} the mass matrix in the Minkowski minimum is diagonal, is a nice feature of our examples in the type IIA case. Meanwhile, in type IIB we will find  much more complicated mass matrices, already at the Minkowski level. The  mass matrices in dS will still be of the form shown in \eqref{corrdS}, however, the matrices $V_{\phi^i \phi^j}^{dS} \approx V_{\theta^i \theta^j}^{dS} $ inside \eqref{corrdS} will be highly non-diagonal, as shown in \eqref{nond}.

To summarize, after the downshift and uplift, the relation \eqref{dSmassN} is still valid and therefore we are constructing models with a stable dS minimum.

 \section{Type IIA examples} 
 \label{sec:IIAexamples}

We present now, in detail, several examples of string theory inspired models in which we apply the mass production procedure. We devote the present section to the type IIA case and the next one to type IIB.

First, we discuss a seven-moduli model, in which the kinetic term has a $[SL(2, \mathbb{R})]^7$ symmetry that is broken only by non-perturbative corrections. One can understand this model from a more fundamental perspective, either from M-theory/string theory or from maximal supergravity in $d=4$. However, given the fact that the moduli space has real dimension 14, the numerical analysis and the resulting plots are generically complicated. Therefore, for this specific model, we will only present a short summary of our results. Instead, a simplified version of such a seven-moduli model with just three moduli, namely where we identify $S$, $T_1= T_2, = T_3\equiv T$ and $U_1= U_2  = U_3\equiv U$, will be presented in more detail, including the relevant plots of the potentials.

\subsection{Seven-moduli  $[SL(2, \mathbb{R})]^7$ model}
The interest in a seven-moduli model in M-theory and type IIA string theory and in $d=4$, $\cN=8$ supergravity was renewed recently, in the context of the future B-mode detection  \cite{Ferrara:2016fwe,Kallosh:2017ced}. Indeed, such a seven-moduli model revisited in these works, was described originally a long time ago, in \cite{Derendinger:2004jn,Villadoro:2005cu}. There, the effective $d=4$ supergravity for the seven main moduli of type IIA D6/O6 orientifolds was derived, starting with $d=10$ supergravity compactified on ${T^6\over Z_2 \times Z_2}$  in the presence of general fluxes. The seven moduli include one axio-dilaton $S$, three complex-structure $T$-moduli, $\{T_1,T_2,T_3\}$ and three \K\,  $U$-moduli, $\{U_1,U_2,U_3\}$. For convenience, we collect all of them in 
\be
 \Phi^i=\{S,  \, T_1,  \, T_2,  \, T_3,  \, U_1,  \, U_2,  \, U_3\}.
\ee

The \K\,  potential for this seven-moduli model is associated with a  $[SL(2, \mathbb{R})]^7$ symmetry, where the seven complex moduli are the coordinates of the  $\Big ({SL(2, \mathbb{R})\over U(1)}\Big ) ^7$ coset space,
 \be
K = -  \sum_i ^7 \log \left(-\rmi (\Phi^i-\bar \Phi^{\bar \imath})\right)\,.
\label{7m}\ee
As motivated in the Introduction, the superpotential which we will use in the present work is of the racetrack type, where the constant part $f_6 $  comes from the six-flux, similarly to  \cite{Cribiori:2019bfx},
\be
W = f_6 + \sum_i^7 \bigl( A_i e^{\rmi a_i \Phi^i}- B_i e^{\rmi b_i \Phi^i}\bigr)\,.
\label{7mW}
\ee

The origin of these seven complex fields from the $d=10$ geometry compactified on ${T^6\over Z_2 \times Z_2}$ can be found in \cite{Derendinger:2004jn,Villadoro:2005cu}. In   \cite{Ferrara:2016fwe}, it is also shown how to get the same model from M-theory/$d=11$ supergravity compactified on a particular $G_2$ manifold, as well as from $d=4$ maximal $\cN=8$ supergravity with duality symmetry
\be
E_{7(7)}( \mathbb{R})  \supset [SL(2, \mathbb{R})]^7.
\ee 
 The kinetic term for the seven scalars has an $[SL(2, \mathbb{R})]^7$ symmetry 
\be
{\cal L}_{kin} =  \sum_i ^7 {\partial \Phi^i \partial \bar \Phi^{\bar \imath}\over (\Phi^i-\bar \Phi^{\bar \imath})^2}.
\ee
Note that the non-perturbative terms in the superpotential, namely the exponential terms in \eqref{7mW}, break this symmetry. The kinetic term for each of the seven scalars corresponds to a unit size Poincar\'e disk, with squared radius $3\alpha=1$, after a Cayley transform
\be
3\alpha{\partial \Phi \partial \bar \Phi\over (\Phi-\bar \Phi)^2}\ \Rightarrow \ 3\alpha {\partial Z \partial \bar Z\over (1- Z\bar Z)^2}.
\ee
This is why the seven moduli  set up  is relevant for observational cosmology and $\alpha$-attractor models of inflation  \cite{Ferrara:2016fwe,Kallosh:2017ced}. Here, however, we are only interested is using this type IIA string theory model for an example of the moduli stabilization, as predicted in \cite{Kallosh:2019zgd}.

It is worth mentioning that one of the recent studies of the minima of the  $[SL(2, \mathbb{R})]^7$ model was performed in \cite{Blaback:2018hdo}. The authors there used the same \K\, potential as the one in \eqref{7m}, however, they did not include non-perturbative exponents in the superpotential, but only polynomial terms in the moduli, as suggested in  \cite{Villadoro:2005cu}.  Within this set up, they stabilized all of the moduli at the origin of the moduli space, $\Phi^i = \bar \Phi^i = 1$, and then used a numerical optimization procedure. In particular, the superpotential contains 26 parameters and the authors of \cite{Blaback:2018hdo} were able to find a de Sitter extremum which was fully metastable up to one single flat direction. They also noted that the classical flat direction of their solution could in principle receive corrections from quantum or non-pertubative effects.

 As one can see from \eqref{7mW}, in our example of the $[SL(2, \mathbb{R})]^7$ model we use only one perturbative term, namely the six-flux $f_6$, while the remaining terms in $W$ include two exponents for each modulus, resulting in $1+4\times 7=29$ parameters:  $A_i$,  $a_i$,$B_i$,  $b_i$. We will choose them to be real. In particular, 21 of these parameters are free, while 8 are obtained by solving 8 equations $W=\partial_i  W=0$ to guarantee the desired properties in the progenitor model.
According to \cite{Kallosh:2019zgd}, performing the three-steps procedure with starting point \eqref{7m} and \eqref{7mW}, leads to stable dS minima. While in some examples analized \cite{Cribiori:2019bfx} it occurs that the interplay between non-perturbative terms and fluxes does not seem to give simple solutions, it might be possible that a combination of our approach with the one of \cite{Blaback:2018hdo}, if applied for example to this $[SL(2,\mathbb{R})]^7$ model, may yield interesting results.

\subsection{Specific example of the 7-moduli  IIA model}

In this subsection, we present some details of the analysis we performed on the seven-moduli $[SL(2,\mathbb{R})]^7$ model. The progenitor model is defined by the \K\ potential and superpotential given respectively in  \eqref{7m} and \eqref{7mW}. Our choice for the parameters are listed in Table \ref{tab:para7mod}.

\begin{table}[H]
\centering
\begin{tabular}{|c|c|c|c|c|c|c|}\hline
$A_S = 1$ & $A_{T_1} = 3.1$& $A_{T_2} = 3.2$& $A_{T_3} = 3.3$ & $A_{U_1} =11$& $A_{U_2} =12$& $A_{U_3} =13$\\\hline
$a_S = 2$ & $a_{T_1} = 2.1$& $a_{T_2} = 2.2$& $a_{T_3} = 2.3$ & $a_{U_1} =0.41$& $a_{U_2} =0.42$& $a_{U_3} =0.43$\\\hline
$b_S = 3$ & $b_{T_1} = 3.1$& $b_{T_2} = 3.2$& $b_{T_3} = 3.3$ & $b_{U_1} =1.1$& $b_{U_2} =1.2$& $b_{U_3} =1.3$\\\hline
$S_0 = 1$ & $T_{1,\,0} = 1.1$  & $T_{2,\,0} = 1.2$  & $T_{3,\,0} = 1.3$ & $U_{1,\,0} = 5.1$& $U_{2,\,0} = 5.2$& $U_{3,\,0} = 5.3$\\\hline
\end{tabular}
\caption{   The choice for the parameters in the seven-moduli model.}
\label{tab:para7mod}
\end{table}

In particular, we are taking the values of the $U$ moduli at the minimum to be close to 5, so that the volume of the compact manifold is not small, whereas the values of the  $S$ and $T$ moduli are close to the origin of the moduli space. Moreover, the values reported in Table \ref{tab:para7mod}  refer  to the imaginary parts of the moduli, since we set the real parts (axions) to zero at the minimum. In other words, unless specified otherwise, in our examples the vacua will be at
\be
S \equiv {\rm i} S_0, \qquad T \equiv {\rm i} T_0, \qquad U \equiv {\rm i}U_0, 
\ee
with $S_0$, $U_0$ and $T_0$ real parameters.
When we impose the 8 equations $D_i W = 0$ and $W = 0$, which yield a supersymmetric Minkowski vacuum, we solve for the parameters $B_i$ and the flux parameter $f_6$. This allows us to freely choose the $A_i$, $a_i$ and $b_i$, as well as the position of the Minkowski vacuum, given by $S_0$, $T_0$ and $U_0$. With this solution, we find that the Minkowski mass matrix has the form predicted in \eqref{Mink}.

Before proceeding with the discussion, we would like to remark that the choice of the parameters is not fine-tuned. In fact, a wide range of values is possible. To confirm this,  we studied examples with different parameters  where the deviation from Minkowski  is small. In all of these cases, a de Sitter minimum was produced. Furthermore, in section \ref{big}  we discuss the complementary case in which the deviation from the progenitor Minkowski model is not small.

Resuming our analysis, the next step, namely the downshift, is implemented by substituting $f_6 \to f_6 + \Delta f_6$, which gives a supersymmetric AdS minimum at a slightly shifted position, with respect to the Minkowski one. In particular, the masses of the fields change only minutely and the AdS mass matrix takes the form predicted in \eqref{corrAdS}. In order to go to de Sitter we introduce anti-D6-branes, which contribute to the scalar potential via a term of the type:
\bea
V^{uplift}_{\overline{D6}} &=& \frac{\mu^4_1}{{\rm Im}(T_1) {\rm Im}(T_2) {\rm Im}(T_3) } +\frac{\mu^4_2}{{\rm Im}(S)  {\rm Im}(T_2)  {\rm Im}(T_3) } \nonumber \\
\cr
&+& \frac{\mu^4_3}{{\rm Im}(S)  {\rm Im}(T_1)  {\rm Im}(T_3)}+\frac{\mu^4_4}{{\rm Im}(S)  {\rm Im}(T_1)  {\rm Im}(T_2)}.
\label{eq:7ModUplift}
\eea
This can be generated in supergravity by using a nilpotent chiral superfield $X$ and by including appropriate terms in $K$ and $W$, as explained in \cite{Kallosh:2018nrk,Cribiori:2019bfx}. For what concerns the parameters, in our example the downshift is
\begin{equation}
\Delta f_6 = -10^{-5}
\end{equation}
and the uplift was chosen to be
\begin{equation}
\mu_1^{\,4} = \mu_2^{\,4} = \mu_3^{\,4} = \mu_4^{\,4} = 5.49028 \cdot 10^{-15}.
\end{equation}
The mass matrix is diagonal before the downshift and remains block diagonal afterwards, with the diagonal entries changing only slightly.

As an example, we present below  the upper left corner of the $7\times 7$ mass matrix, $V_{\phi^i \phi^j}$ \eqref{corrdS}, in the dS case. It shows  the masses of scalars after the dS uplift. The diagonal values are very close to the ones in Minkowski and in AdS minima, with values of order $10^{-3}- 10^{-5}$. All of the off-diagonal terms are many orders smaller, ranging from $10^{-9}$ to $10^{-10}$.
{\tiny
\begin{equation}
 \left(
\begin{array}{ccccccc}
1.89809 \cdot 10^{-5} & -6.19837 \cdot 10^{-10} &- 5.36624 \cdot 10^{-10}& -4.56280 \cdot 10^{-10} & -1.30375 \cdot 10^{-9}& -1.52470 \cdot 10^{-9} & -1.74268 \cdot 10^{-9}\\
\\
-6.19837 \cdot 10^{-10} & 1.30911 \cdot 10^{-4} & -7.38721 \cdot 10^{-10} & -6.46383 \cdot 10^{-10} & -1.24516 \cdot 10^{-9} & -1.44398 \cdot 10^{-9} & -1.64104 \cdot 10^{-9}\\
\\
-5.36624 \cdot 10^{-10} & -7.38721 \cdot 10^{-10} & 9.41667 \cdot 10^{-5} & -5.68241 \cdot 10^{-10} & -1.13520 \cdot 10^{-9} &-1.31758 \cdot 10^{-9} &-1.49834 \cdot 10^{-9}\\
\\
-4.56280 \cdot 10^{-10} & -6.46383 \cdot 10^{-10} & -5.68241 \cdot 10^{-10} &6.37888 \cdot 10^{-5}&-1.04022 \cdot 10^{-9}& -1.20871 \cdot 10^{-9}&-1.37571 \cdot 10^{-9}\\
\\
-1.30475 \cdot 10^{-9} & -1.24516 \cdot 10^{-9} & -1.13520 \cdot 10^{-9} & -1.04022 \cdot 10^{-9} & 9.96472 \cdot 10^{-4} &-5.36645 \cdot 10^{-10} & -5.74900 \cdot 10^{-10}\\
\\
-1.52470 \cdot 10^{-9}&-1.44398 \cdot 10^{-9}& -1.31758 \cdot 10^{-9}& -1.20871 \cdot 10^{-9 }&-5.36646 \cdot 10^{-10} & 1.37262 \cdot 10^{-3}& -6.10079 \cdot 10^{-10}\\
\\
-1.74268 \cdot 10^{-9}& -1.64104 \cdot 10^{-9}& -1.49834 \cdot 10^{-9} & -1.37571 \cdot 10^{-9}& -5.74900 \cdot 10^{-10} &-6.10079 \cdot 10^{-10}&1.80465 \cdot 10^{-3}
\end{array}
\right)\nonumber
\end{equation}
}
\noindent  The values of this matrix match exactly with what has been predicted from the mass production procedure. 
The mass production mechanism explains in this example why all eigenvalues of the mass matrix are positive in dS, as in the progenitor Minkowski stage:  the  off-diagonal entries, which where absent at the  Minkowski and appeared in AdS and dS stages, and which we named   $\Delta_{\phi^i \phi^j}  + \tilde \Delta_{\phi^i \phi^j}$ in  \eqref{corrdS}, 
are too small to affect  the positivity of the original eigenvalues.

In table \ref{tab:7modev}, we present the eigenvalues of Minkowski and dS mass squares (second derivatives of the potential), both for the scalars and the pseudoscalars. As one can see, the difference between Minkowski masses and dS masses is very small. In addition, in the Minkowski case, as predicted, all scalar masses coincide with the pseudoscalar ones. However, in dS, already at the order of the digits we kept in the table, some of the scalar and pseudoscalar masses are slightly different, as expected.

At this point, we decided not to present the plots of the potential that we produced for this seven-moduli example (mainly for convenience, since there are many of them). Instead, we will show them in the next subsection, for a simpler version of the model. As we mentioned before, this is an STU model constructed by identifying $T_1=T_2=T_3  \equiv T$ and $U_1=U_2=U_3 \equiv U$.

\begin{table}[H]
\centering
\begin{tabular}{|c|c|c|c|}\hline
& Mink  & dS \\\hline
$m_1^{\,2}$   & $\;1.80473 \cdot 10^{-3}\;$  & $\;1.80465 \cdot 10^{-3}\;$\\\hline
$m_2^{\,2}$   & $1.80473 \cdot 10^{-3}$  & $1.80465 \cdot 10^{-3}$\\\hline
$m_3^{\,2}$   & $1.37269 \cdot 10^{-3}$  & $1.37262 \cdot 10^{-3}$\\\hline
$m_4^{\,2}$   & $1.37269 \cdot 10^{-3}$  & $1.37262 \cdot 10^{-3}$\\\hline
$m_5^{\,2}$   & $9.96519 \cdot 10^{-4}$  & $9.96472 \cdot 10^{-4}$\\\hline
$m_6^{\,2}$   & $9.96519 \cdot 10^{-4}$  & $9.96471 \cdot 10^{-4}$\\\hline
$m_7^{\,2}$   & $1.30924 \cdot 10^{-4}$  & $1.30911 \cdot 10^{-4}$\\\hline
$m_8^{\,2}$   & $1.30924 \cdot 10^{-4}$  & $1.30911 \cdot 10^{-4}$\\\hline
$m_9^{\,2}$   & $9.41773 \cdot 10^{-5}$  & $9.41667 \cdot 10^{-5}$\\\hline
$m_{10}^{\,2}$& $9.41773 \cdot 10^{-5}$  & $9.41660 \cdot 10^{-5}$\\\hline
$m_{11}^{\,2}$& $6.37973 \cdot 10^{-5}$  & $6.37888 \cdot 10^{-5}$\\\hline
$m_{12}^{\,2}$& $6.37973 \cdot 10^{-5}$  & $6.37883 \cdot 10^{-5}$\\\hline
$m_{13}^{\,2}$& $1.89843 \cdot 10^{-5}$  & $1.89809 \cdot 10^{-5}$\\\hline
$m_{14}^{\,2}$& $1.89843 \cdot 10^{-5}$  & $1.89806 \cdot 10^{-5}$\\\hline
\end{tabular}
\caption{  The eigenvalues of the mass matrix  for the seven-moduli type example. The mass shift is small, but noticeable, when going from Minkowski to dS. One can also notice, as predicted by the mass production procedure, that in dS the masses of scalars and pseudoscalars are not exactly equal anymore, as was the case in Minkowski. }
\label{tab:7modev}
\end{table}

\subsection{The  STU model}
\label{sec:stumodel}
As a more detailed test for the mass production procedure, we employ a simple STU model that has also been used in \cite{Cribiori:2019bfx} to illustrate the $\overline{D6}$-brane uplift and in \cite{Kallosh:2019zgd} to test some aspects of the mass production mechanism. In particular, we extend here the analysis of \cite{Kallosh:2019zgd}.

The \K\ potential and superpotential of this model are
\begin{equation}
\begin{aligned}
\label{eq:IIA3modPot}
K=& - \log\left(-\rmi (S-\bar{S})\right) - 3 \log\left(-\rmi (T-\bar{T})\right)- 3 \log\left(-\rmi (U-\bar{U})\right) ,\\
W=& f_6 + \sum_{\Phi = S, T, U} (A_\Phi e^{\rmi a_\Phi \Phi} - B_\Phi e^{\rmi b_\Phi \Phi}).
\end{aligned}
\end{equation}

\noindent Again, $f_6$ is a tree-level constant contribution to the superpotential from six-form flux, while the terms with exponents are  non-perturbative corrections. The contributions in $S$ and $T$ directions can arise, for example, from either gaugino condensation of stacks of $D6$-branes or from euclidean $D2$-branes \cite{Palti:2008mg}. 
One possible motivation for the contribution in the $U$ direction, which is discussed in \cite{Cribiori:2019bfx}, comes from string theory U-duality. In short, string theory explicitly only exhibits $S$- and $T$-duality but in M-theory this is expected to combine with $U$-duality, mixing all three types of moduli into each other. Another motivation comes from instantons. In fact, it was shown in \cite{Kachru:2000ih,Blumenhagen:2009qh} that there are string theory world sheet instantons in $\mathcal{N}=1$ orientifold compactifications of type IIA string theory that can give rise to such terms.

We recall, once more, that the first step in the mass production procedure is to find a stable supersymmetric Minkowski vacuum, where all masses are non-negative. %In fact, the existence of this solution will only depend on the superpotential $W$, while it will not pose particular restrictions on the  choice of $K$. %for as long as the \K potential is \emph{reasonable}, meaning that the kinetic terms are positive definite. Then, a solution for the dS vacuum is given, for example, 
In the specific model we are considering, the solution will be given in terms of the three parameters $B_\Phi$, with $\Phi=(S,T,U)$ and of the flux parameter $f_6$. Moreover, we choose to set the axions contained in $(\Phi + \bar{\Phi})$ to zero. This assumption is justified as long as all masses are positive.
%In order to find a supersymmetric Minkowski vacuum of the scalar potential we only need to simultaneously solve the equations $D_\Phi W=0$, which gives a supersymmetric vacuum, and $W=0$, which enforces vanishing vacuum energy. This will always yield a supersymmetric solution of $\partial_\Phi V = 0$ with $V=0$. 
We have already shown the general solution for these equations for a KL-type superpotential in section \ref{KL}.

\begin{figure}[H]
\center
\includegraphics[scale=0.52]{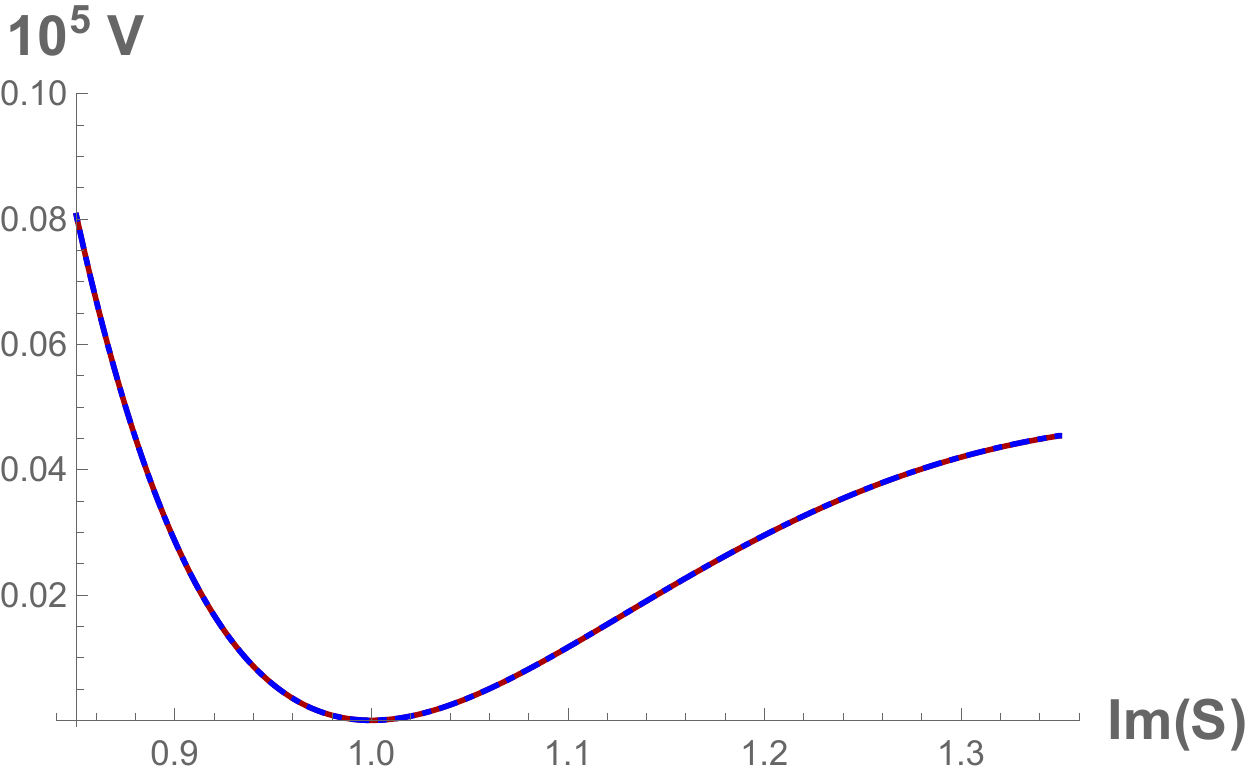}\qquad
\includegraphics[scale=0.52]{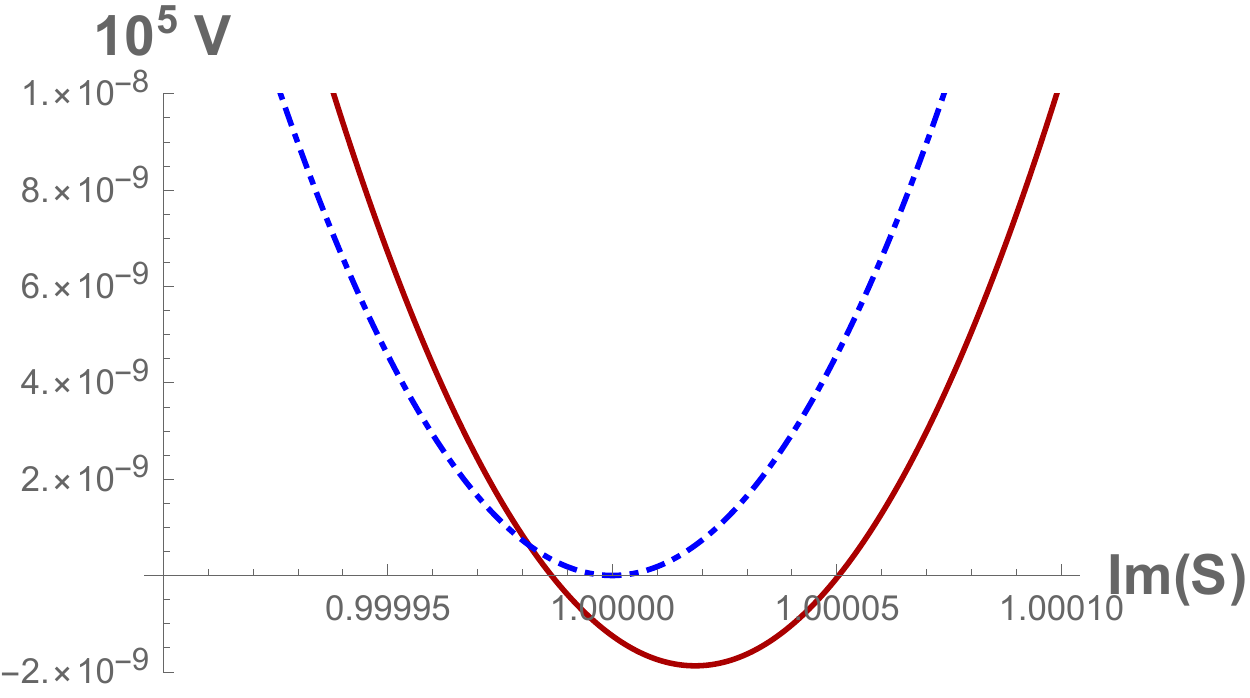}
\caption{\footnotesize The S-direction for one choice of parameters, before and after the downshift. From the graphic on the left, it is evident that the general behaviour of the potential does not change significantly. The zoomed in region on the right shows that the vacuum energy at the minimum becomes negative and that the point of the minimum shifts by a small amount.}
\label{fig:3modshift}
\end{figure}

In the second step we introduce a shift in the superpotential which will change the value of the scalar potential from zero to a negative value, meaning that we go from a Minkowski solution to AdS. Since this shift is parametrically small by  construction, the masses will stay positive as outlined in \cite{Kallosh:2019zgd} and as explained in the previous sections. Explicitly, the shift is given by
\begin{equation}
\label{eq:Wdownshift}
f_6\, \to \, f_6 + \Delta f_6\, \Rightarrow \, W \to W + \Delta f_6 \ .
\end{equation}
Note that, for small $\Delta f6$, the sign of the shift can slightly change the result but not the general behaviour. Indeed, both a positive or negative value for $\Delta f_6$ will lead to AdS. Besides the change in the value of $V$, we also expect the position of the minimum to slightly shift in moduli space, as given by \eqref{DeltaPhi}. The shift is illustrated in figure \ref{fig:3modshift}, for one of the three moduli. 

\begin{figure}[H]
\center
\includegraphics[scale=0.52]{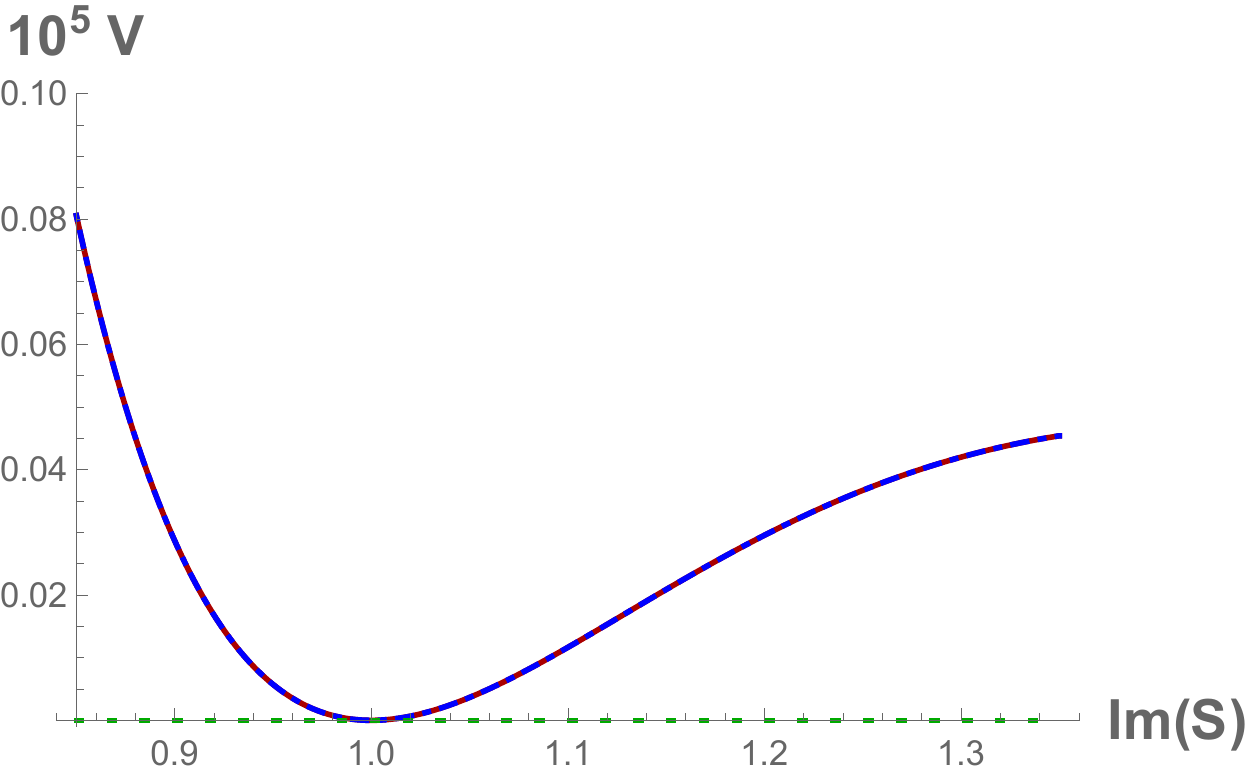}\qquad
\includegraphics[scale=0.52]{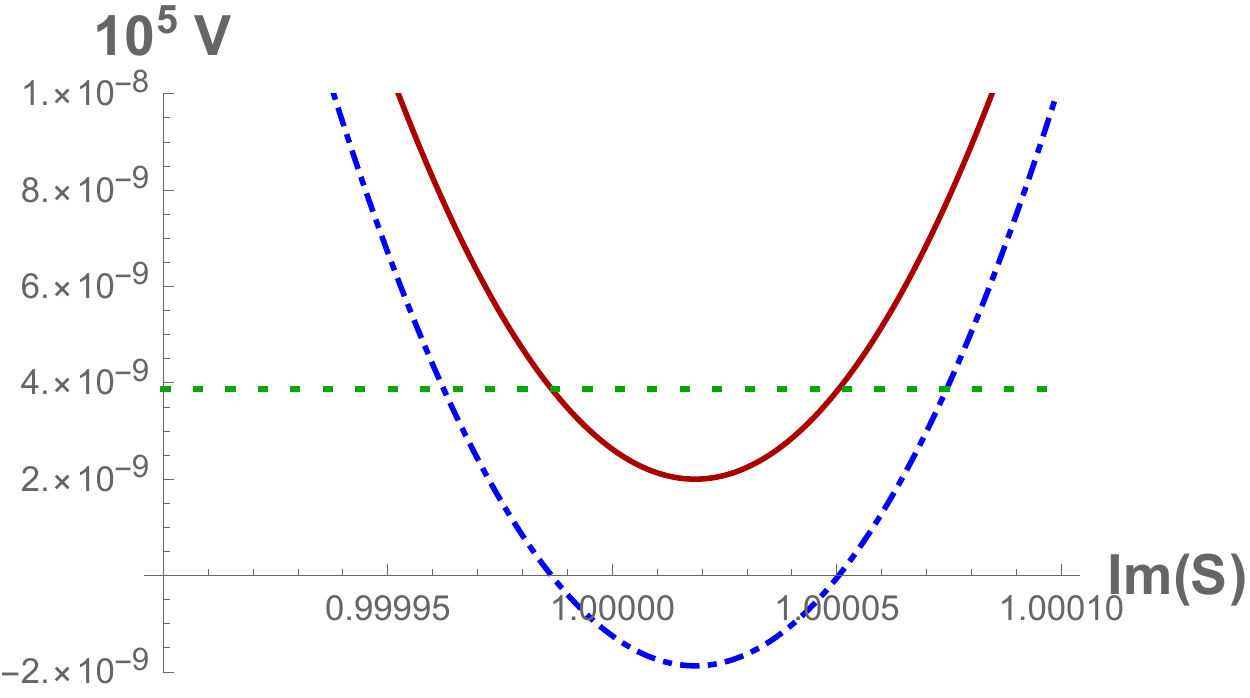}
\caption{\footnotesize  The S-direction for one choice of parameters, both for AdS and dS. Again, the overall shape of the potential does not change visibly. In the zoomed region on the right, the AdS minimum before the uplift is clearly visible and the constant contribution from the anti-D6-brane lifts the potential to dS.  The uplifting contribution appears constant due to the smallness of the parameters $\mu_i$ and the plot range. The overall shape of the contribution follows from equation \eqref{eq:3ModAntiD6Up}.}
\label{fig:3modshiftdS}
\end{figure}

Finally, as a third step, we use an anti-D6-brane uplift \cite{Cribiori:2019bfx} in order to go from AdS to dS. The inclusion of anti-branes introduces a term of the form
\begin{equation}
\label{eq:3ModAntiD6Up}
V_{\overline{D6}} = \frac{\mu^4_1}{{\rm Im} \, S\,  {\rm Im} \, T ^2} + \frac{\mu^4_2}{{\rm Im} \, T^3}
\end{equation}
in the scalar potential, which will lift the value of the vacuum energy above zero. Again, this will also shift the position of the minimum slightly \cite{Kallosh:2019zgd}. The result is shown in figure \ref{fig:3modshiftdS}. Note that we have chosen the value of the potential after the uplift for illustration purposes. Likewise, by appropriate choice of the parameters $\mu_i\; (i=1,2)$ it is possible to match the observed cosmological constant.

\begin{figure}[H]\hskip 2cm 
%\begin{center}
\includegraphics[scale=0.6]{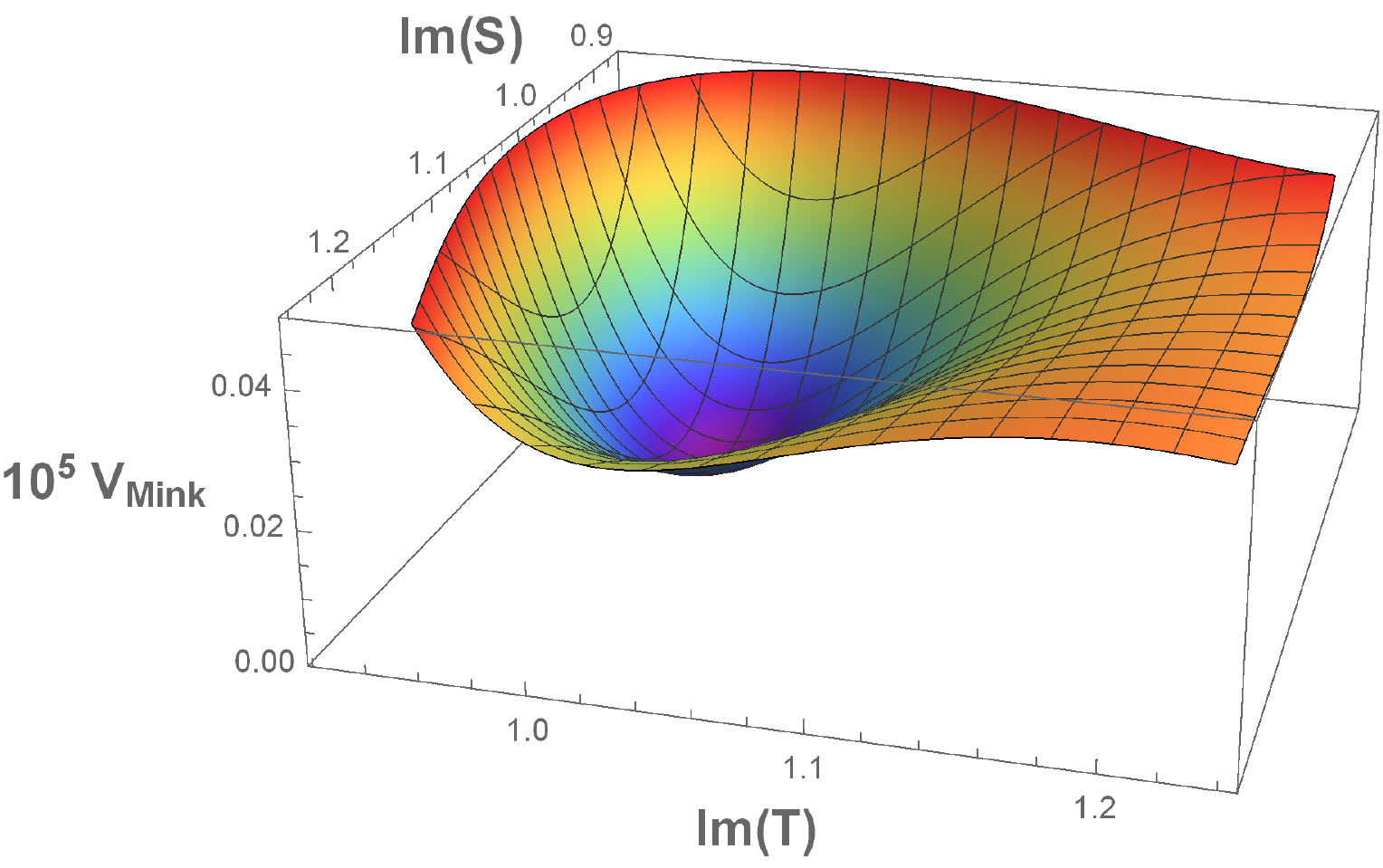}
%\end{center}
\caption{\footnotesize The overall form of the potential in the directions $Im(S)$ and $Im(T)$. This shape does not change significantly along the 3 steps.}
\label{fig:3mod3DLarge}
\end{figure}

\begin{figure}[H]
\includegraphics[scale=0.457]{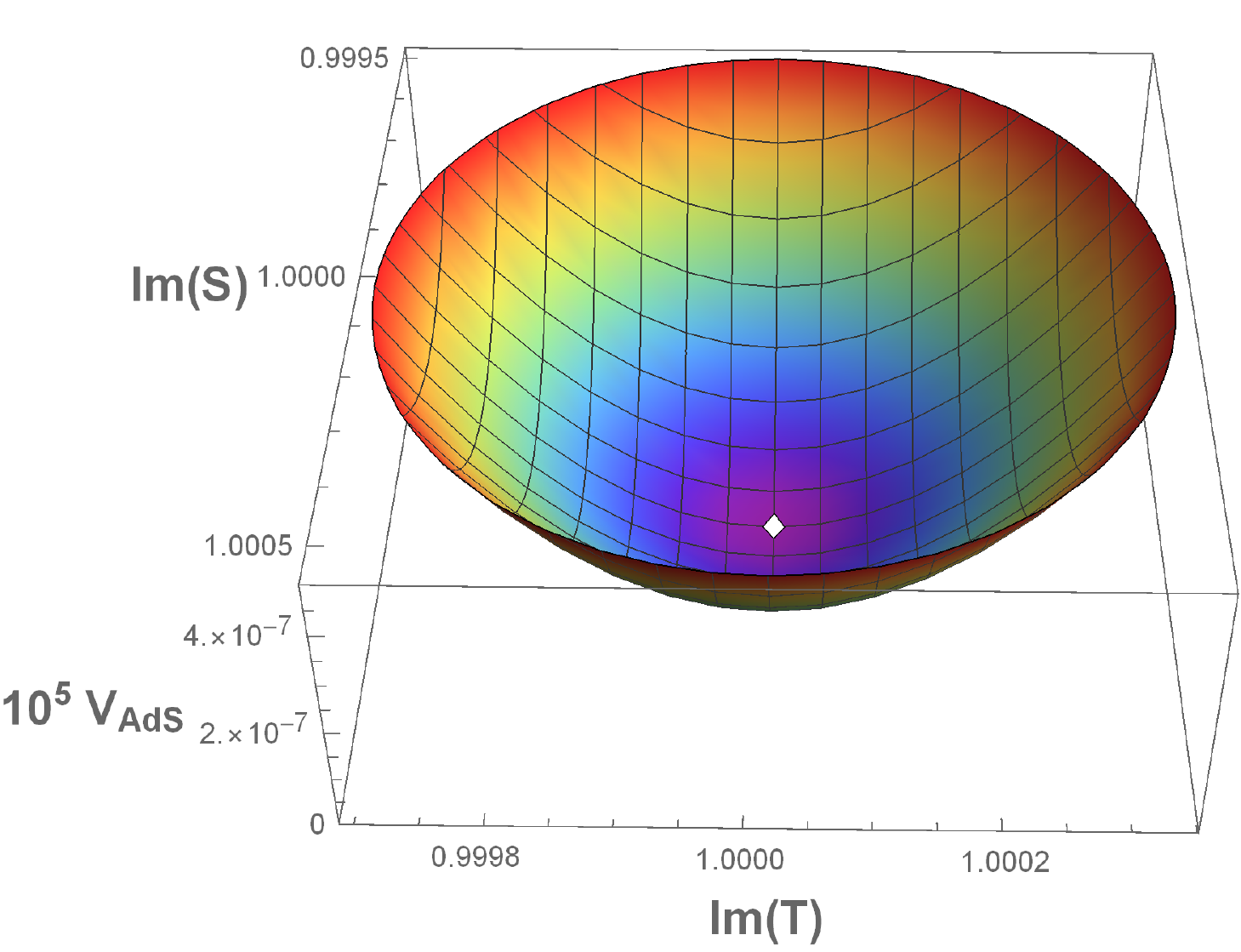} 
\includegraphics[scale=0.46]{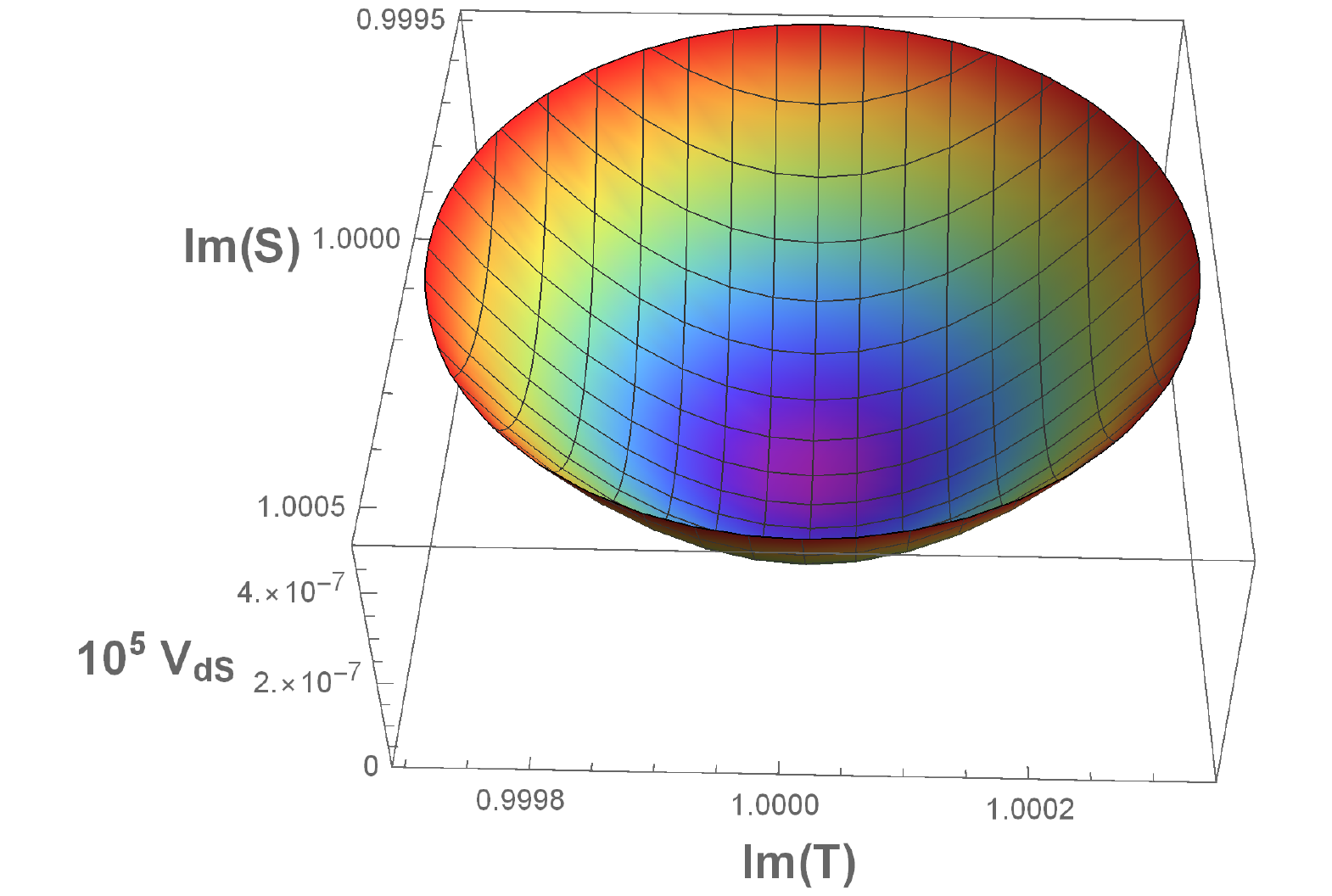}
\caption{\footnotesize  A close-up picture of the minimum in the directions ${\rm Im}(S)$ and ${\rm Im}(T)$. In the left picture, we see the situation in AdS. The visible hole is where the potential is below zero. On the right, we have presented the dS potential within the same plot range. We see that the potential is now positive everywhere. }
\label{fig:3mod3DClose}
\end{figure}

Since the masses remain positive during all of the steps of this procedure, we expect that the minima are stable. In general, however, by checking only 2D-plots, one might miss a possible diagonal runaway direction. For this reason, it is instructive to also look at 3D-plots, in order to make sure that there are no inconsistencies. These are shown, for the directions ${\rm Im}(S)$ and ${\rm Im}(T)$, in figures \ref{fig:3mod3DLarge} and \ref{fig:3mod3DClose}.
 
An important feature of the mass production procedure is that, if the masses are block diagonal in Minkowski, they remain block diagonal during all of the subsequent steps. In this example, it is indeed easy to see that the change in the masses is small, when the downshift is small. To illustrate the point, we can look at the mass matrix after the downshift to AdS, which is
\begin{equation}
{\footnotesize
\hskip -0.3 cm
\mathcal{M}^2_{AdS}=\begin{pmatrix}
3.662 \cdot 10^{-5} & -1.979 \cdot 10^{-9} & -2.955 \cdot 10^{-10} &0&0&0\\
\\
-1.979 \cdot 10^{-9} & 9.916 \cdot 10^{-5} & -8.562 \cdot 10^{-10} &0&0&0\\
\\
-2.956 \cdot 10^{-10} & -8.562 \cdot 10^{-10} & 9.146 \cdot 10^{-7} &0&0&0\\
\\
0&0&0&3.662 \cdot 10^{-5} & -1.979 \cdot 10^{-9} & -2.955 \cdot 10^{-10} \\
\\
0&0&0&-1.979 \cdot 10^{-9} & 9.916 \cdot 10^{-5} & -8.562 \cdot 10^{-10} \\
\\
0&0&0&-2.956 \cdot 10^{-10} & -8.562 \cdot 10^{-10} & 9.146 \cdot 10^{-7} \\
\end{pmatrix} .\nonumber }
\end{equation}
%}

The diagonal entries are the same as in the Minkowski vacuum, up to the third relevant digit, while the off-diagonal entries are around 3-4 orders of magnitude smaller. Likewise, the numbers do not change much after the uplift to dS. This is expected from \eqref{corrAdS}. A comparison of the eigenvalues of the mass matrix for  Minkowski and dS minima is given in table \ref{tab:3modev}.
\begin{table}[H]
\centering
\begin{tabular}{|c|c|c|c|}\hline
& Mink  & dS \\\hline
$m_1^{\,2}$ & $\; 9.91957 \cdot 10^{-5} \;$ & $\; 9.91570 \cdot 10^{-5} \;$ \\\hline
$m_2^{\,2}$ & $\; 9.91957 \cdot 10^{-5} \;$ & $\; 9.91551 \cdot 10^{-5} \;$ \\\hline
$m_3^{\,2}$ & $\; 3.66313 \cdot 10^{-5} \;$ & $\; 3.66189 \cdot 10^{-5} \;$ \\\hline
$m_4^{\,2}$ & $\; 3.66313 \cdot 10^{-5} \;$ & $\; 3.66183 \cdot 10^{-5} \;$ \\\hline
$m_5^{\,2}$ & $\; 9.15565 \cdot 10^{-7} \;$ & $\; 9.14587 \cdot 10^{-7} \;$ \\\hline
$m_6^{\,2}$ & $\; 9.15565 \cdot 10^{-7} \;$ & $\; 9.14550 \cdot 10^{-7} \;$ \\\hline
\end{tabular}
\caption{   The eigenvalues of the mass matrix in the Minkowski and dS minima. It is evident that they only change by a small amount both during the downshift and after the anti-D6-brane uplift. 
%The reason the uplift has no visible effect to the given precision is due to its smallness. <- This makes no sense if we do not show the masses in AdS
}
\label{tab:3modev}
\end{table}
\noindent For what concerns the free parameters in the model, the above plots and masses were achieved with the set of parameters reported in table \ref{tab:paraIIA3mod}.
\begin{table}[H]
\centering
\begin{tabular}{|c|c|c|}\hline
$A_S = 1$ & $A_T = 3$ & $A_U =11$\\\hline
$a_S = 2$ & $a_T = 2.1$ & $a_U = 1$\\\hline
$b_S = 3$ & $b_T = 3.1$ & $b_U = 1.2$\\\hline
$S_0 = 1$ & $T_0 = 1$ & $U_0 = 5$\\\hline
\end{tabular}
\caption{   Our choice of parameters for the 3 moduli IIA example.}
\label{tab:paraIIA3mod}
\end{table}
The downshift to AdS was chosen to be
\begin{equation}
\Delta f_6 = -10^{-5},
\end{equation}
and, subsequently, the uplift as
\begin{equation}
\mu_1^{\,4} = \mu_2^{\,4}= 1.93752 \cdot 10^{-14},
\end{equation}
for illustrative purposes.

As explained previously, no particular care is necessary when choosing these values. This suggests that the examples are rather robust and they possess a considerably large parameter space, which can potentially accommodate string theory restrictions as well. The only conscious choice that was made is $U_0 =5$ in order to have a large volume for the compact manifold.  In some type IIA examples, in section \ref{big}, we will consider other values of the parameters, including larger values of the volume modulus.
 
 %\newpage
 
% \parskip 8pt

\section{Type IIB models}\label{IIB}

In this section we present examples of string theory inspired models in the type IIB case. In particular, the \K\ potentials associated with Calabi-Yau three-folds that we will discuss, have been already used before, in \cite{Cicoli:2008va,Cicoli:2008gp,Burgess:2016owb,Kallosh:2017wku,Bobkov:2010rf}, for cosmological applications.  However, in these works, stabilization of the volumes of the four-cycles was performed either using the LVS, or some versions of KKLT stabilization. Instead, as motivated in the Introduction, the choice of the superpotential in the present work will be different from the examples in \cite{Cicoli:2008va,Cicoli:2008gp,Burgess:2016owb,Kallosh:2017wku,Bobkov:2010rf}, since we will be using a double set of exponentials for each modulus.

Before entering the details of the analysis, we would like to recall some basic facts of flux compactification in IIB string theory.
Type IIB string theory flux compactifications on Calabi-Yau three-folds, when dimensionally reduced to $d=4$, $\cN = 1$ supergravity are described by a  \K\, potential  and a superpotential of the form \cite{Grimm:2004uq}
\begin{align} \label{Gen}
K &= - \log\left({\rm i}\int \Omega \wedge \bar \Omega \right) -  \log\left(-{\rm i}(\tau-\bar \tau)\right)-2 \log\left(\mathcal{V}_6\right),\\
W &= \int G_3\wedge \Omega.
\label{Gen1}\end{align}
Here, $\Omega$ is a function of the CS moduli, $\tau$ is the axion-dilaton and $G_3$ is the complex three-form flux. The six-dimensional volume is defined as a cubic polynomial 
\be
\mathcal{V}_6={1\over 3!}  \int J \wedge J \wedge J = {1\over 3!}d_{ijk} t_i t_j t_k,
\ee
where $d_{ijk}$ are CY intersection numbers and $t_i$'s are the volumes of the two-cycles. One can switch to a complexified \K\, moduli space, using holomorphic coordinates
\be
T_i= \tau_i +i\chi_i,
\label{T}\ee
where the
\be
\tau_i = {\partial \mathcal{V}_6\over \partial t_i}
\label{4c}\ee
are the volumes of the four-cycles, which are quadratic in the volumes of the two-cycles.
This allows us to rewrite the six-dimensional volume as a function of the four-cycle volumes $\tau_i $. As a result of the procedure of replacing the regular expression that is cubic in two-cycles,  $\mathcal{V}_6= {1\over 3!}d_{ijk} t_i t_j t_k$, via a function of the four-cycle volumes $\tau_i $ satisfying \eqref{4c}, one typically finds somewhat complicated expressions for the \K\, potential as a functional of the volumes of the four-cycles
\be
\mathcal{V}_6= {1\over 3!}d_{ijk} t_i t_j t_k = \mathcal{V}_6 (\tau_i)\,.
\ee
For this reason, as explained in subsection \ref{subsec:IIBgen}, the analysis of the stability of the vacuum configuration will be generally more complicated in our type IIB examples than the one we performed in the type IIA case. In particular, 
we will study only stabilization of the \K\, moduli $T_i$. The complex structure moduli in $\Omega$ and the axion-dilaton $\tau$ in \rf{Gen}, \rf{Gen1} are fixed at  an  earlier stage by fluxes, as it is usually the case in type IIB.

The \K\, potential  for the \K\, moduli will depend on the four-cycle volumes ${\rm Re} \, T_i = \tau_i$ and, additionally, on the nilpotent chiral multiplet $X$.  In particular, the  uplift in these IIB models will be realized with anti-D3-branes \footnote{For convenience we will label the different \K\ moduli as $S$, $T$ and $U$, when discussing explicit models in the following sections. These should not be confused with the $S$, $T$,  $U$ moduli in type IIA.}. We recall that two different possibilities can occur  \cite{Kachru:2003sx}: one where stabilization is in the bulk, the other when it is within a warped throat. In $d=4$ language, one can associate to these different regimes the following \K\ potentials:
\be
 K_{bulk}  =   -2 \log\left(\mathcal{V}_6(\tau_i) \right) +  X \bar X\,,
\ee
\be
 K_{throat}  = -3 \log\left(\mathcal{V}^{2/3}_6(\tau_i) - {1\over 3}   X \bar X\right)\,. 
\ee

We will stabilize the \K\ moduli as suggested in the mass production method. Namely, we will use the KL type double exponents and follow the three-steps procedure to get dS minima. We recall once more the form of the superpotential in a three-moduli problem, namely
\begin{equation}
\label{eq:KLpotIIB}
W = W_0 + \sum_{\Phi = S, T, U} A_\Phi e^{i a_\Phi \Phi} - B_\Phi e^{i b_\Phi \Phi} + \mu^2 X,
\end{equation}
which encompasses all of the examples we will analyze in the following.
Given the \K\ potential $K$ and the superpotential $W$, one can use the standard rules of $\cN =1$ supergravity to calculate the scalar potential, including also the contribution of the anti-D3-branes. When moduli stabilization occurs in the bulk, the uplift potential is
\begin{equation}
\label{eq:antiD3upliftBulk}
V_{\overline{D3}\,bulk} = e^{ K_{bulk}  } DW_X K^{X \bar X} \overline{D_{ X} W}  |_{X=0}= \frac{\mu^4}{(\mathcal{V}_6)^2},
\end{equation}
where $\mathcal{V}_6$ is the volume of the compactification.
Alternatively, we can place the anti-D3-brane at the bottom of a warped throat, where the uplifting term will scale differently with the volume \cite{Kachru:2003sx}, giving
\begin{equation}
\label{eq:antiD3upliftThroat}
V_{\overline{D3}\,\text{warped throat}} = e^{ K_{throat}  } DW_X K^{X \bar X} \overline{D_{ X} W}  |_{X=0}= \frac{\mu^4}{(\mathcal{V}_6)^{4/3}}.
\end{equation}

Before proceeding with the examples, we would like to comment briefly on the difference between the notation used in \cite{Cicoli:2008va,Cicoli:2008gp,Burgess:2016owb,Kallosh:2017wku,Bobkov:2010rf} and the one we adopt in the present work. Indeed, notice that in \eqref{T} the fields appearing explicitly in the six-dimensional volume are the real parts of the chiral multiplets, $\tau_i={\rm Re} \,T_i$. This is in agreement with the notation used in \cite{Cicoli:2008va,Cicoli:2008gp,Burgess:2016owb,Kallosh:2017wku,Bobkov:2010rf}. On the other hand, in the previous sections, we used a slightly different notation, in which the volume $\mathcal{V}_6$ potential is a function of the imaginary parts of the chiral multiplets, $\mathcal{V}_6=\mathcal{V}_6({\rm Im} T_i)$. However, it is easy to show that the difference is merely due to conventions and does not change the physical results. First, we recall that $\mathcal{V}_6$, appearing in  \eqref{Gen}, in type IIB models is a homogeneous function of degree $3/2$ in the $\tau_i$. As a consequence, a constant rescaling of the $T_i$ by a multiplicative factor $-2\rmi$,
 \be
 \tau_i = {\rm Re }T_i \quad \rightarrow \quad  \tau_i = 2{\rm Im} T_i,
\ee 
will give
 \be
 \,\,\,\,\,\log \mathcal{V}_6(\tau_i) \quad \rightarrow \quad \log(2^{\frac32}\mathcal{V}_6(\tau_i)).
 \ee
 Then, the total \K\ potential \eqref{Gen} changes only by a constant factor
 \be
 \label{Ktransf}
 \quad\quad\,\, K \quad \rightarrow \quad   K - \log 8. 
 \ee
This difference is not physical, since it can be absorbed in a \K\ transformation, by rescaling also the superpotential as $W\to W e^{-C}$, with $C =\frac12 \log 8$. Alternatively, one can think of such an additional constant term in the \K\ potential as being originated from $- \log\left({\rm i}\int \Omega \wedge \bar \Omega \right) -  \log\left(-{\rm i}(\tau-\bar \tau)\right)$ in \eqref{Gen}, due to a different stabilization of the CS moduli and the axion-dilaton. 
Therefore, without loss of generality, in the following we will set
\be
\label{ourtau}
\tau_1 =  -\rmi (S-\bar{S})\, ,  \qquad \tau_2 =  -\rmi (T-\bar{T})\, ,  \qquad \tau_3 =  -\rmi (U-\bar{U}).
\ee

%\subsection{Explicit type IIB Examples}
\label{sec:IIBexamples1}

\subsection{K3-fibration with two parameters}

The first type IIB model we look at is a K3 fibration of $\mathbb{C}P^4_{[1,1,2,2,6]} $, discussed in \cite{Cicoli:2008va}. As shown in \cite{Cicoli:2008va}, it occurs that there is no moduli stabilization in this model in the context of  a large volume scenario (LVS) \cite{Balasubramanian:2005zx}. The volume is given in equation (3.14) of \cite{Cicoli:2008va}, in terms of the volumes of 4-cycles. In particular, it is a function of $\tau_1$ and $\tau_2$ only
\begin{equation}
\mathcal{V}_6(\tau_i) = \frac{1}{2} \sqrt{\tau_1} \left[ \tau_2 - \frac{2}{3}\tau_1\right]\,,
\end{equation} 
which, in our conventions \eqref{ourtau}, becomes
\begin{equation}
\mathcal{V}_6(S,T) = \frac{1}{2} \sqrt{\left(-\rmi (S-\bar{S})\right)} \left[ \left(-\rmi(T-\bar{T})\right) - \frac{2}{3} \left(-\rmi(S-\bar{S})\right)\right]\,.
\end{equation}

The parameters we used in the analysis are given in table \ref{tab:para2paraK3}. 
\begin{table}[H]
\centering
\begin{tabular}{|c|c|}\hline
$A_S = 1.1$ & $A_T = 1.2$ \\\hline
$a_S = 2.1$ & $a_T = 2.2$ \\\hline
$b_S = 3.1$ & $b_T = 3.2$ \\\hline
$S_0 = 1$ & $T_0 = \pi$ \\\hline
\end{tabular}
\caption{   Set of parameters for the K3 fibration of $\mathbb{C}P^4_{[1,1,2,2,6]} $.}
\label{tab:para2paraK3}
\end{table}
\noindent The downshift is achieved with the value
\begin{equation}
\Delta W_0 = -10^{-5} \ ,
\end{equation}
and the uplifts in the two different regions, namely bulk and throat, are realized with parameters
\begin{equation}
\begin{aligned}
\mu_{bulk}^{\,4}=& 3.61516 \cdot 10^{-10},\\
\mu_{throat}^{\,4} =& 1.56758 \cdot 10^{-10}\,.
\end{aligned}
\end{equation}
For what concerns the masses and the plots, we will only give the results for the case in which the anti-D3 brane is placed in the bulk. Indeed, at least for the choice of parameters we made, the difference between the uplift with an anti-D3-brane in the bulk or at the bottom of a throat cannot be really appreciated.  Due to the simplicity of the model, we display also both 2D and 3D plots in figures \ref{fig:2paraK32D} and \ref{fig:2paraK33D} respectively.

\

\begin{figure}[H]
\includegraphics[scale=0.52]{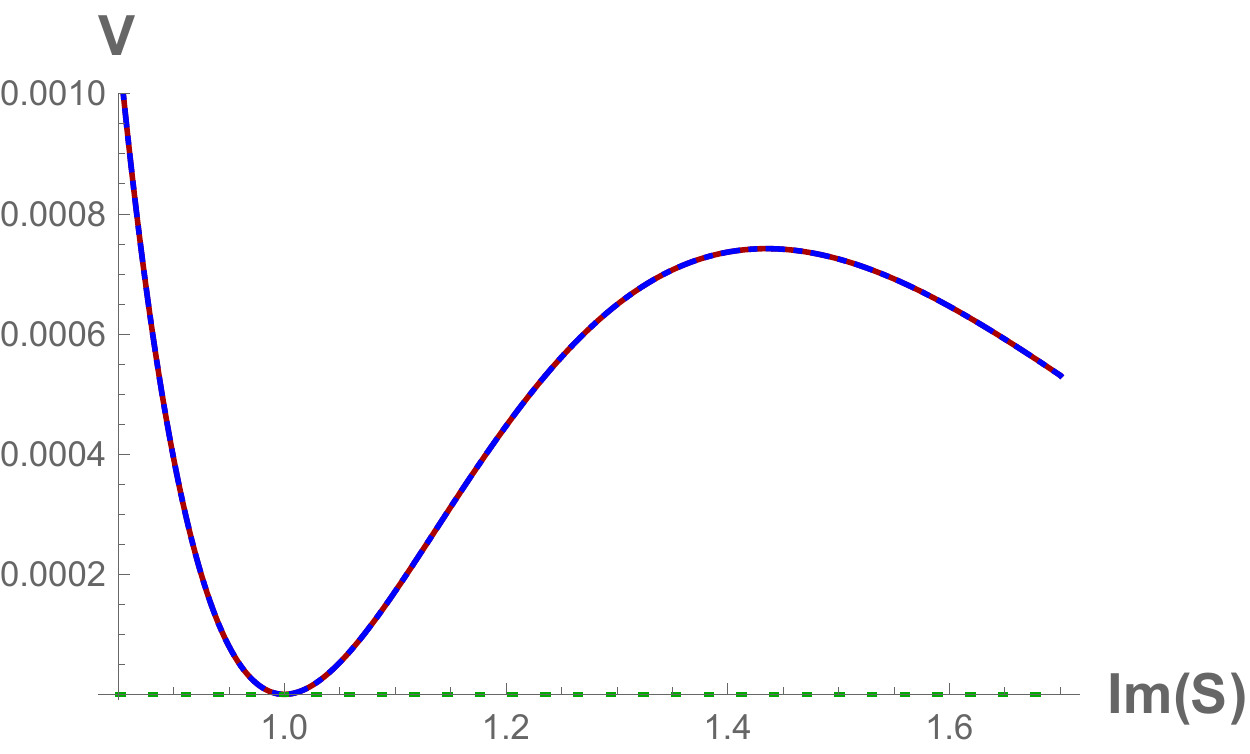}\qquad\includegraphics[scale=0.58]{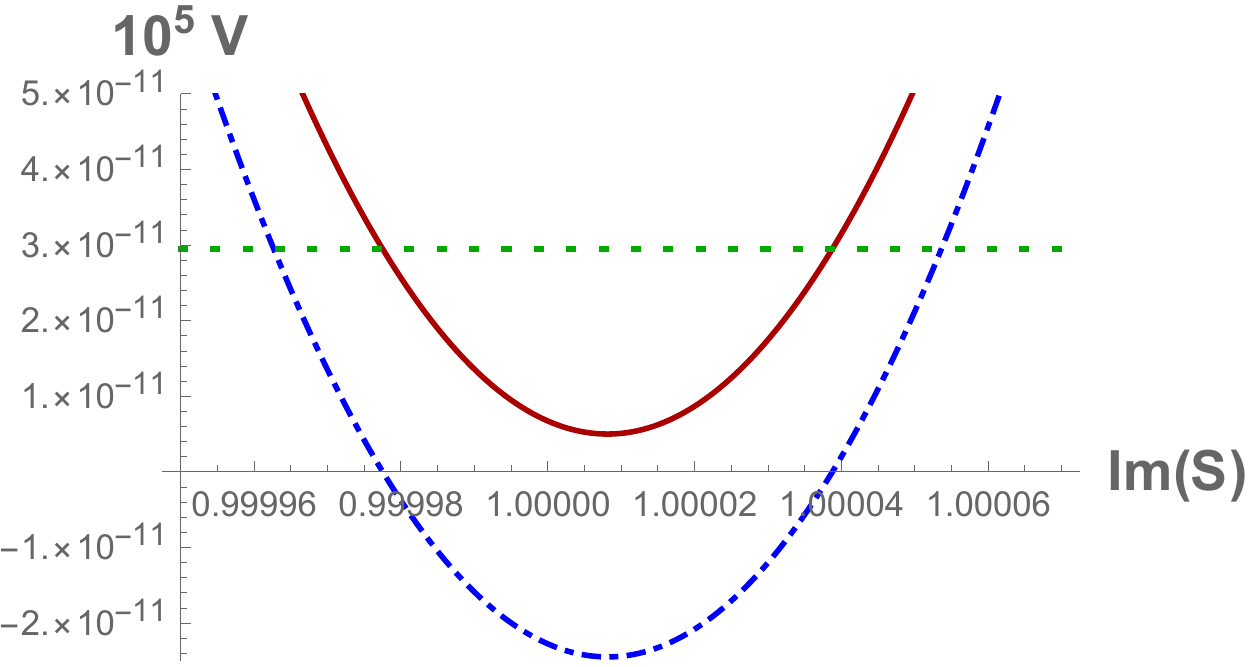}
\end{figure}
\begin{figure}[H]
\includegraphics[scale=0.52]{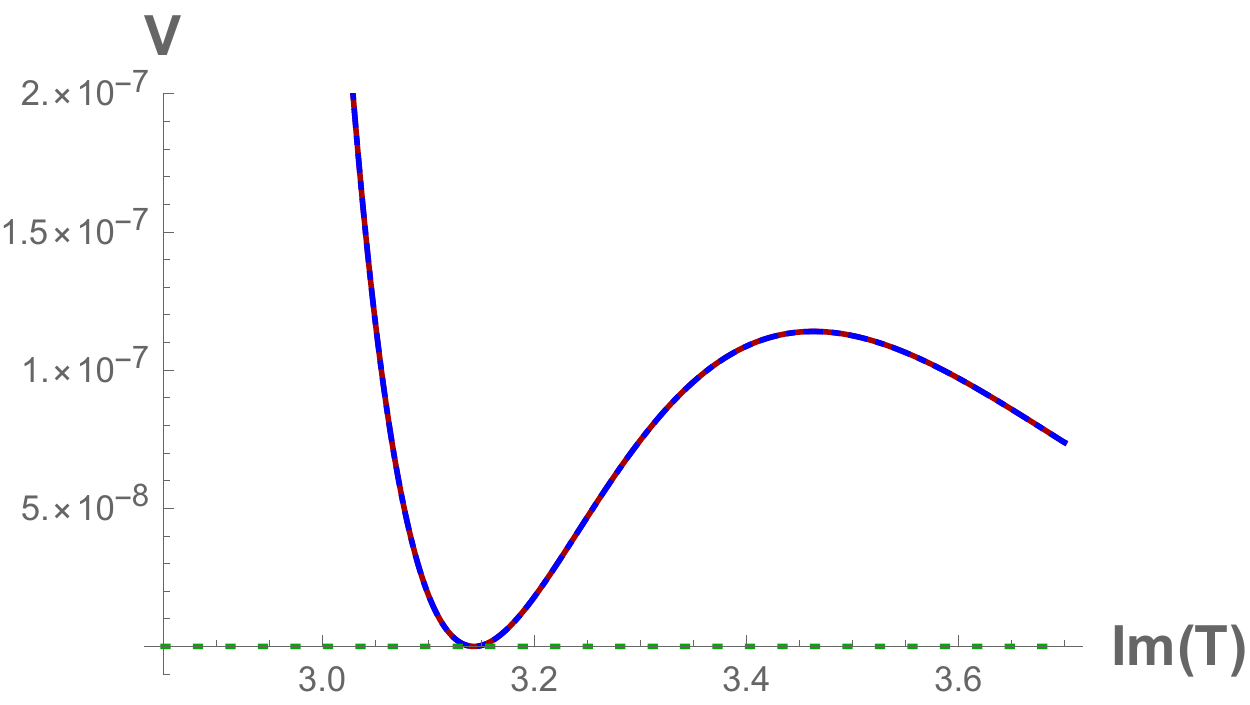}\qquad\includegraphics[scale=0.58]{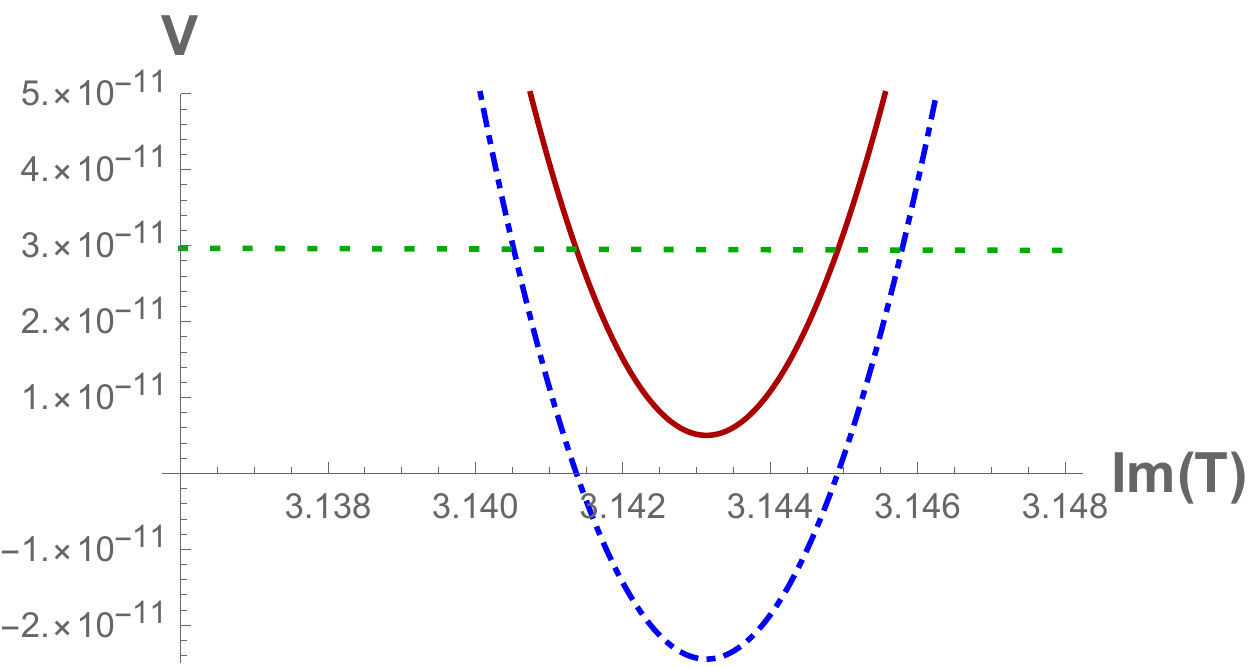}
\caption{\footnotesize All 2D plots for the K3 fibration of $\mathbb{C}P^4_{[1,1,2,2,6]} $. Notice that the ${\rm  Im}(T)$-direction is considerably flatter than the ${\rm Im}(S)$-direction.}
\label{fig:2paraK32D}
\end{figure}

\

\

\begin{figure}[H]
\centering
\includegraphics[scale=0.68]{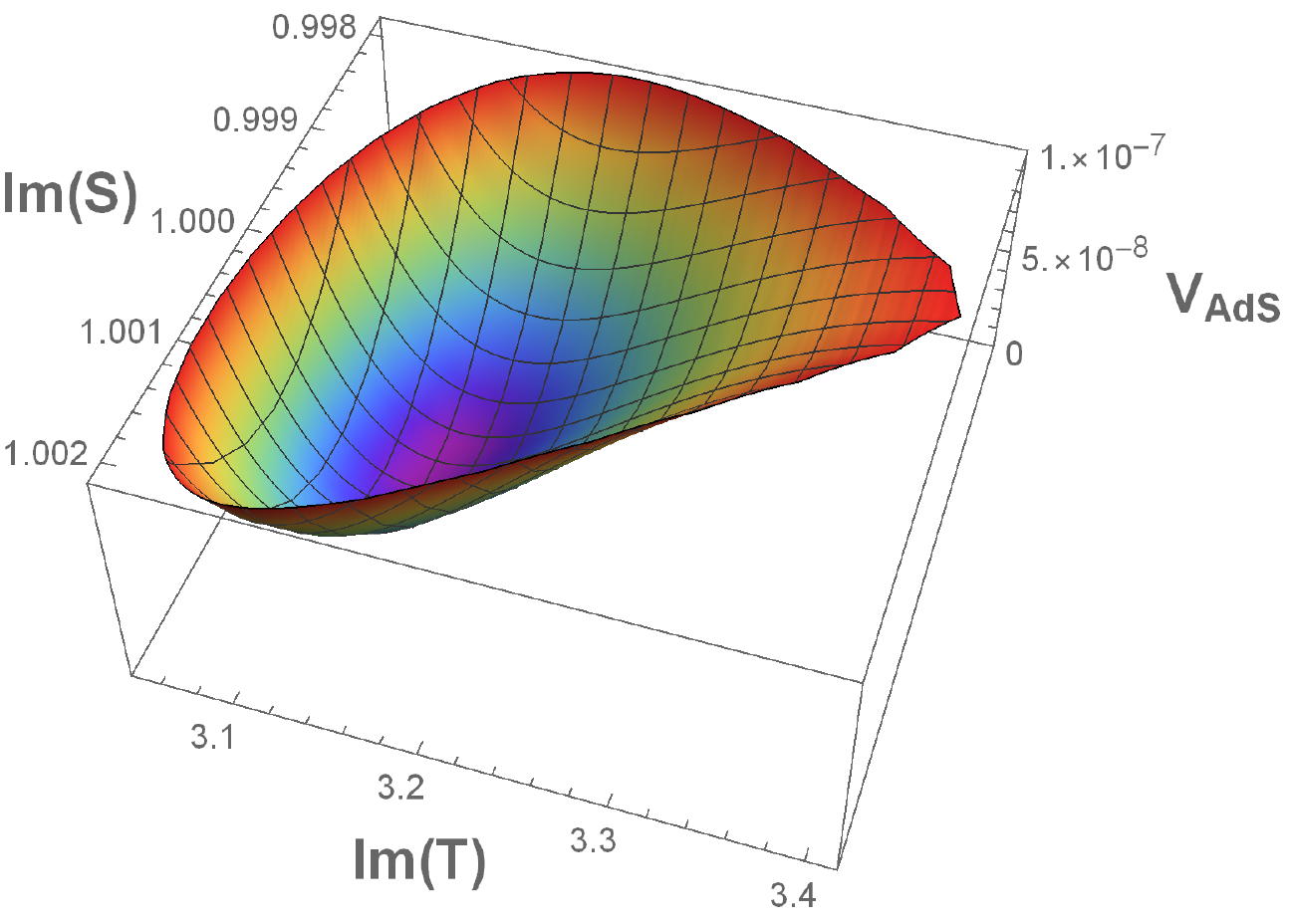}
\end{figure}
\begin{figure}[H]
\centering
\includegraphics[scale=0.55]{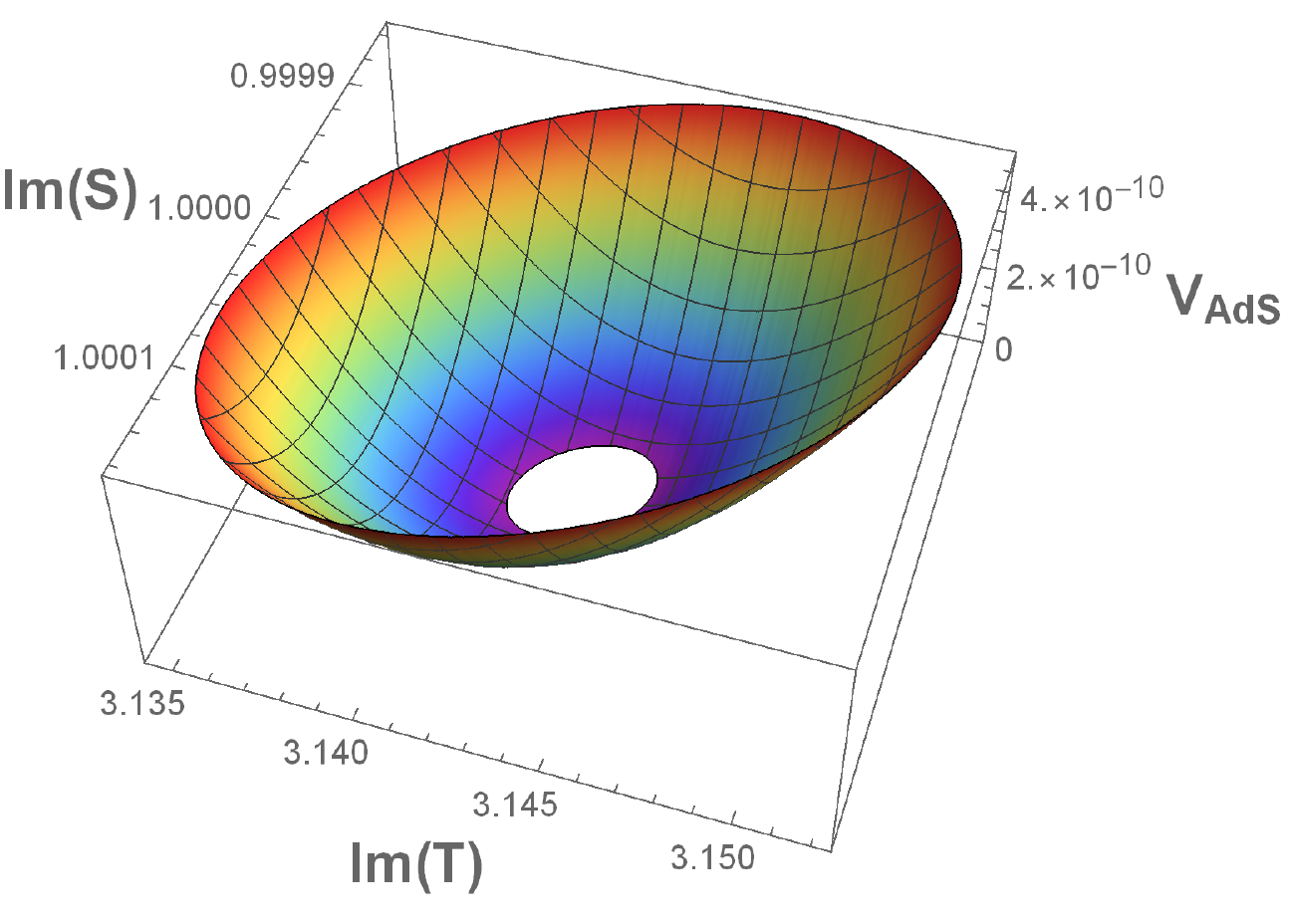} \qquad
\includegraphics[scale=0.53]{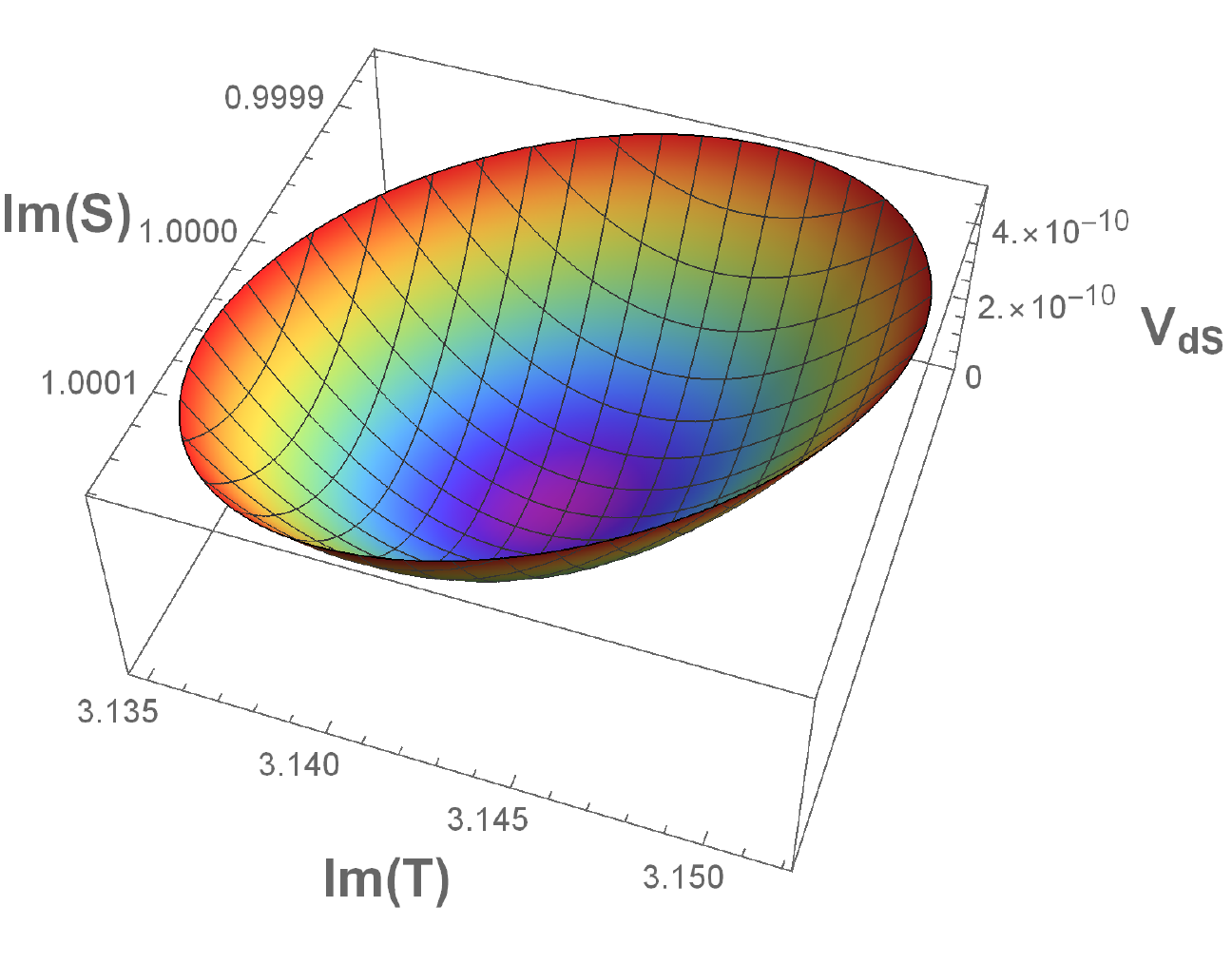}
\caption{\footnotesize  A complete set of 3D plots of the scalar potential for the K3 fibration of $\mathbb{C}P^4_{[1,1,2,2,6]} $. On the top we show the overall shape of the potential, and below a close-up of the minimum in AdS and dS case.}
\label{fig:2paraK33D}
\end{figure}
The eigenvalues of the masses in Minkowski and dS are given in table \ref{tab:2paraK3ev}. Since the model has only two moduli, we give also the complete mass matrices for Minkowski and dS.
\begin{equation}
\begin{aligned}
M_{K3fib,\,Mink}=&\begin{pmatrix}
5.22544 \cdot 10^{-2} & 3.23902 \cdot 10^{-4} & 0 & 0\\
3.23902 \cdot 10^{-4} & 1.58428 \cdot 10^{-2} & 0 & 0\\
0 & 0 & 5.22544 \cdot 10^{-2} & 3.23902 \cdot 10^{-4}\\
0 & 0 & 3.23902 \cdot 10^{-4} & 1.58428 \cdot 10^{-2}\\
\end{pmatrix},\\
M_{K3fib,\,dS}=&\begin{pmatrix}
5.21864 \cdot 10^{-2} & 3.20421 \cdot 10^{-4} & 0 & 0\\
3.20421 \cdot 10^{-4} & 1.5569 \cdot 10^{-2} & 0 & 0\\
0 & 0 & 5.2155 \cdot 10^{-2} & 3.20421 \cdot 10^{-4}\\
0 & 0 & 3.20421 \cdot 10^{-4} & 1.55606 \cdot 10^{-2}\\
\end{pmatrix}.\\
\end{aligned}
\end{equation}
\begin{table}[H]
\centering
\begin{tabular}{|c|c|c|}\hline
& Mink &  dS   \\\hline
$m_1^{\,2}$ & $\; 5.22564 \cdot 10^{-2} \;$ & $\; 5.21884 \cdot 10^{-2} \;$\\\hline
$m_2^{\,2}$ & $\; 5.22564 \cdot 10^{-2} \;$ & $\; 5.21875 \cdot 10^{-2} \;$\\\hline
$m_3^{\,2}$ & $\; 1.38346 \cdot 10^{-5} \;$ & $\; 1.36014 \cdot 10^{-5} \;$\\\hline
$m_4^{\,2}$ & $\; 1.38346 \cdot 10^{-5} \;$ & $\; 1.35926 \cdot 10^{-5} \;$\\\hline
\end{tabular}
\caption{   Eigenvalues of the mass matrices in Minkowski and dS. In this model and for this choice of the parameters, the mass splitting between the fields and their axionic partners is clearly visible, after moving away from Minkowski space.}
\label{tab:2paraK3ev}
\end{table}

\subsection{K3-fibration with three parameters}
This model is a generalization of the previous one, with the inclusion of a blow-up mode $\tau_3$. The six-dimensional volume is given in equation (3.28) of \cite{Cicoli:2008va}
\begin{equation}
\label{eq:VX3parK3tau}
\mathcal{V}_6(\tau_i)= \alpha \left( \sqrt{\tau_1} \left(\tau_2 - \beta \tau_1\right) - \gamma \tau_3^{3/2} \right)\,,
\end{equation}
and, in our conventions, is 
\begin{equation}
\label{eq:VX3parK3}
\mathcal{V}_6 (S,T,U)= \alpha \left( \sqrt{\left(-\rmi(S-\bar{S})\right)} \left[\left(-\rmi(T-\bar{T})\right) - \beta \left(-\rmi(S-\bar{S})\right)\right] - \gamma \left(-\rmi(U-\bar{U})\right)^{3/2} \right)\,.
\end{equation}
The parameters $\alpha$, $\beta$ and $\gamma$ are positive and model dependent. ${\rm Im}(U)$ is the blow up mode, that turns the model we investigate here into  $\mathbb{C}P^4_{[1,1,2,2,6]} $. 

\begin{figure}[H]
\centering
\includegraphics[scale=0.6]{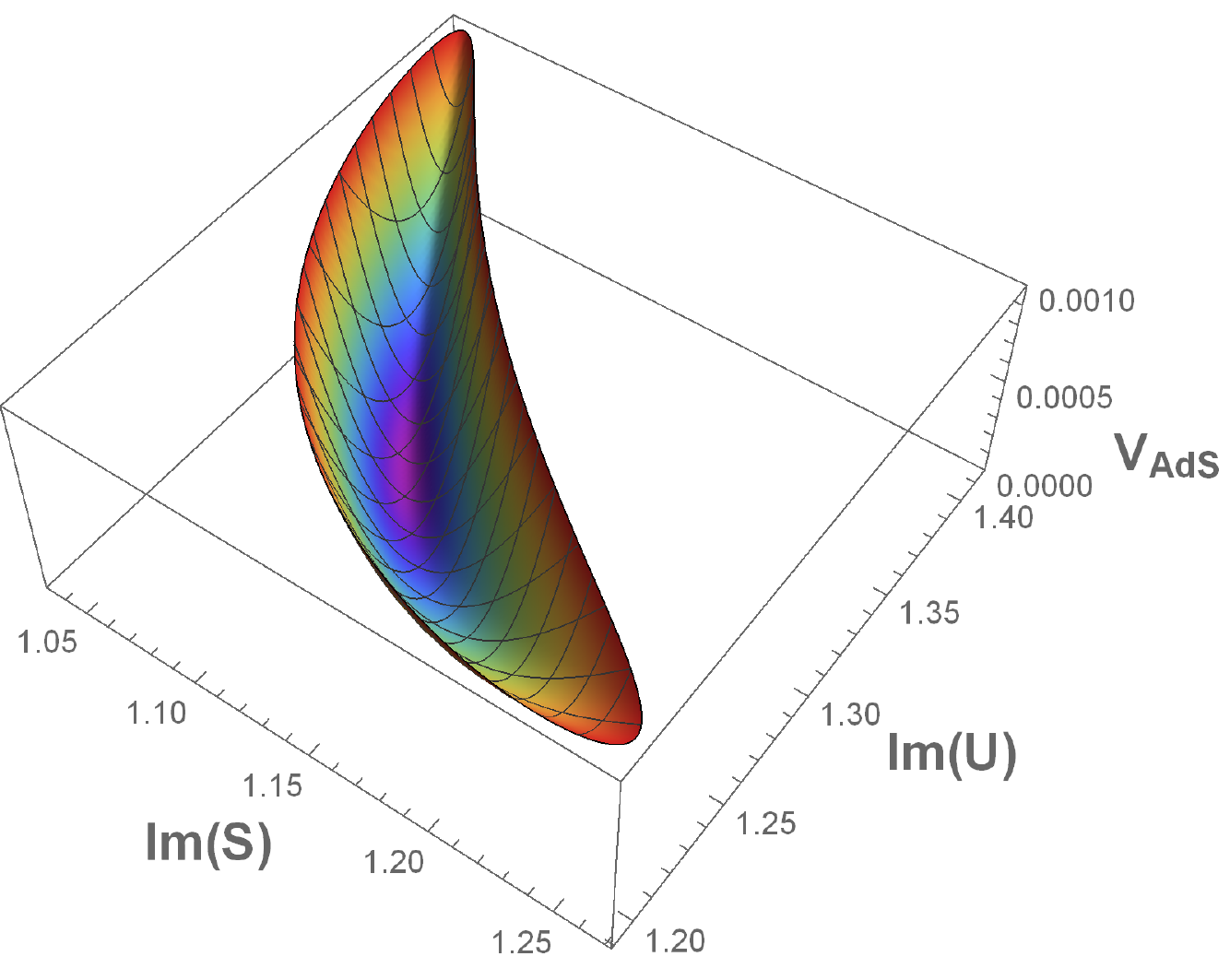}
\caption{\footnotesize  A 3D slice in the ${\rm Im}(U)$ and ${\rm Im}(S)$ direction of the three-parameters K3 fibration model.}
\label{fig:3paraK33D}
\end{figure}

The parameters we use in the analysis are given in table \ref{tab:para3parK3}.
\begin{table}[H]
\centering
\begin{tabular}{|c|c|c|}\hline
$A_S = 1.1$ & $A_T = 1.2$ & $A_U =1.3$\\\hline
$a_S = 2.1$ & $a_T = 2.2$ & $a_U = 2.3$\\\hline
$b_S = 3.1$ & $b_T = 3.2$ & $b_U = 3.3$\\\hline
$\alpha = 1$ & $\beta = \frac{1}{4}$ & $\gamma = \frac{1}{2}$\\\hline
$S_0 = 1.1$ & $T_0 = 1.2$ & $U_0 = 1.3$\\\hline
\end{tabular}
\caption{  The parameters used in the three-parameters K3 fibration.}
\label{tab:para3parK3}
\end{table}
Moreover, we take
\begin{equation}
\Delta W_0 = -5 \cdot 10^{-7}
\end{equation}
for the downshift, while for the uplift parameters we have 
\begin{equation}
\begin{aligned}
\mu^4_{bulk}= 1.16982 \cdot 10^{-12} \qquad \text{or} \qquad
\mu^4_{throat} = 3.80480 \cdot 10^{-12}\,.
\end{aligned}
\end{equation}
In figure \ref{fig:3paraK33D} we show a 3D slice of the scalar potential. We give the eigenvalues of the mass matrix in table \ref{tab:3paraK3ev}.
\begin{table}[H]
\centering
\begin{tabular}{|c|c|c|}\hline
& Mink  & dS \\\hline
$m_1^{\,2}$ & $\; 2.76956 \;$ & $\; 2.76898 \;$  \\\hline
$m_2^{\,2}$ & $\; 2.76956 \;$ & $\; 2.76898 \;$ \\\hline
$m_3^{\,2}$ & $\; 1.38664 \cdot 10^{-1} \;$ & $\; 1.38648 \cdot 10^{-1} \;$ \\\hline
$m_4^{\,2}$ & $\; 1.38664 \cdot 10^{-1} \;$ & $\; 1.38647 \cdot 10^{-1} \;$ \\\hline
$m_5^{\,2}$ & $\; 7.64496 \cdot 10^{-3} \;$ & $\; 7.64471 \cdot 10^{-3} \;$ \\\hline
$m_6^{\,2}$ & $\; 7.64496 \cdot 10^{-3} \;$ & $\; 7.64391 \cdot 10^{-3} \;$ \\\hline
\end{tabular}
\caption{  Masses in three-parameters K3 fibration. The mass splitting between the real scalar and the pseudoscalar is very small and does not appear within the digits we are showing explicitly. The dS case represents the anti-D3 brane in the bulk.}
\label{tab:3paraK3ev}
\end{table}

\subsection{K3 fibered CY model used for fibre inflation}
An interesting subcase of the previous model is obtained by taking $\beta=0$ in \eqref{eq:VX3parK3}. This results in
\begin{equation}
\label{eq:VXfibreX}
\mathcal{V}_6(\tau_i) =  \alpha \left[\sqrt{\tau_1}\,  \tau_2 - \gamma \tau_3^{3/2}\right],
\end{equation}
or
\begin{equation}
\label{eq:VXfibre}
\mathcal{V}_6(S,T,U) =  \alpha \left[\sqrt{\left(-\rmi (S - \bar{S})\right)} \left(-\rmi (T - \bar{T})\right) - \gamma \left(-\rmi (U - \bar{U})\right)^{3/2}\right],
\end{equation}
in our own notation. In particular, ${\rm Im}(S)$ corresponds to the volume of the K3-fibre, ${\rm Im}(T)$ controls the overall volume and ${\rm Im}(U)$ is related to the blow-up mode. Again, $\gamma$ and $\alpha$ are positive constants. This explicit example of type IIB  \K\, potential was used  for  \emph{fibre inflation} \cite{Cicoli:2008gp,Burgess:2016owb,Kallosh:2017wku}. Fibre inflation is based on the LVS scenario \cite{Balasubramanian:2005zx}, which can be used to produce a potential with a flat plateau, suitable for inflation. While constructing a model for inflation is not the goal of this work, it is still interesting to study such a setup in the context of mass production of dS vacua. 

%With the KL-superpotential (\ref{eq:KLpotIIB}) we will follow our 3 step procedure once more. Again, we can use the solution of $D_\Phi W = 0$ and $W=0$ for the KL-superpotential to arrive at a Minkowski vacuum. Subsequently shifting $W \to W + \Delta W$ will give a supersymmetric AdS vacuum where the change in the position of the minimum and in the masses is set by $\Delta W$ and thus parametrically small. At last, we can use an anti-D3-brane for the uplift,   either in the form (\ref{eq:antiD3upliftBulk}), or (\ref{eq:antiD3upliftThroat}), in order to obtain a meta-stable dS vacuum.

For our analysis, we take the parameters reported in table \ref{tab:parafibre}.
\begin{table}[H]
\centering
\begin{tabular}{|c|c|c|}\hline
$A_S = 1.1$ & $A_T = 1.2$ & $A_U =1.3$\\\hline
$a_S = 2.1$ & $a_T = 2.2$ & $a_U = 2.3$\\\hline
$b_S = 3.1$ & $b_T = 3.2$ & $b_U = 3.3$\\\hline
$S_0 = 1$ & $T_0 = 1$ & $U_0 = 1$\\\hline
$\alpha = 1$ & $\gamma = \frac{1}{2} $& \\\hline
\end{tabular}
\caption{  One possible set of parameters for the fibre inflation model.}
\label{tab:parafibre}
\end{table}
As predicted in \eqref{nond1} and \eqref{nond}, the mass matrices have a block diagonal form. We recall that, in contrast to our type IIA examples, the Minkowski masses are no longer diagonal, due to the more intricate form of the \K\,  potential given by $\mathcal{V}_6$ in \eqref{eq:VXfibre}. 
The downshift we used in this case was
\begin{equation}
\Delta W_0 = -10^{-5}
\end{equation}
while the value for the uplift parameter was chosen, again for illustrative purposes, to be 
\begin{equation}
\mu^4_{bulk} = 3.10079 \cdot 10^{-10}
\end{equation}
in the bulk or, for an anti-D3-brane at the bottom of a warped throat
\begin{equation}
\mu^4_{throat} = 2.46069 \cdot 10^{-10}.
\end{equation}
Below, we will show the plots and the results related to the case with the anti-D3 brane in the bulk. Indeed, the placement of the anti-D3 brane at the bottom of a throat gives very similar results in all investigated examples, with the uplift parameters we give here.

\begin{figure}[H]
\includegraphics[scale=0.55]{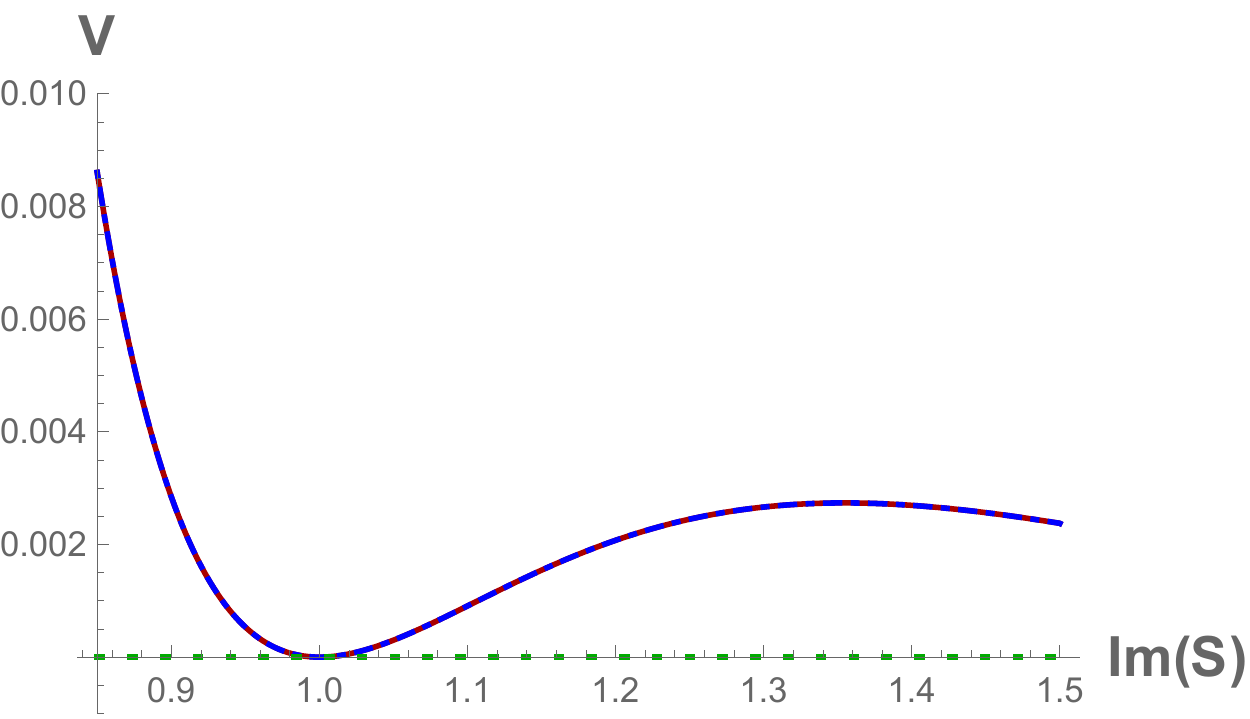} \qquad\includegraphics[scale=0.6]{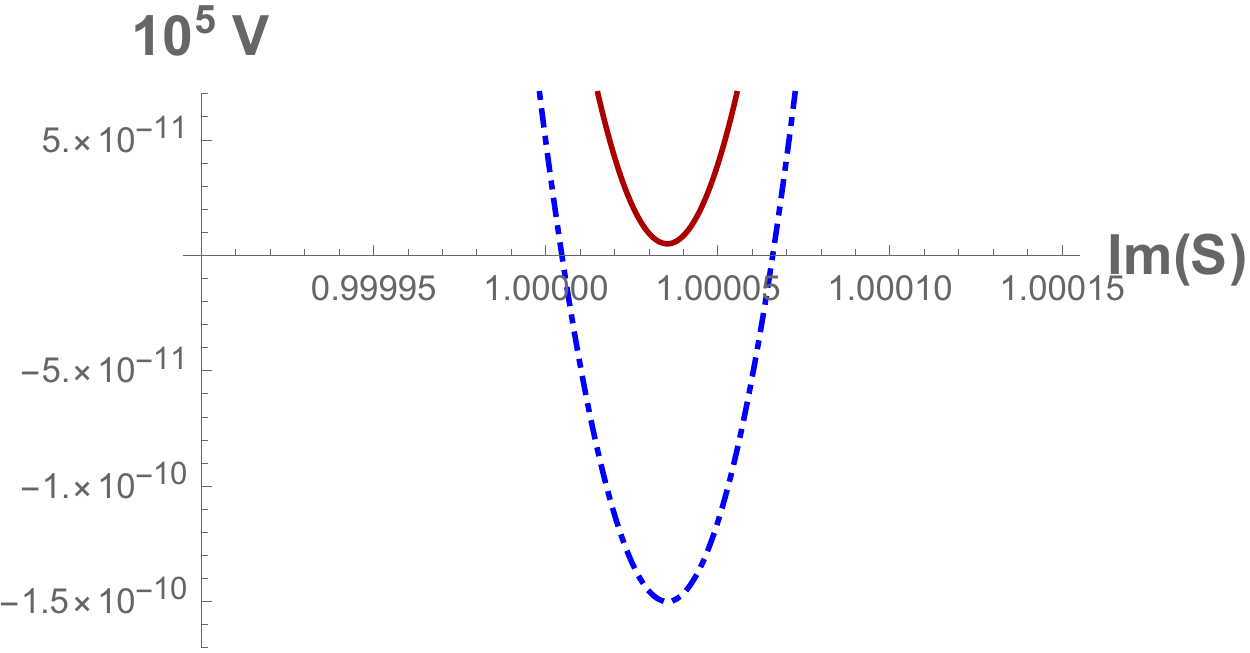}
\end{figure}
\begin{figure}[H]
\includegraphics[scale=0.55]{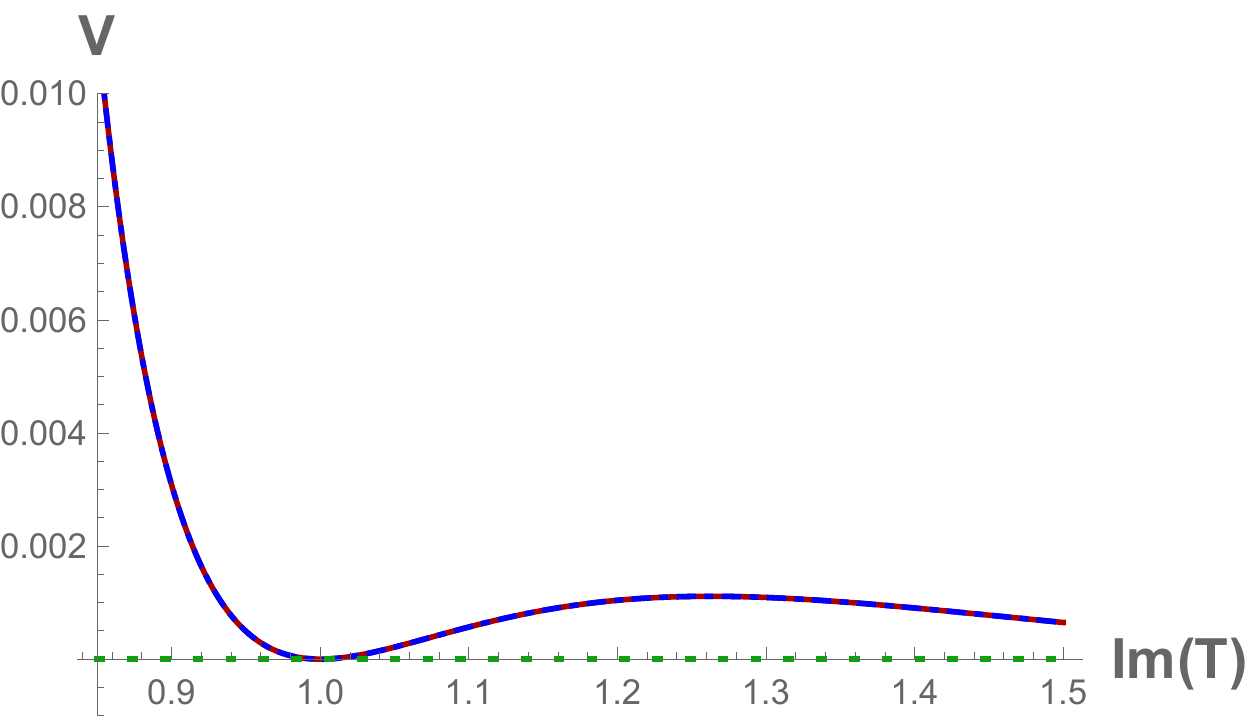} \qquad\includegraphics[scale=0.6]{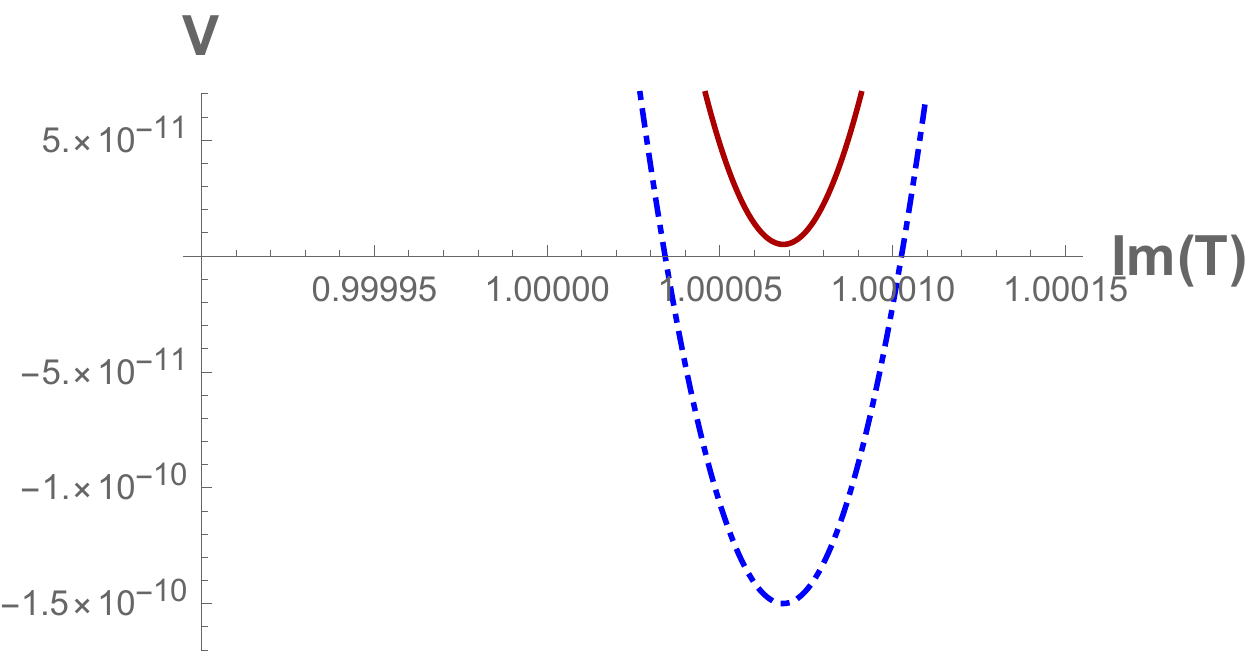}\\
\end{figure}
\begin{figure}[H]
\includegraphics[scale=0.55]{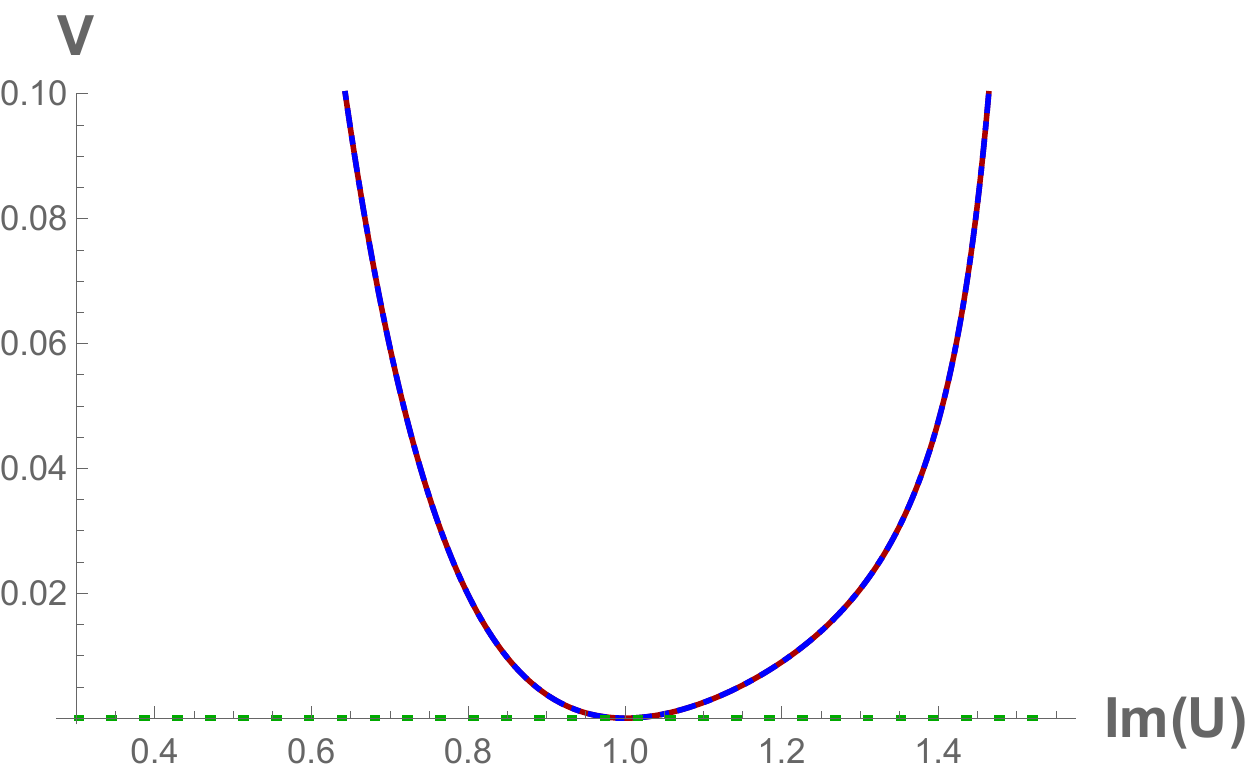} \qquad\includegraphics[scale=0.6]{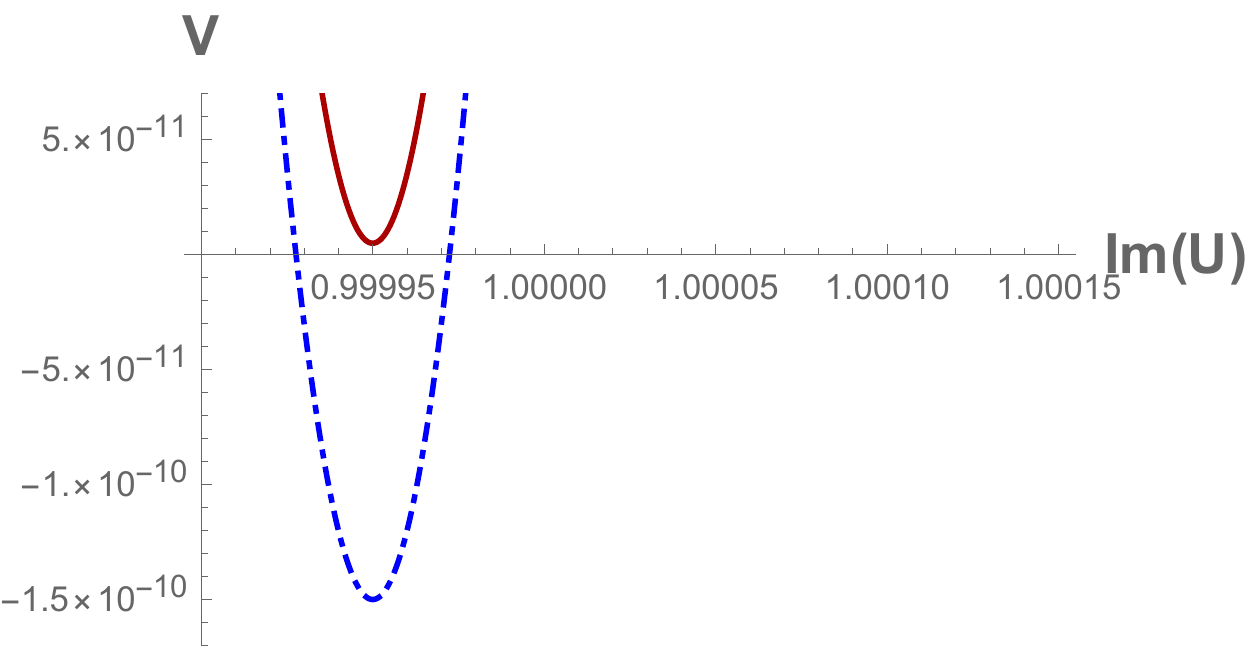}
\caption{\footnotesize From top to bottom, we have the overall form of the potential on the left, for AdS and dS as well as a close-up of the minimum on the right for the directions Im$(S)$, Im$(T)$ and Im$(U)$. The shift of the minimum from the initial point, with the imaginary part of the moduli set at one, is also visible.}
\label{fig:fibre2D}
\end{figure}

\begin{figure}[H]
\centering
\includegraphics[scale=0.7]{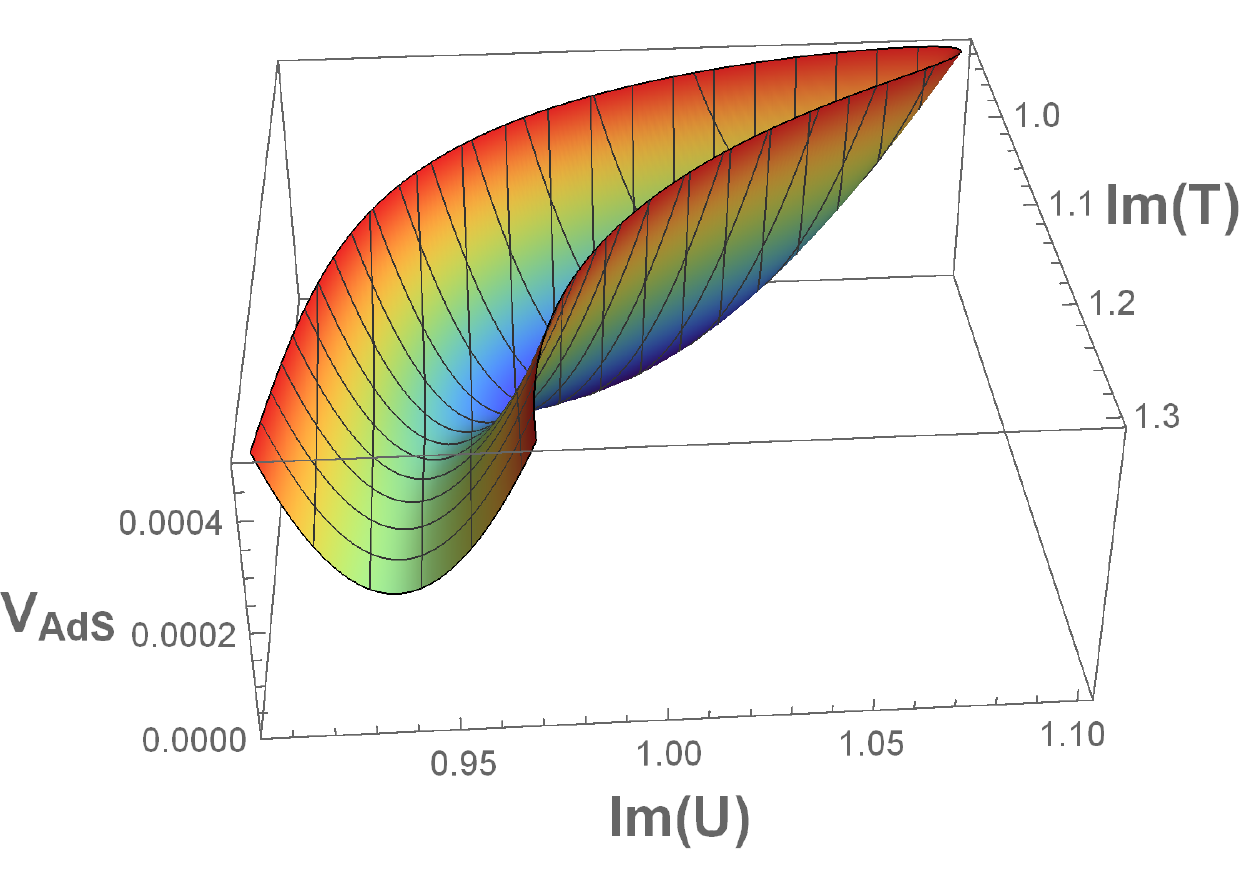}
\caption{\footnotesize A 3D slice of the scalar potential in the Im$(T)$ and Im$(U)$ directions.}
\label{fig:fibre3D}
\end{figure}

In order to illustrate this model better, in figures \ref{fig:fibre2D} and \ref{fig:fibre3D} we include 2D plots for all directions as well as  3D plots for the Im$(T)$ and Im$(U)$ slice.
For the set of parameters in table \ref{tab:parafibre}, we get the following entries for the upper left corner of the mass matrix in the Minkowski vacuum as given in \eqref{nond} and in dS situation as given in \eqref{corrdS}:
\begin{equation}\label{nondiag}
\begin{aligned}
V_{ij}^{Mink}=&\begin{pmatrix}
0.3201 & 0.1655 & 0.3392 \\
0.1655 & 0.2567 & 0.3508 \\
0.3392 & 0.3508 & 0.5991 \\
\end{pmatrix},
\hskip 1 cm V_{ij}^{dS ( bulk)}=&\begin{pmatrix}
0.3198 & 0.1653 & 0.3391 \\
0.1653 & 0.2564 & 0.3506 \\
0.3391 & 0.3506 & 0.5992 \\
\end{pmatrix}\\
\end{aligned}
\end{equation}
The eigenvalues of the mass matrix in Minkowski and dS are given in table \ref{tab:fibev}.
\begin{table}[H]
\centering
\begin{tabular}{|c|c|c|}\hline
& Mink &  dS  \\\hline
$m_1^{\,2}$ & $\; 1.01997 \;$ & $\; 1.01957 \;$\\\hline
$m_2^{\,2}$ & $\; 1.01997 \;$ & $\; 1.01957 \;$\\\hline
$m_3^{\,2}$ & $\; 1.31424 \cdot 10^{-1} \;$ & $\; 1.31344 \cdot 10^{-1} \;$\\\hline
$m_4^{\,2}$ & $\; 1.31424 \cdot 10^{-1} \;$ & $\; 1.31338 \cdot 10^{-1} \;$\\\hline
$m_5^{\,2}$ & $\; 2.44807 \cdot 10^{-2} \;$ & $\; 2.44724 \cdot 10^{-2} \;$\\\hline
$m_6^{\,2}$ & $\; 2.44807 \cdot 10^{-2} \;$ & $\; 2.44665 \cdot 10^{-2} \;$\\\hline
\end{tabular}
\caption{  The eigenvalues of the mass matrix for the fibre inflation model, for Minkowski and dS. %Similar to the type IIA examples we see that the larger shift in the mass eigenvalues occurs during the downshift to IIA. The last two columns show that we can achieve a similar result by putting the brane in the bulk or at the bottom of a warped throat.
}
\label{tab:fibev}
\end{table}

\subsection{A complete intersection Calabi-Yau (CICY)  model}
Another interesting example is associated with a  Complete Intersection Calabi-Yau (CICY) manifold, presented in \cite{Bobkov:2010rf}.  There, the authors consider a Calabi-Yau three-fold with $h^{1,1} = h^{1,2} = 19$ and focus on a three moduli part of it, arguing that, under certain conditions, the 16 remaining moduli can be ignored. Here, we will simply use the three moduli part of their \K\, potential and show how the stabilization of moduli in the dS minimum can be realized. The overall volume is given by
\be
\label{eq:VXbob1}
\left(\mathcal{V}_6(\tau_i)\right) ^2 = \frac{1}{108} \tau_1 \left[6 \tau_2-\tau_1\right]\left[2\tau_3 -\tau_1\right]
\end{equation}
and in our choice of basis we have
\begin{equation}
\label{eq:VXbob}
\mathcal{V}_6^2 = \frac{1}{108}  \left(-\rmi (S-\bar{S})\right)\left[6\left(-\rmi (T-\bar{T})\right)-\left(-\rmi (S-\bar{S})\right)\right]\left[2\left(-\rmi (U-\bar{U})\right)-\left(-\rmi (S-\bar{S})\right)\right]\,.
\end{equation}
Following our procedure and using the solution for the KL-model that we described above, we arrive at a meta-stable dS vacuum in exactly the same manner as in our previous examples.

The parameters we chose are listed in table \ref{tab:parabob}.
\begin{table}[H]
\centering
\begin{tabular}{|c|c|c|}\hline
$A_S = 1.1$ & $A_T = 1.2$ & $A_U =1.3$\\\hline
$a_S = 2.1$ & $a_T = 2.2$ & $a_U = 2.3$\\\hline
$b_S = 3.1$ & $b_T = 3.2$ & $b_U = 3.3$\\\hline
$S_0 = 1$ & $T_0 = 1$ & $U_0 = 1$\\\hline
\end{tabular}
\caption{  Parameters for the Calabi-Yau model with $h^{1,1} = h^{1,2} = 19$.}
\label{tab:parabob}
\end{table}
\noindent A downshift with
\begin{equation}
\Delta W_0 = -5 \cdot 10^{-6}
\end{equation}
yields the following upper left corner of the AdS mass matrix
\begin{equation}
\begin{aligned}
M_{\text{CICY, AdS}}=&\begin{pmatrix}
\; 1.7284 \;&\; 0.2979 \;& 0.9157 \;\\
0.2979 & 1.3347 & 0.1578 \\
 0.9157 & 0.1578 & 0.9703 \\
\end{pmatrix}.
\end{aligned}
\end{equation}
The uplift in the bulk was performed with a parameter
\begin{equation}
\mu_{bulk}^4 = 7.68537 \cdot 10^{-11}
\end{equation}
and a similar result can be achieved by placing the anti-D3-brane at the bottom of a warped throat with
\begin{equation}
\mu_{throat}^4 = 1.07015 \cdot 10^{-10}.
\end{equation}
The eigenvalues of the mass matrix in all scenarios can be found in table \ref{tab:bobev}. 
\begin{table}[H]
\centering
\begin{tabular}{|c|c|c|}\hline
& Mink  & dS \\\hline
$m_1^{\,2}$ & $\; 2.44220 \;$ & $\; 2.44196 \;$\\\hline
$m_2^{\,2}$ & $\; 2.44220 \;$ & $\; 2.44195 \;$\\\hline
$m_3^{\,2}$ & $\; 1.23452 \;$ & $\; 1.23438 \;$\\\hline
$m_4^{\,2}$ & $\; 1.23452 \;$ & $\; 1.23436 \;$\\\hline
$m_5^{\,2}$ & $\; 3.57064 \cdot 10^{-1} \;$ & $\; 3.57041 \cdot 10^{-1} \;$\\\hline
$m_6^{\,2}$ & $\; 3.57064 \cdot 10^{-1} \;$ & $\; 3.57026 \cdot 10^{-1} \;$ \\\hline
\end{tabular}
\caption{   Mass eigenvalues for the CY-model with $h^{1,1} = h^{1,2} = 19$. Due to the smallness of the downshift and subsequent uplift, the values remain almost the same through the entire process.}
\label{tab:bobev}
\end{table}
 \noindent We illustrate the model in figure \ref{fig:bob2D2},  where we plot the situation for the uplift in the bulk for AdS and dS. Both the downshift as well as the uplift are visible and one can see how the minimum of the potential moves away slightly from the chosen original values we give in table \ref{tab:parabob}. In figure \ref{fig:bob3D}, we present a 3D slice of the potential in the dS case. 
\begin{figure}[H]
\includegraphics[scale=0.48]{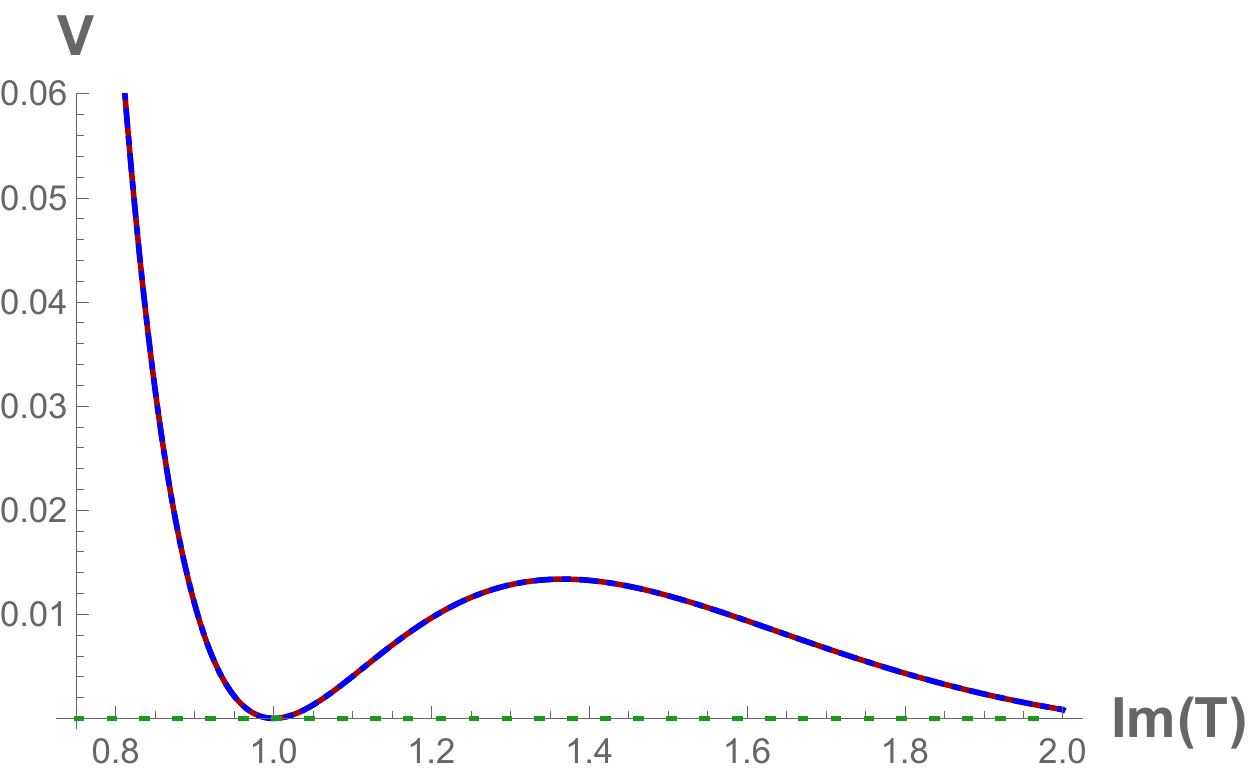}\qquad
\includegraphics[scale=0.6]{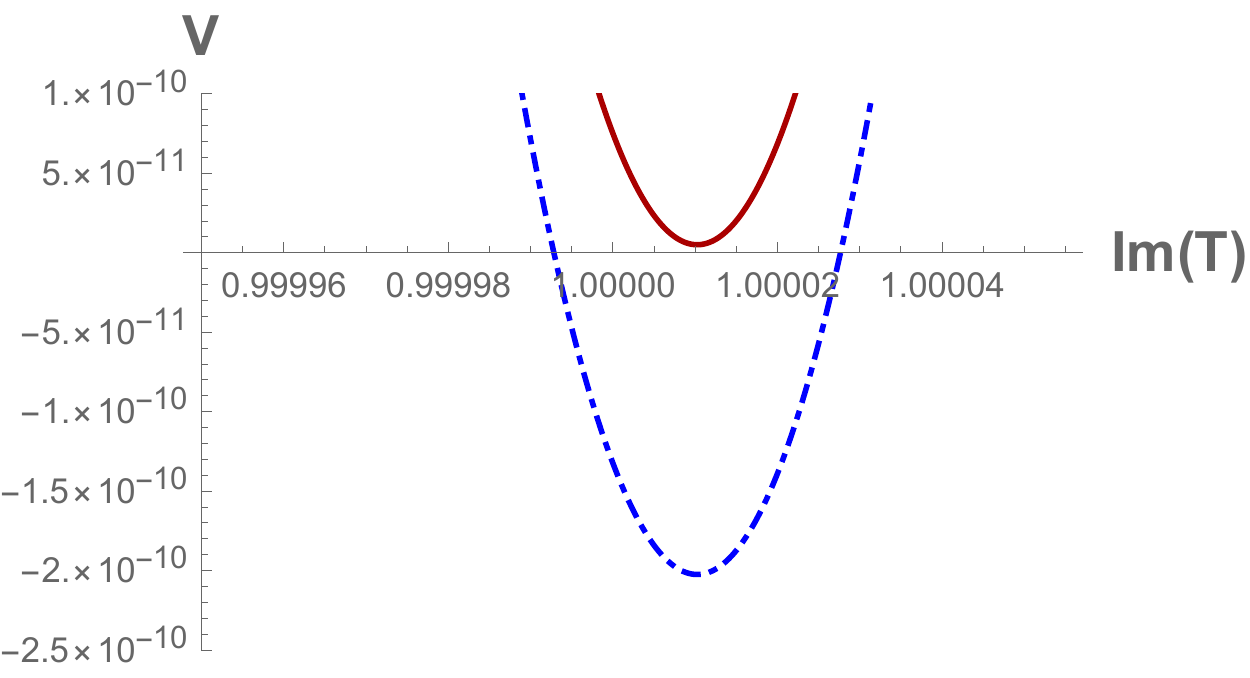}
\end{figure}
\begin{figure}[H]
\includegraphics[scale=0.48]{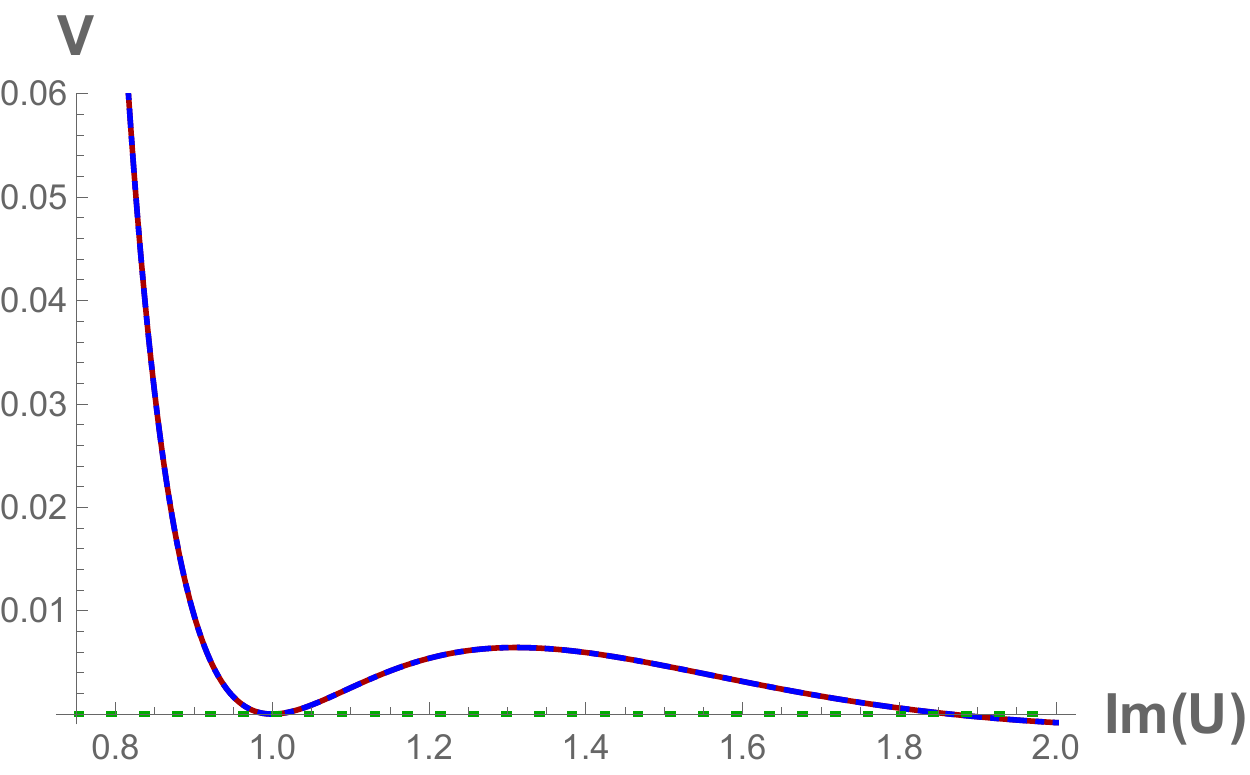}\qquad
\includegraphics[scale=0.6]{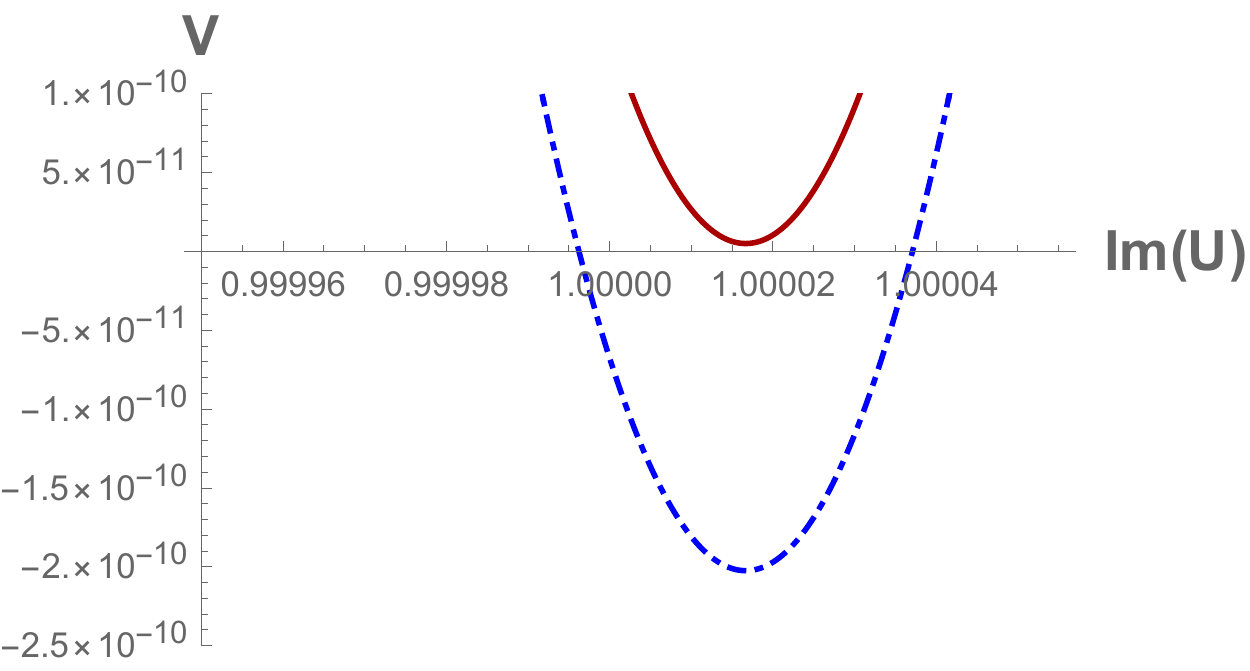}\\
\includegraphics[scale=0.48]{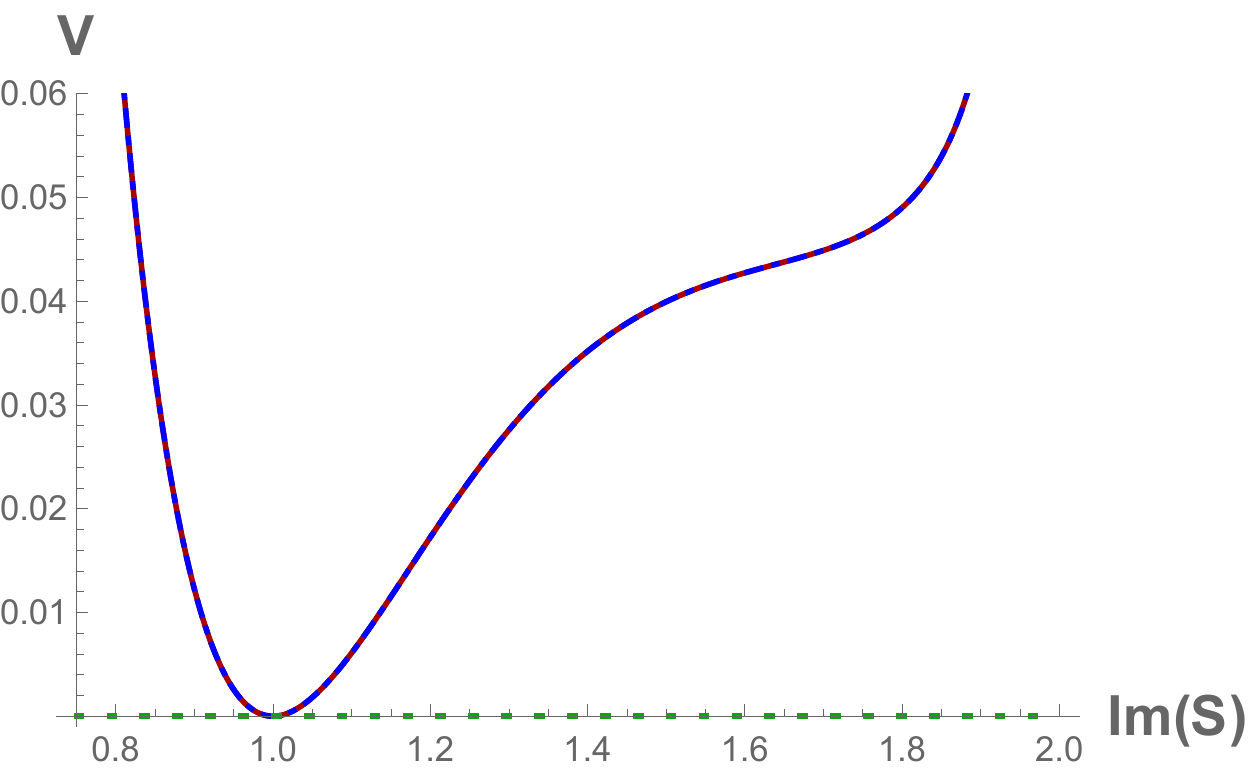}\qquad
\includegraphics[scale=0.6]{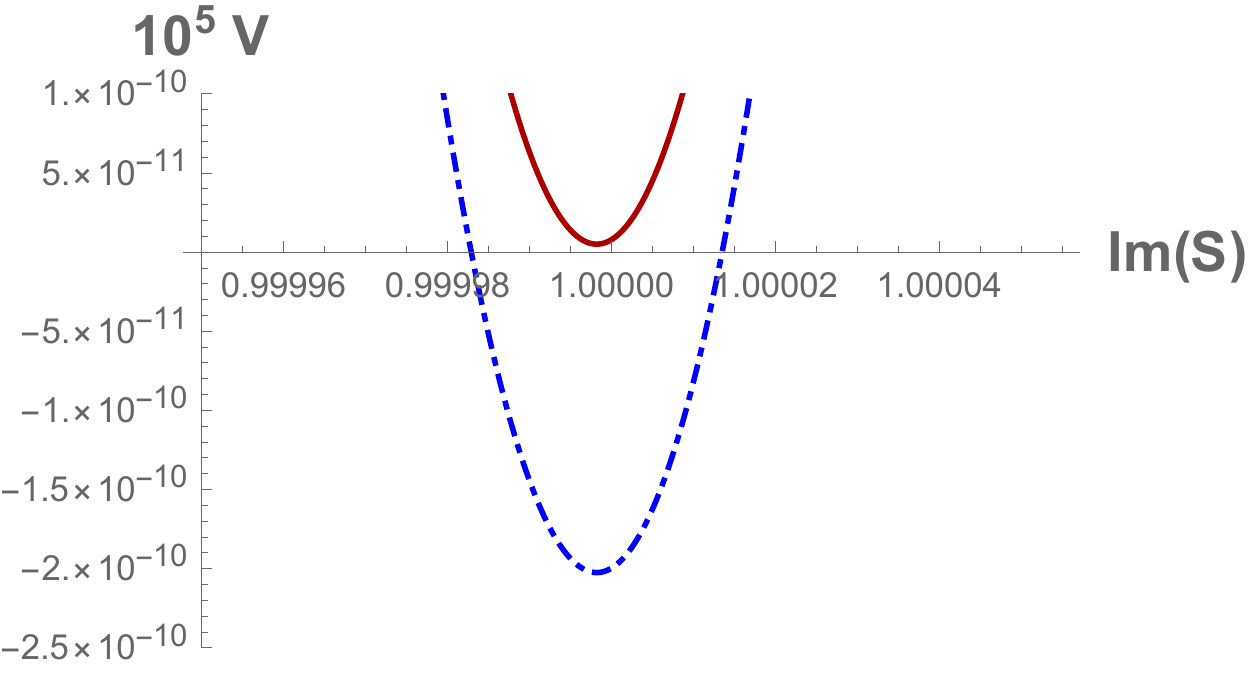}
\caption{\footnotesize The 2D plots for the Calabi-Yau model with $h^{1,1} = h^{1,2} = 19$. For all 3 moduli we have the overall form of the potential on the left and a close up of the minimum in AdS and dS (bulk) on the right.}
\label{fig:bob2D2}
\end{figure}
\begin{figure}[H]
\centering
\includegraphics[scale=0.5]{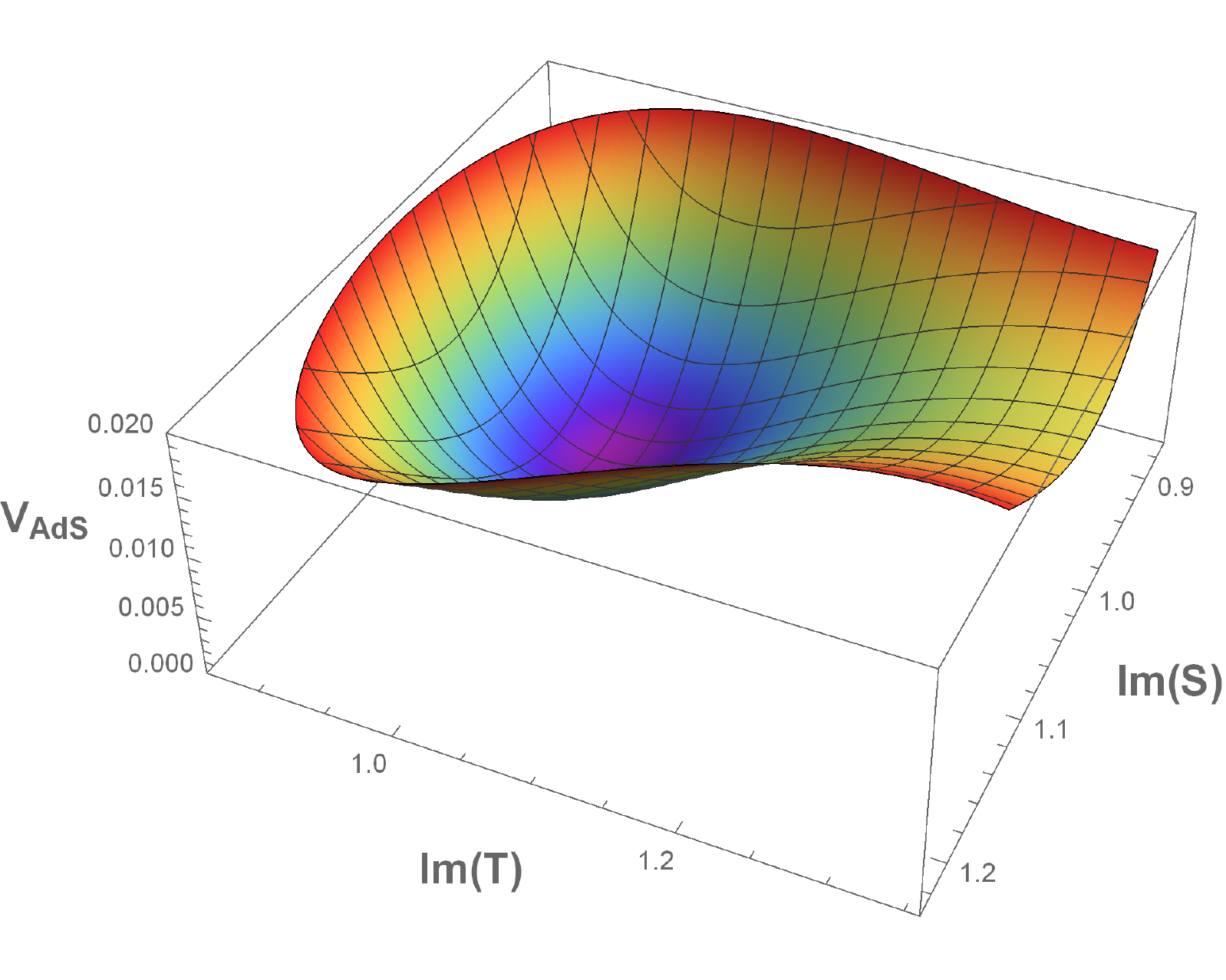}
\caption{\footnotesize 3D plot of the Im$(S)$ and Im$(T)$ slice of the Calabi-Yau model with $h^{1,1} = h^{1,2} = 19$.}
\label{fig:bob3D}
\end{figure}

\subsection{Multi-hole Swiss cheese model}
One more model, given in equation (3.8) in \cite{Cicoli:2008va}, is called a \emph{multi-hole swiss cheese} model. It is based on the Fano three-fold $\mathcal{F}_{11}$ which is explained in detail in \cite{Denef:2004dm}. This three-fold is topologically equivalent to a Calabi-Yau three-fold with $h^{1,1} = 3$ and $h^{2,1} = 111$. The volume $\mathcal{V}_6$ is given as
\begin{equation}
\mathcal{V}_6 (\tau_i) = \frac{1}{3 \sqrt{2}} \left( 2 \left[ \tau_1 + \tau_2 + 2 \tau_3 \right]^{3/2} - \left[ \tau_2 + 2 \tau_3 \right]^{3/2} - \tau_2^{3/2} \right)
\end{equation}
and  in our parametrisation, it reads
\begin{equation}
\begin{aligned}
\label{eq:volmultihole}
\mathcal{V}_6 = \frac{1}{3 \sqrt{2}} \Big(& 2\left[ \left(-\rmi (S-\bar{S}) \right) + \left(-\rmi (T-\bar{T}) \right) + 2 \left(-\rmi (U-\bar{U})\right) \right]^{3/2} \\
\;-& \left[ \left(-\rmi (T-\bar{T})\right) + 2\left(-\rmi (U-\bar{U})\right) \right]^{3/2} -  \left(-\rmi (T-\bar{T})\right)^{3/2} \Big)\,.
\end{aligned}
\end{equation}

The set of parameters that we chose is listed in table \ref{tab:parafano}. 
\begin{table}[H]
\centering
\begin{tabular}{|c|c|c|}\hline
$A_S = 1.1$ & $A_T = 1.2$ & $A_U =1.3$\\\hline
$a_S = 2.1$ & $a_T = 2.2$ & $a_U = 2.3$\\\hline
$b_S = 3.1$ & $b_T = 3.2$ & $b_U = 3.3$\\\hline
$S_0 = 1$ & $T_0 = 1$ & $U_0 = 1$\\\hline
\end{tabular}
\caption{  Our choice of parameters for the model based on the Fano three-fold.}
\label{tab:parafano}
\end{table}
\noindent For the downshift we chose 
\begin{equation}
\Delta W_0 = -5 \cdot 10^{-6},
\end{equation}
while the uplift parameters for the placement of the anti-D3-brane in the bulk and at the bottom of a warped throat are
\begin{equation}
\begin{aligned}
\mu_{bulk} ^4&= 9.62862 \cdot 10^{-11}\;,\\
\mu_{throat} ^4&= 2.75437 \cdot 10^{-11} \;.
\end{aligned}
\end{equation}
The eigenvalues of the mass matrix are given in table \ref{tab:fanoev} and we illustrate the model in the figures \ref{fig:fano2D} and \ref{fig:fano3D}.
\begin{table}[H]
\centering
\begin{tabular}{|c|c|c|}\hline
& Mink  & dS \\\hline
$m_1^{\,2}$ & $\; 1.85578 \cdot 10^{-1} \;$ & $\; 1.85557 \cdot 10^{-1} \;$\\\hline
$m_2^{\,2}$ & $\; 1.85578 \cdot 10^{-1} \;$ & $\; 1.85557 \cdot 10^{-1} \;$\\\hline
$m_3^{\,2}$ & $\; 1.00760 \cdot 10^{-1} \;$ & $\; 1.00753 \cdot 10^{-1} \;$\\\hline
$m_4^{\,2}$ & $\; 1.00760 \cdot 10^{-1} \;$ & $\; 1.00753 \cdot 10^{-1} \;$\\\hline
$m_5^{\,2}$ & $\; 1.30646 \cdot 10^{-2} \;$ & $\; 1.30627 \cdot 10^{-2} \;$\\\hline
$m_6^{\,2}$ & $\; 1.30646 \cdot 10^{-2} \;$ & $\; 1.30626 \cdot 10^{-2} \;$\\\hline
\end{tabular}
\caption{  The masses for the Fano three-fold model. The shift from Minkowski to  dS is small but visible with this precision. The discrepancy of field-axion pairs is tiny and only appears in the last pair.}
\label{tab:fanoev}
\end{table}

\begin{figure}[H]
\center
\includegraphics[scale=0.528]{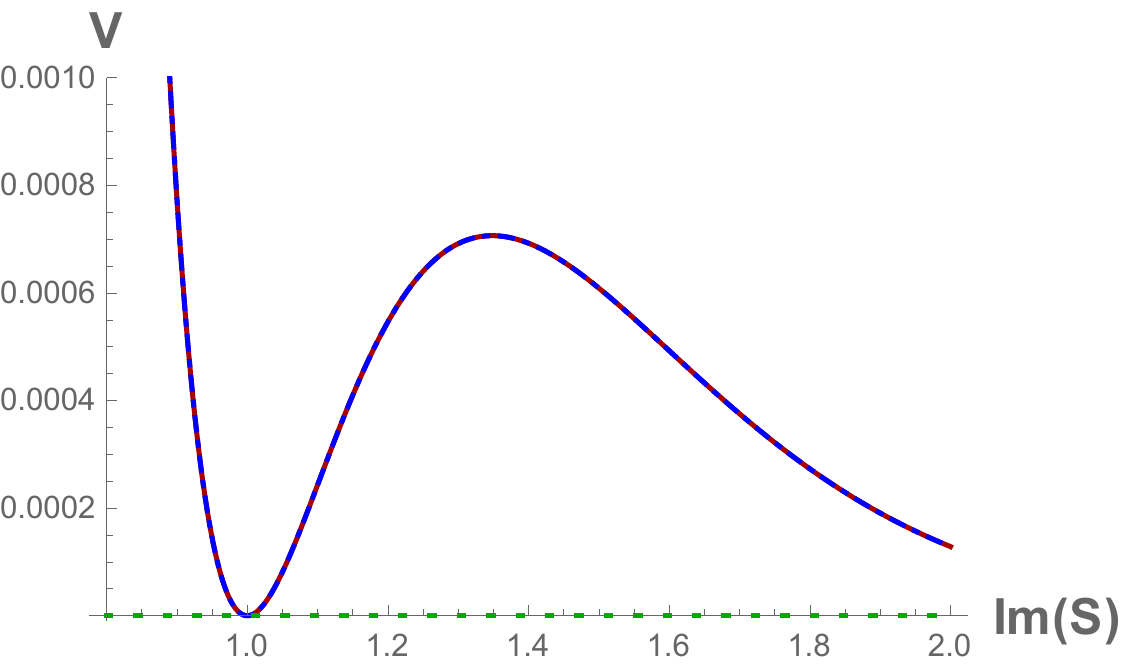}\qquad\includegraphics[scale=0.568]{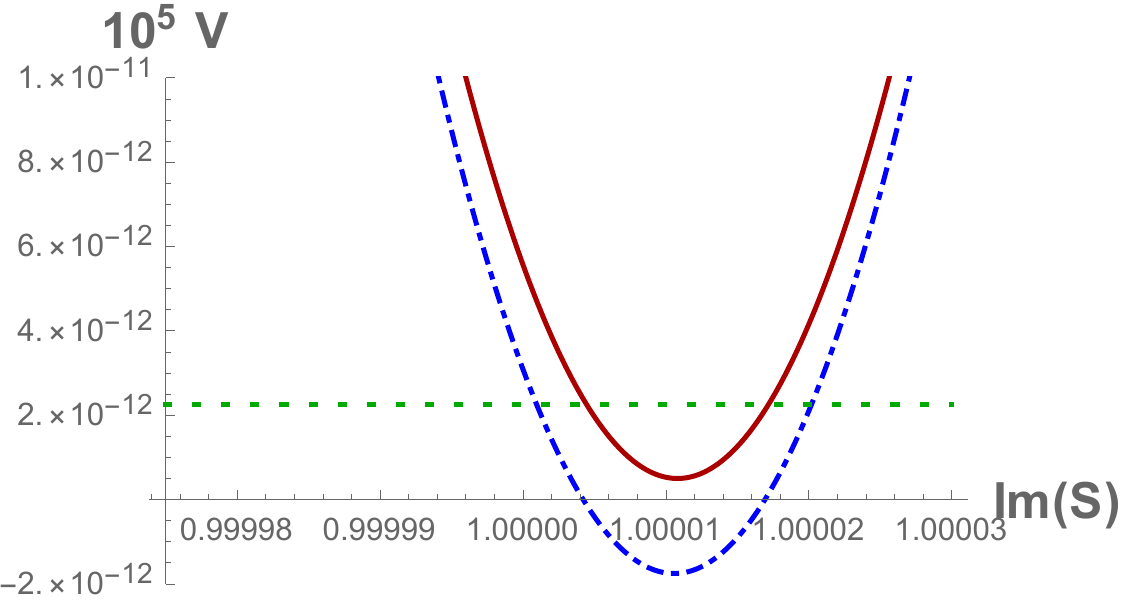}
\end{figure}
\begin{figure}[H]
\center
\includegraphics[scale=0.528]{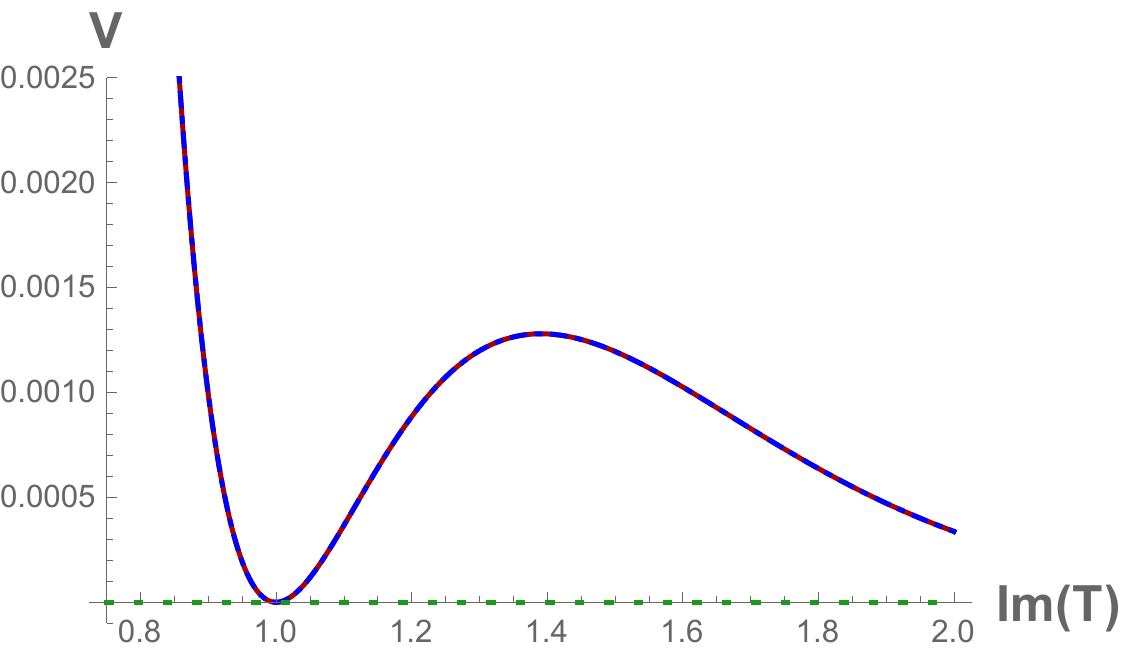}\qquad\includegraphics[scale=0.568]{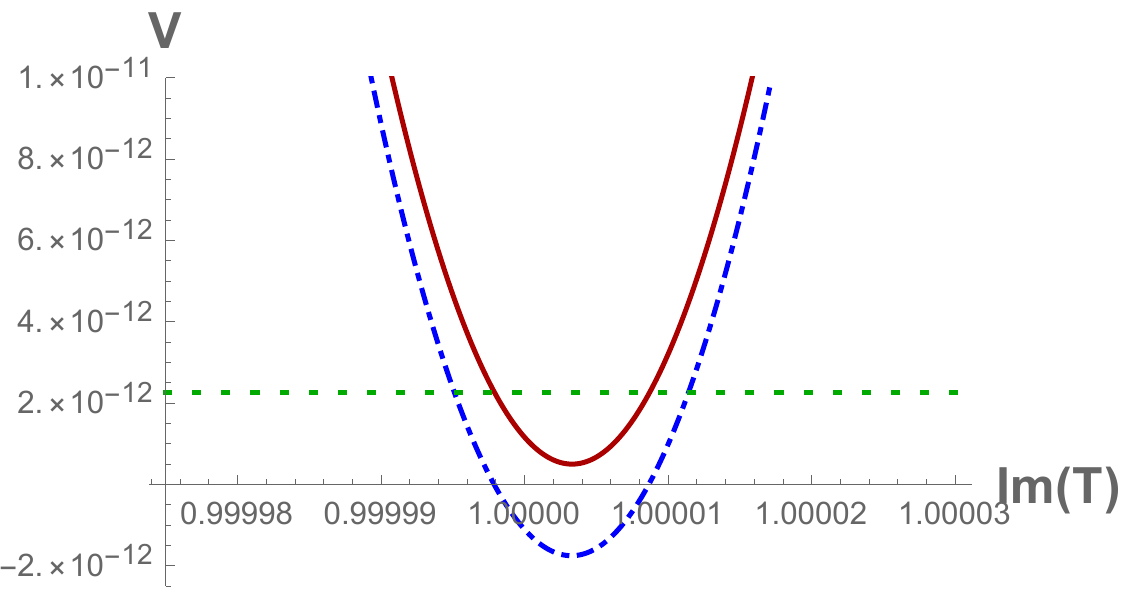}\\
\includegraphics[scale=0.528]{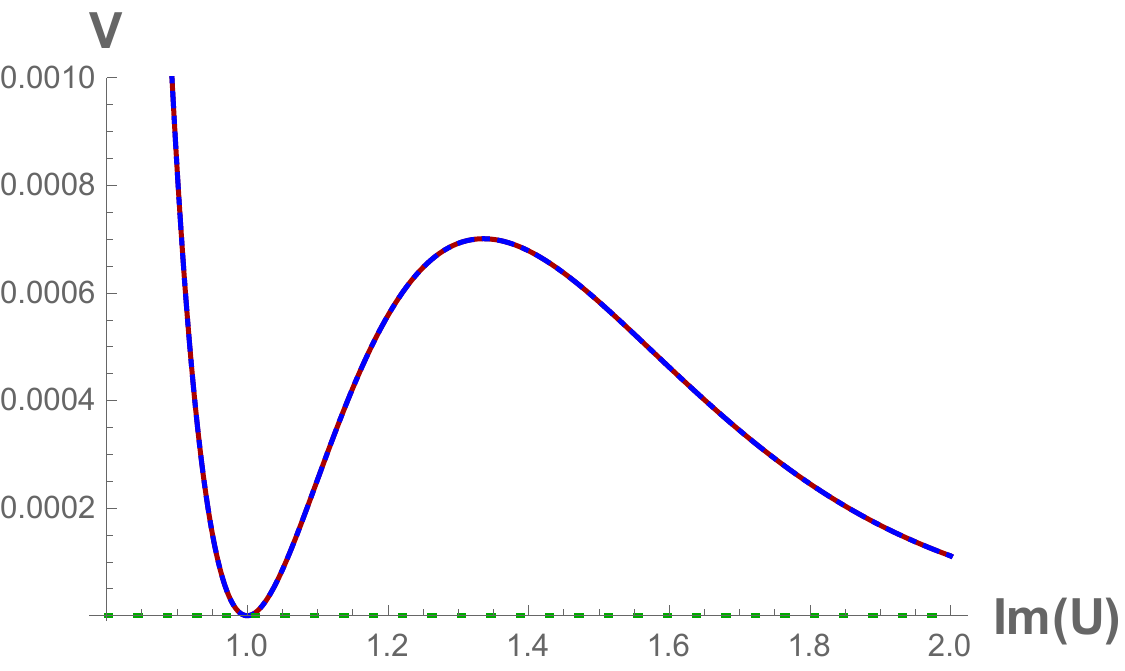}\qquad\includegraphics[scale=0.568]{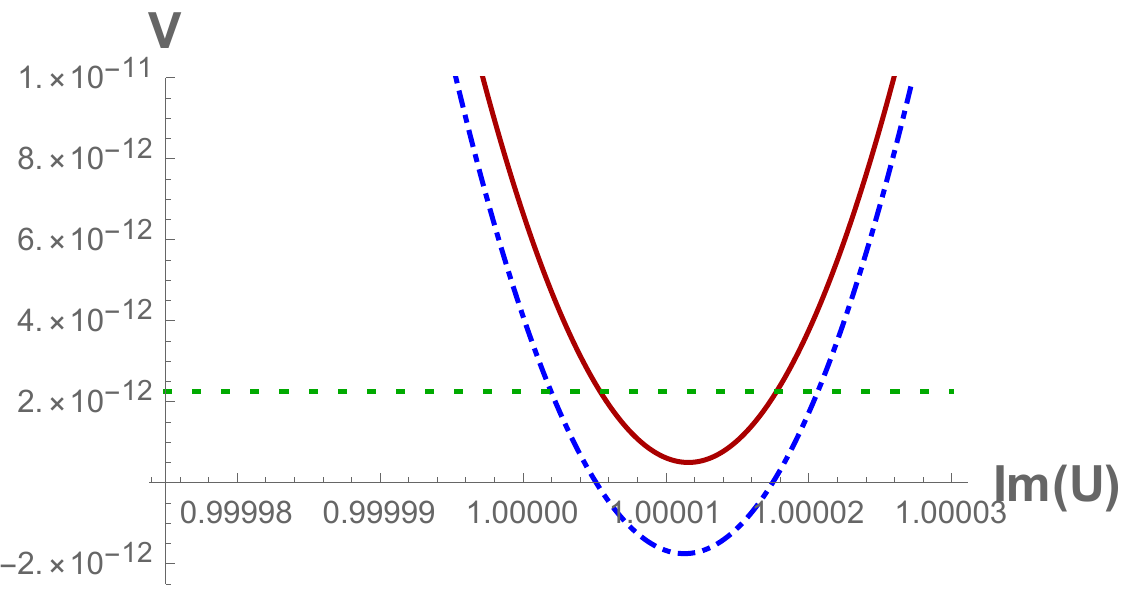}
\caption{\footnotesize  2D plots for the Fano model. Again, the overall behaviour does not change much during the process and indeed the lines for AdS and dS are indistinguishable, on the left. On the right, a zoomed region is present in which not only the shift of the minima is visible, but also the constant contribution of the anti-D3-brane.}
\label{fig:fano2D}
\end{figure}

\begin{figure}[H]
\centering
\includegraphics[scale=0.65]{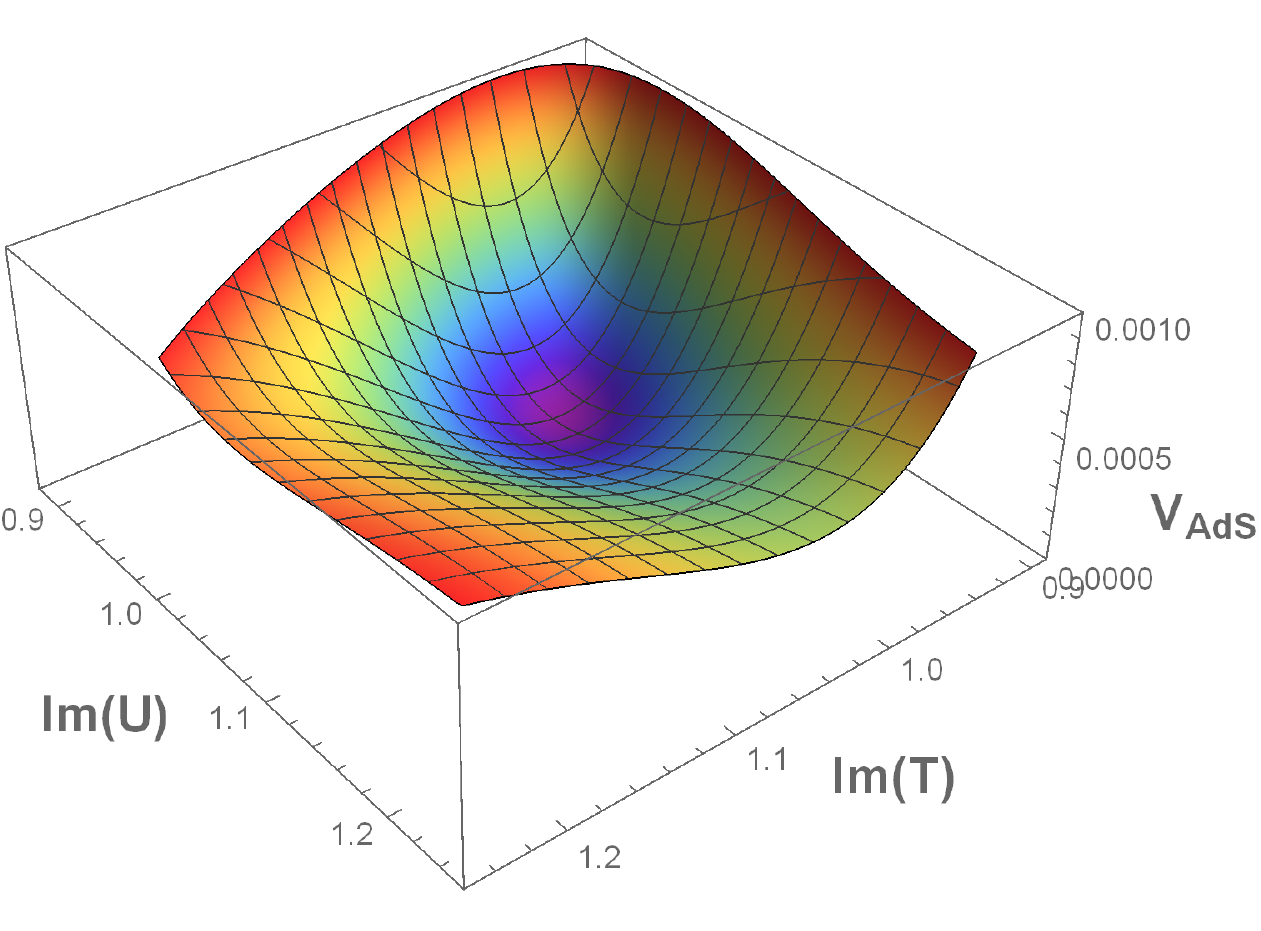}
\caption{\footnotesize One 3D slice of the Fano model scalar potential. Depicted are the Im$(T)$ and Im$(U)$ directions.}
\label{fig:fano3D}
\end{figure}

 \section{Stable dS states obtained by a large downshift and uplift}
\label{big} 

In this work, we have made the general prediction, confirmed by all of the examples studied thus far, that sufficiently small deformations (such as downshift and uplift) of the original supersymmetric Minkowski vacuum, without flat directions, preserve its stability and they do result in a  stable dS vacuum. An approximate criterion of the range of deformations which lead to a stable dS vacuum is that the gravitino mass in the uplifted vacuum should be much smaller than the mass of the lightest moduli field in the original Minkowski vacuum. More precisely formulated conditions can be deduced from section 5 of  \cite{Kallosh:2019zgd} and from Appendix \ref{appA} of this paper.

Interestingly, in our investigation of some particular models, we encountered certain situations where the dS vacua obtained by downshift and uplift remain stable even after very large deviations from the original supersymmetric Minkowski state. As instructive examples, we will consider here the simplest single field KL model described in section \ref{KLBASIC}, as well as a particular version of the STU model.

\subsection{KL model with large downshift and uplift}

According to \rf{vads}, a small addition $\Delta W$ to the superpotential of the KL model with a supersymmetric Minkowski vacuum leads to  the  formation of an AdS vacuum with negative vacuum energy proportional to $(\Delta W)^{2}$. In other words, for small $\Delta W$ the magnitude of the downshift does not depend on the sign of $\Delta W$. 

\begin{figure}[H]
\begin{center}
%\hspace{-4mm}
\includegraphics[scale=0.7]{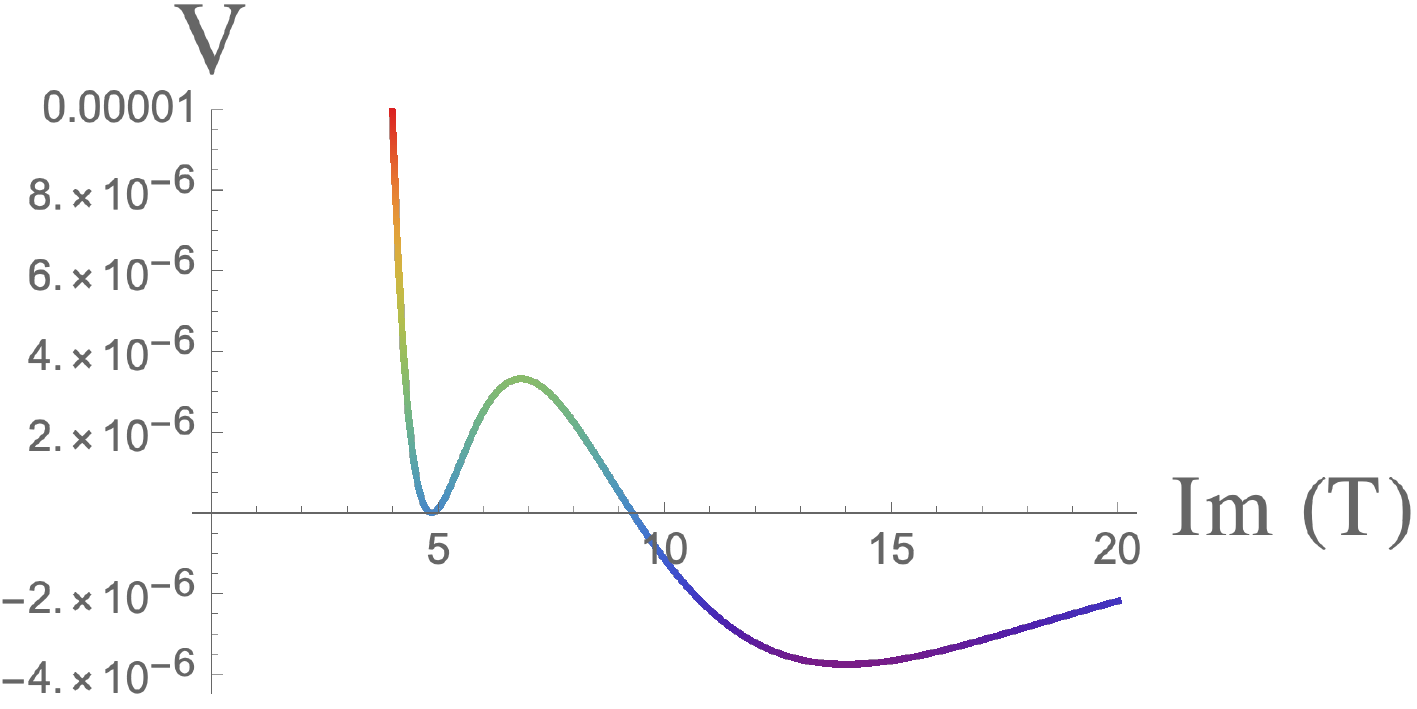}
\end{center}
\vskip -0.5cm 
\caption{\footnotesize  The potential of the KL model \rf{adssupKeith} in Planck units.}
\label{t1}
\end{figure}

However, if one wants to investigate actual limits of dS stability with respect to very large downshift controlled by $\Delta W$, and very large uplift controlled by $\mu$, one can no longer consider $\Delta W$ small, and  the sign of $\Delta W$ becomes important. As we will see, the downshift with $\Delta W> 0$ and subsequent uplift can  still give strongly stabilized dS vacua. In this section, we will  investigate this issue in the simplest KL model, with $\Phi = T$ and the parameters used in \cite{Kallosh:2011qk}, translated to our notations.  In particular, the \K\ potential is
\be\label{KLK1} 
K  = -3 \ln \left(-\rmi (T-\bar T)\right) + X\bar X  \ ,
\ee
where $X$ is a nilpotent chiral multiplet, which should be set to zero after the calculations, and the superpotential is
\be
W(T,X)   = e^{\frac{\rmi{\pi T}}{25}}- e^{\frac{\rmi{\pi T}}{ 10}} + W_{0} + \Delta W + \mu^2 X \ .
\label{adssupKeith}
\ee
 The potential of the field $T$ has a supersymmetric Minkowski minimum with $W_{0} = -{3\over 5} \left({2\over 5}\right)^{2/3} \approx -0.32573$, $ \Delta W =\mu =0$,  $V =0$, Re$\, (T) = 0$ and Im$\, (T) \approx 5$, as  shown in Fig. \ref{t1}. The mass of the field Im$\, (T)$ at the minimum is  $m_{T} =0.013$. 

\begin{figure}[h!]
\begin{center}
%\hspace{-4mm}
\includegraphics[scale=0.58]{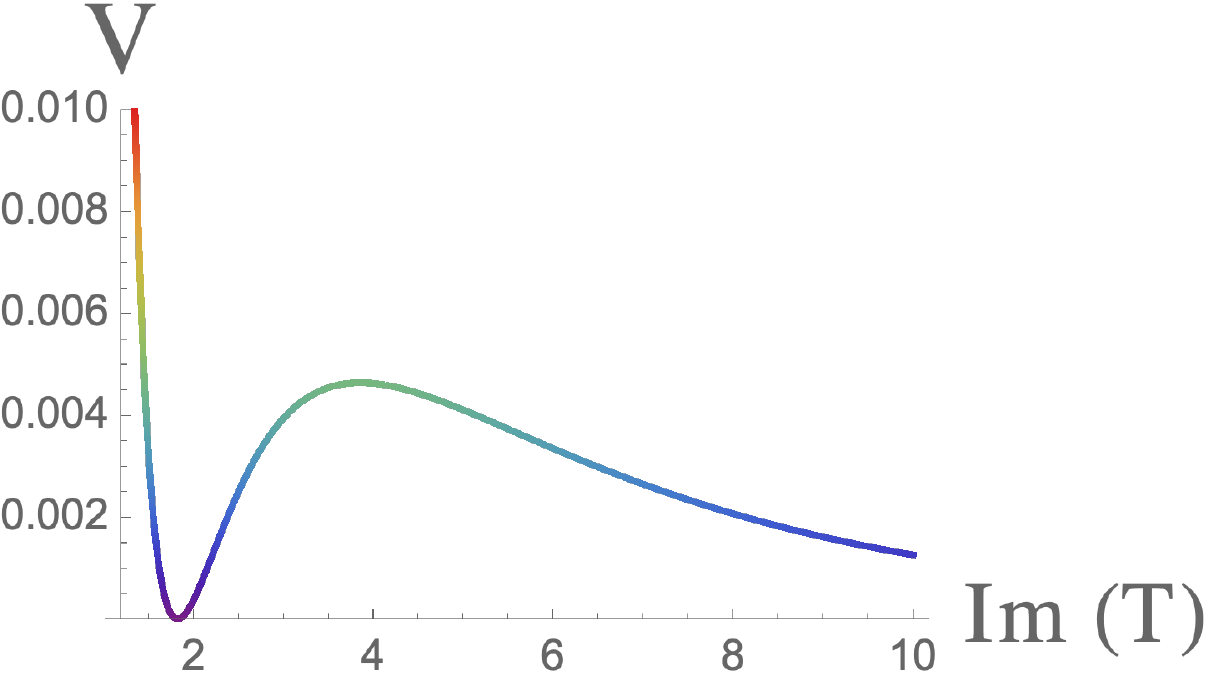} \hspace{10pt} \includegraphics[scale=0.58]{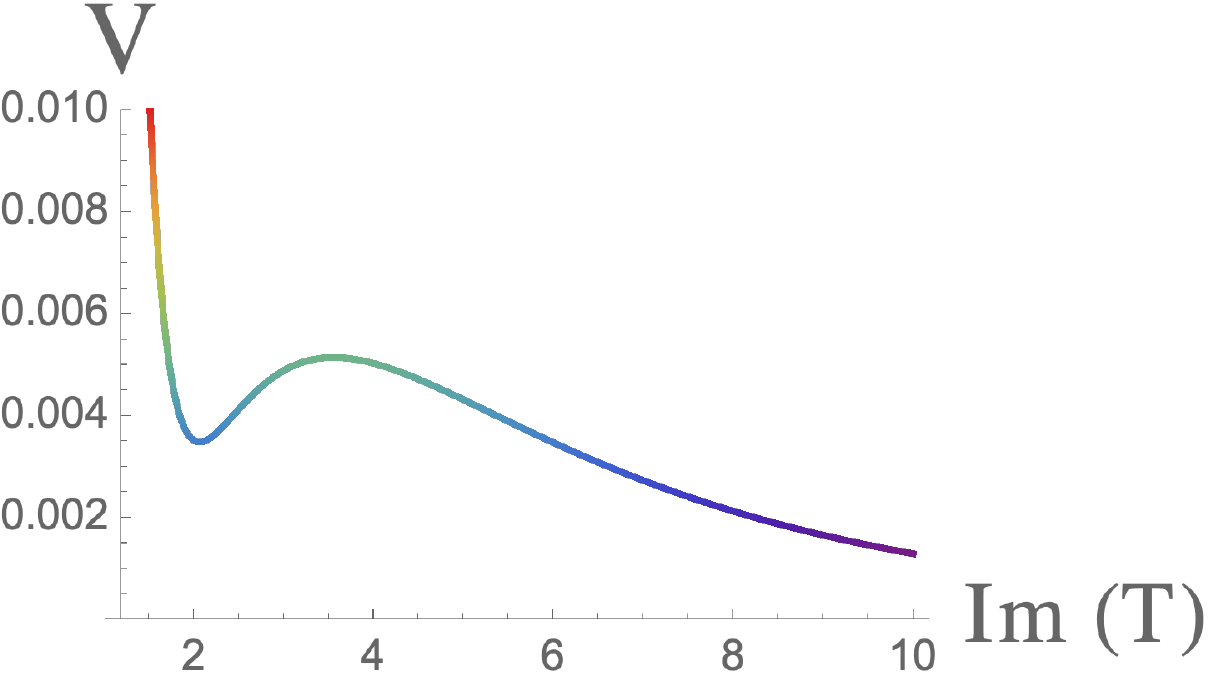}
\end{center}
\vskip -0.5cm 
\caption{\footnotesize  The potential of the KL model \rf{adssupKeith} after the downshift  obtained  by adding $\Delta W = 7$ to the superpotential. The left panel shows the potential uplifted by $\mu = 1.40199$. The right panel shows the potential uplifted by $\mu = 1.42$.}
\label{t2}
\end{figure}

The left panel of Fig. \ref{t2} shows the potential after a very large downshift, with $\Delta W = 7$, and uplift with $\mu = 1.40199$. It has a stable dS minimum, with a tiny positive vacuum energy and a gravitino mass $m_{3/2} = 1$.  It is remarkable that, by increasing $\Delta W$ from $\Delta W=0$ to $\Delta W = 7$, and compensating it by the corresponding increase of $\mu$, one can continuously interpolate between the supersymmetric Minkowski vacuum with $m_{3/2} = 0$ and %the dS vacuum with the gravitino mass in a stable dS state with strongly broken supersymmetry and with the Planckian value of the gravitino mass $m_{3/2} = 1$.
 a stable dS state with strongly broken supersymmetry and with a Planckian value of the gravitino mass, $m_{3/2} = 1$. 

The height of the barrier stabilizing the dS minimum at $\Delta W = 7$,  $\mu = 1.40199$ shown in Fig.~\ref{t2} is three orders of magnitude higher than in the original model with $\Delta W = 0$, $\mu = 0$, shown in Fig.~\ref{t1}. The mass of the field Im$(T)$ in this minimum is $0.27$, which is approximately 20 times greater than in the original Minkowski vacuum. Thus, instead of destabilizing the dS state, an increase of $\Delta W $ and $\mu$ increases the level of its stability.

\begin{figure}[h!]
\begin{center}
%\hspace{-4mm}
\includegraphics[scale=0.39]{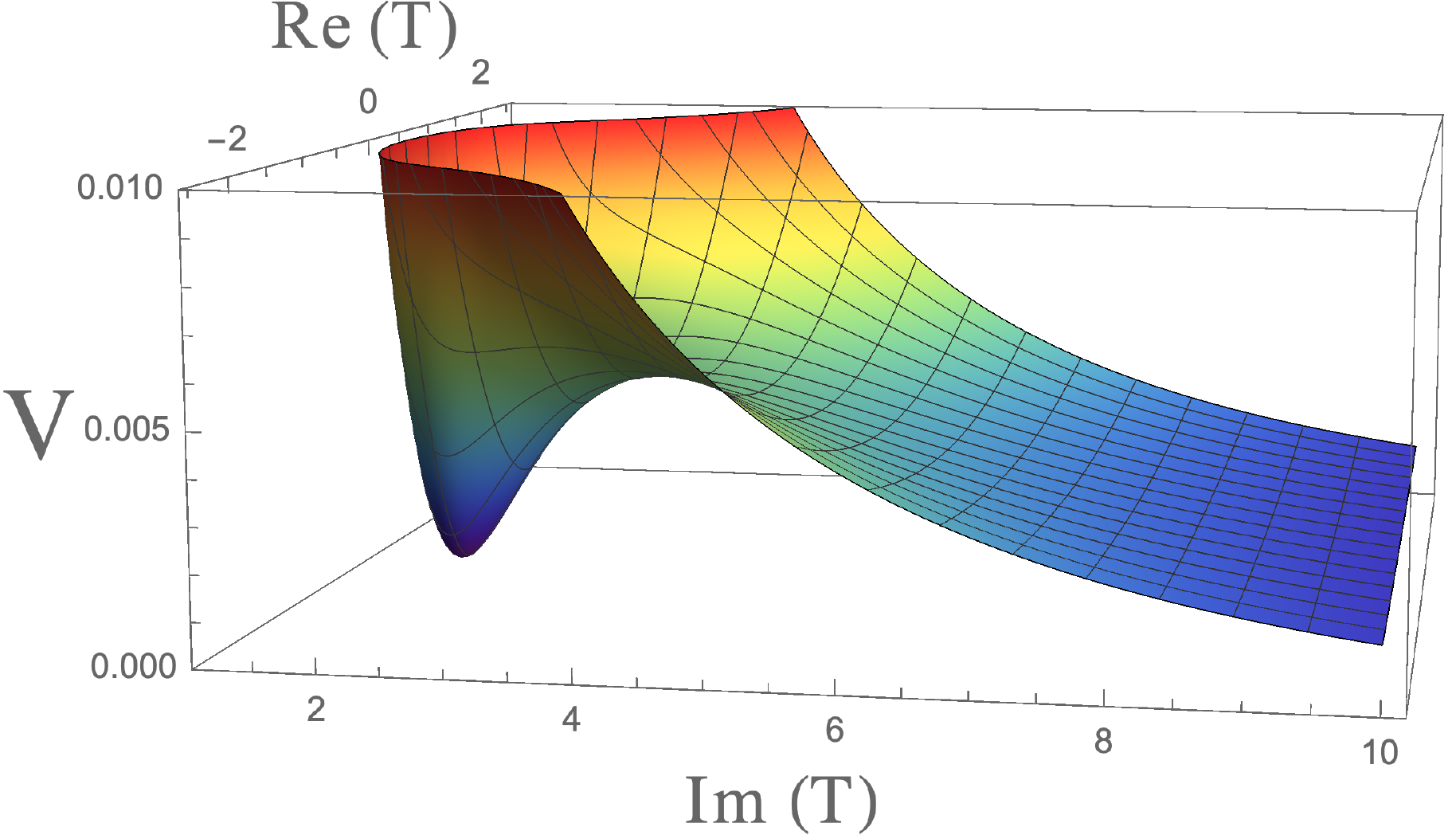}  \hspace{10pt}  \includegraphics[scale=0.39]{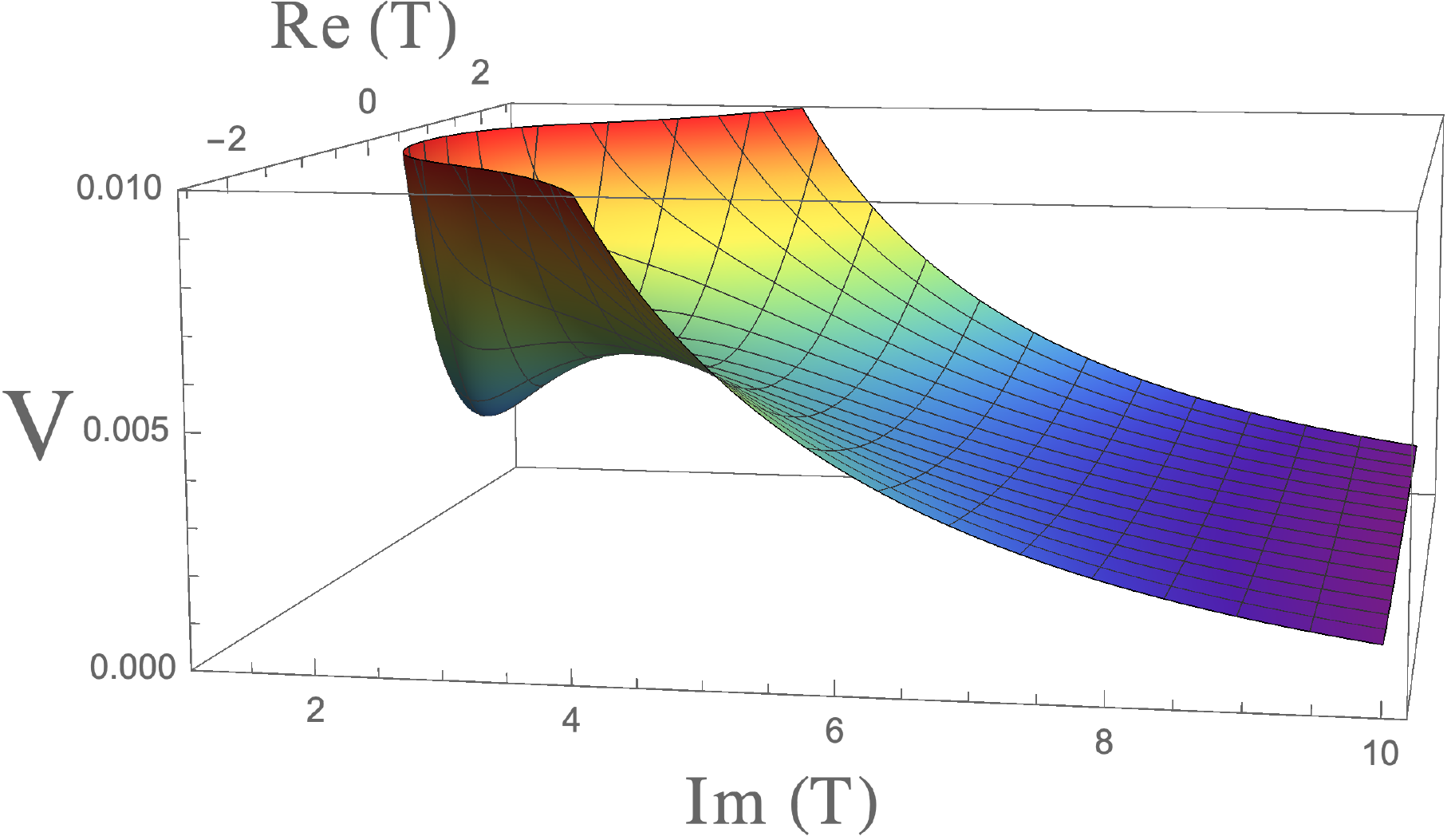}
\end{center}
\vskip -0.5cm 
\caption{\footnotesize  The same potential as in Fig. \ref{t2} at $\Delta W = 7$, $\mu = 1.40199$, and at  $\Delta W = 7$, $\mu = 1.42$, shown as a function of $ {\rm Im}\, (T)$ and ${\rm Re}\, (T)$.}
\label{t3}
\end{figure}

 One can raise the vacuum energy at the dS minimum by further increasing the value of $\mu$. This is illustrated by the right panel of figure \ref{t2}, which shows the same potential with $\Delta W = 7$, but with $\mu = 1.42$, which results in the uplift of the dS state to the level of $V_{dS} \sim 3.5\times 10^{-3}$. 
The minima of these potentials  are stable with respect to fluctuations of $ {\rm Im}\, (T)$ and ${\rm Re}\, (T)$, as shown in figure \ref{t3}.

Thus, we see that, with a proper choice of parameters of the KL model, one can obtain stable dS vacua in type IIB string theory with \emph{any desirable degree of supersymmetry breaking} and any height of the dS minimum. As we will show later, similar results are valid in other models, including the STU model in type IIA string theory.  

 For completeness, we should mention also that,  in addition to the dS vacua produced by the downshift and uplift  of  a Minkowski vacuum, a large increase of $\Delta W$ and $\mu$  may produce many other dS vacua, some of which appear at Re\,$(T) \not = 0$. For example, in the model considered above, at $\Delta W = 0$ and $\mu = 0.61243$ one has a deep dS  minimum with a tiny $V_{dS}$ at Im\,$(T) = 20.7$, Re\,$(T)  = 100n$, where $n$ is any integer. Meanwhile, at $\Delta W = 0$ and $\mu = 0.6202$, one has a deep dS  minimum at Im\,$(T) = 20.6$, Re\,$(T)  = 50+100n$.
 
Increasing $\Delta W$ produces other stable dS minima. For example, at $\Delta W = 7$,  $\mu = 1.64369$, one has a stable dS minimum at  Im\,$(T)  = 2.95$, Re\,$(T)  = 38.817$. At  $\Delta W = 7$,  $\mu =  1.93426$, the potential has a stable dS minimum at Im\,$(T) = 4.905$, Re\,$(T)  = 20.848$. Some of these additional dS minima may be interpreted as a result of uplifting of pre-existing  AdS vacua, whereas some of them do not have any AdS progenitors at all. An investigation of such dS vacua could be a subject of a separate  future  work. 

\subsection{Other  examples related to the KL construction, and general comments}

Before turning to the discussion of the STU model with large downshift and uplift, we would like to make some general comments, which  apply to many models studied in this paper, but can be especially easily illustrated using the basic KL model as an instructive example. 

 The models that we study here are formulated in string theory inspired supergravity models. String theory may bring certain additional requirements on such models and their interpretation.  For example, one can increase the downshift only for as long as this does not conflict with a large volume requirement from string theory. One may wonder whether this requirement can be satisfied in the model discussed above, where we started with Im\,$ (T) \approx 5$, and then gradually shifted towards Im\,$ (T) \approx 2$, while increasing $\Delta W$ and $\mu$. Yet another question is whether the results can be sensitive to the interpretation of the uplift. For example, as we discussed in section \ref{IIB}, in the context of type IIB models, the \K\ potential \rf{KLK1} would correspond to placing the anti-D3 brane in the bulk. 
 
To address some of these issues, we will briefly discuss here the results of the investigation of the KL model with a much greater value of  Im\,$(T)$ and with the \K\ potential corresponding to the anti-D3 brane in the throat:
\be\label{KLK2} 
K  = -3 \ln \left(-\rmi (T-\bar T)  - \frac13 X\bar X\right)  \ .
\ee
As an example, we will consider the superpotential \rf{adssup}, \rf{adssupX} with the parameters $A = 1$, $B=2$, $a = \pi/50$, $b= \pi/40$. For $W_{0} \approx -0.00512$, $ \Delta W = 0$, $\mu = 0$, this theory has a supersymmetric Minkowski minimum of the potential at  Im\,$(T) \approx 58$.  The height of the barrier stabilizing this minimum is about $3 \times 10^{{-11}}$.

With the downshift by $\Delta W = 0.1$, which is 20 times greater than $W_{0}$, and the uplift by $\mu = 0.127$, the dS minimum with a tiny cosmological constant shifts to Im\,$(T) \approx 36$, and the height of the stabilizing barrier becomes $2\times 10^{-8}$, i.e. 3 orders of magnitude higher than the original one.  

After the downshift by $\Delta W = 1$, which is 200 times greater than $W_{0}$, and after the uplift by $\mu = 0.44$, the dS minimum 
shifts to Im\,$(T) \approx 9$, and the height of the stabilizing barrier becomes $10^{-5}$, i.e. 6 orders of magnitude higher than the original one.

The conclusion is that the effect of strengthening of stabilization during the significant deviation from the supersymmetric Minkowski minimum  occurs with either of the two \K\ potentials \rf{KLK1} or \rf{KLK2}, although we found that  in the first case it may be easier to achieve strong dS stabilization. By a proper choice of the parameters, one can make the volume modulus arbitrarily large. One can continuously increase the strength of stabilization by increasing $\Delta W$ and $\mu$, while continuously decreasing the values of Im\,$(T)$ in the dS minimum.

\subsection{STU model with large downshift and uplift}

 In the previous subsection it was shown that the dS minimum in the single field KL model may remain  stable not only after small deformations of the theory, but even after an extremely large downshift and uplift,  if the downshift is due to the addition of a positive term $\Delta W$. As we will see, the STU models share this property.  

As an example, we will consider the STU model \rf{eq:IIA3modPot} with   parameters shown in Table~\ref{tab:paraIIA3mod2large}.
\begin{table}[H]
\centering
\begin{tabular}{|c|c|c|}\hline
$a_S = 1$ & $a_T = 3/4$ & $a_U = 1/2$\\\hline
$b_S = 3/2$ & $b_T =5/3$ & $b_U = 2/3$\\\hline
$A_S = -17.8577$ & $A_T = 3$ & $A_U =20$\\\hline
$B_S = -19.6283$ & $B_T = 3.37627$ & $B_U =34.5146$\\\hline
$S_0 = 1$ & $T_0 = 1$ & $U_0 = 5$\\\hline
\end{tabular}
\caption{  Our choice of parameters for the 3 moduli IIA STU example used to study large downshifts and uplifts.}
\label{tab:paraIIA3mod2large}
\end{table}

With these parameters, the potential has a supersymmetric Minkowski minimum at  Im$(S) = 1$, Im$(T) = 1$, and Im$(U) = 5$.  The masses of the real and imaginary components of the   $S$, $T$ and $U$ fields at the minimum are  0.104,  0.104,  0.036,  0.036,  0.010, 0.010,  see  also  table \ref{tab:3paraK3evLARGE}.

Then, we introduced a large downshift term, $\Delta f_{6} = 1$. In addition, we introduced the uplifting term \rf{eq:3ModAntiD6Up} with $\mu_{1} = \mu_{2} = 0.000081479$, which uplifts the minimum to make it just barely above zero.  Note that the absolute value of the parameter $\Delta f_{6} = 1$ which we use here  is much greater than $\Delta f_{6} = -10^{-5}$ used in the  section \ref{sec:stumodel}, where we discussed the STU model with a small downshift and uplift. 
\begin{table}[H]
\centering
\begin{tabular}{|c|c|c|}\hline
& Mink  & dS \\\hline
$m_1 $ & $\; 0.104 \;$ & $\; 0.363 \;$  \\\hline
$m_2 $ & $\; 0.104 \;$ & $\; 0.350 \;$ \\\hline
$m_3 $ & $\; 0.036   \;$ & $\; 0.310   \;$ \\\hline
$m_4 $ & $\; 0.036  \;$ & $\; 0.300   \;$ \\\hline
$m_5 $ & $\; 0.010   \;$ & $\; 0.046  \;$ \\\hline
$m_6 $ & $\; 0.010   \;$ & $\; 0.043   \;$ \\\hline
\end{tabular}
\caption{  Masses in the STU model  with the parameters given in \rf{tab:paraIIA3mod2large}. }
\label{tab:3paraK3evLARGE}
\end{table}

After the uplift, the mass matrix in  the  dS vacuum is  positive  definite, and the masses, corresponding to the square roots of the eigenstates of the canonically normalized mass matrix, are given  as  0.363, 0.350, 0.310, 0.300, 0.046, 0.043, see table \ref{tab:3paraK3evLARGE}. All of these masses are significantly greater than the masses in the original supersymmetric Minkowski minimum prior to the downshift and uplift. The gravitino mass in  the  dS minimum is $m_{3/2} = 0.02$. The barrier stabilizing the dS minimum has height $\sim 2 \times 10^{-4}$. 

\begin{figure}[H] 
%\begin{center}
\includegraphics[scale=0.46]{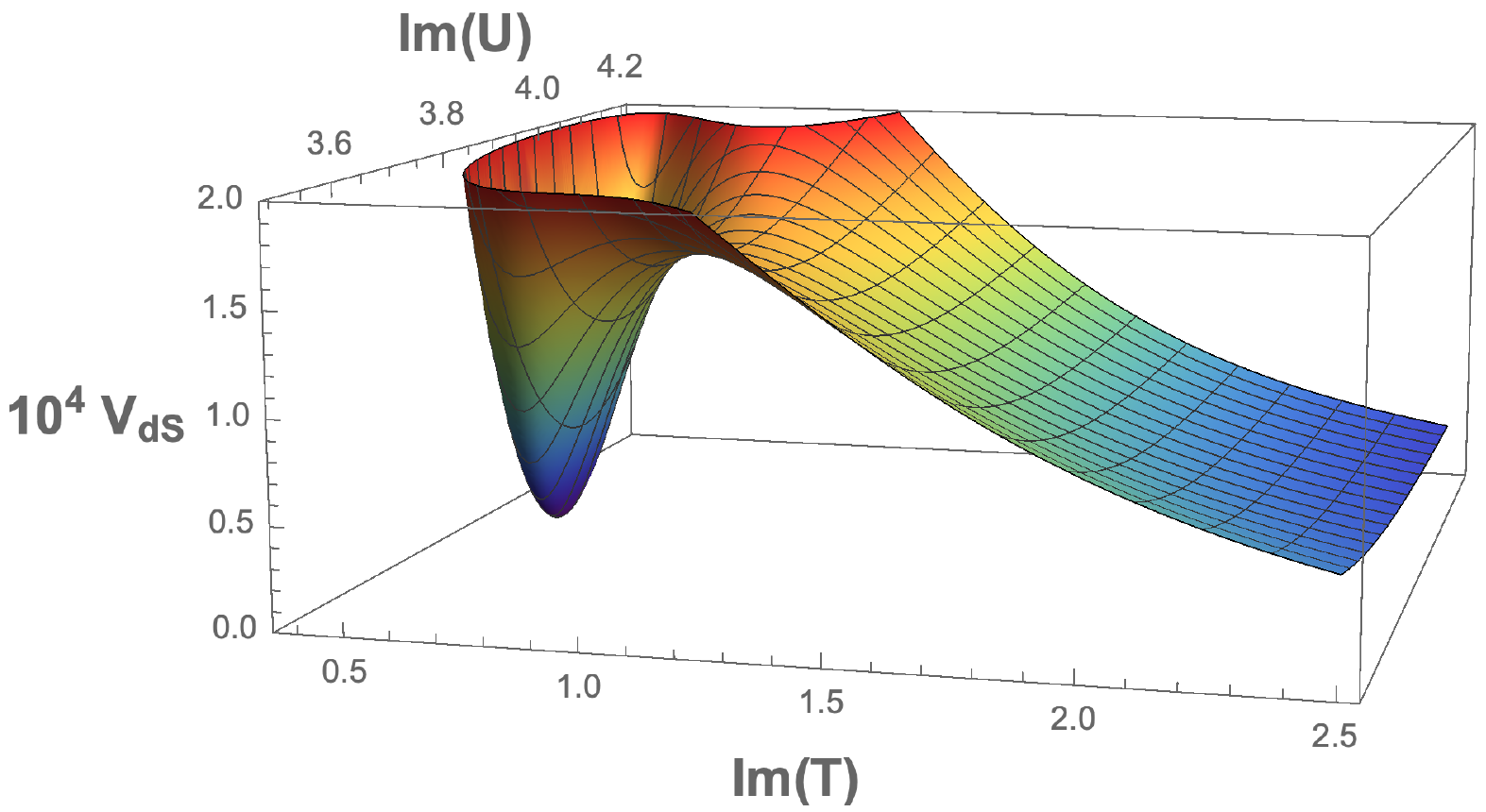}\quad \includegraphics[scale=0.46]{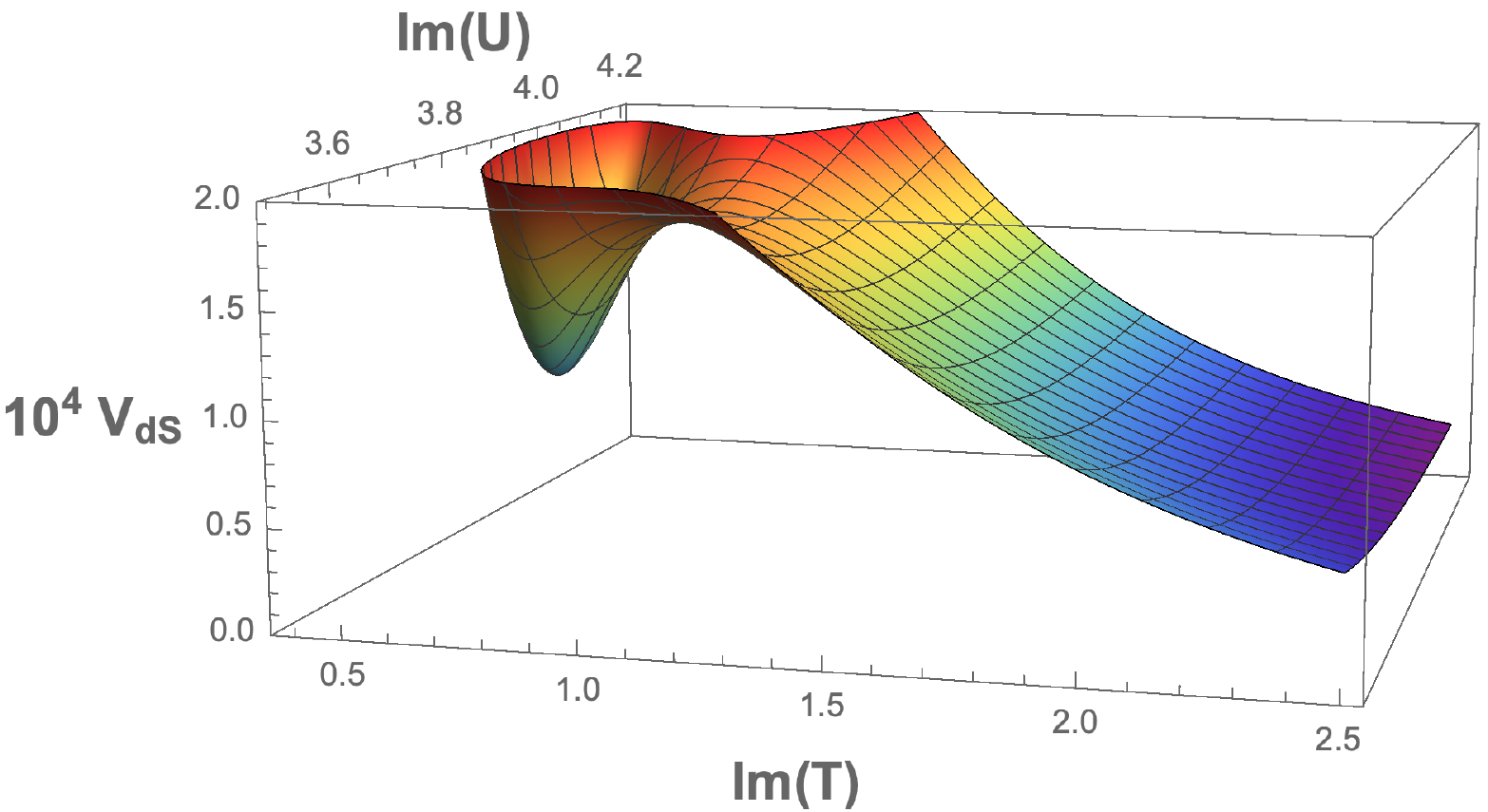}
%\end{center}
\caption{\footnotesize The overall form of the STU potential in the directions Im$(S)$ and Im$(T)$ after a very large downshift and uplift. }
\label{fig:3mod3DLargeSTU}
\end{figure}

To illustrate the properties of the potential, we show the plot of the potential after the downshift and uplift with respect to Im$(S)$ and Im$(T)$ in Fig. \ref{fig:3mod3DLargeSTU}. The left panel shows it for $\Delta f_{6} = 1$, $\mu_{1} = \mu_{2} = 0.000081479$, the right panel shows it for a slightly greater uplift with $\mu_{1} = \mu_{2} = 0.00009$. 
This potential looks strikingly similar to the potential of the simplest KL model after a large downshift and uplift  shown in figure \ref{t3}. Of course, the STU potential is a function of 3 complex variables, and therefore figure \ref{fig:3mod3DLargeSTU} shows only a part of the story, but investigation of the potential in other directions, as well as the positivity of the mass matrix which we evaluated, confirms stability of the dS minima shown in figure \ref{fig:3mod3DLargeSTU}. %Just as in the KL scenario, one may expect that a greater value of $\Delta f_{6} = 1$ and of $\mu_{1}$,  $\mu_{2}$ may further increase the strength of the dS vacuum stabilization.

%\newpage

 \section{Summary and Discussion}
 \label{concl}
 
In this last section we would like to summarize our main results and relate them to the existing literature. 

In sections  \ref{intro}, \ref{KL} and \ref{stability}  we gave a brief review of  our approach and described some general   predictions, which we confirmed in the main body of the paper.
In  section  \ref{sec:IIAexamples} we analyzed several type IIA motivated $d=4$ supergravity models. 
We gave examples of the mass production of dS minima  \cite{Kallosh:2019zgd}  in the case of the seven- and three-moduli model of type IIA orientifolds with D6 branes, compactified on ${T^6\over Z_2 \times Z_2}$. In particular, the \K\,  potential for this seven-moduli model is associated with a  $[SL(2, \mathbb{R})]^7$ symmetry. It takes the form:
 \be
K = -  \sum_i ^7 \log \left(-\rmi (\Phi^i-\bar \Phi^{\bar \imath})\right)\,.
\label{7mG}\ee
Such a model  was studied before in \cite{Blaback:2018hdo}, in the presence of perturbatively-induced superpotentials, polynomial in the moduli, of the form
$
W= a^{(0)}  +  \sum_{i, j}  ^7 \Big ( a^{(1)}_i \Phi^i + a^{(2)}_{ij} \Phi^i \Phi^j +a^{(3)}_{ijk} \Phi^i \Phi^j\Phi^k\Big ).
$
There, it was possible to find a   metastable de Sitter extremum,   but it had a flat direction. This single solution was found with the help of a numerical technique, known as differential evolution.

The  seven-moduli model with the same \K\, potential \rf{7mG}  was also  studied  in \cite{Cribiori:2019bfx}, but with a non-perturbative KKLT-type superpotential
\be
W = f_6 + \sum_i^7  A_i e^{\rmi a_i \Phi^i}\,.
\label{7m1}
\ee 
In addition, the model used uplifting anti-D6 branes, introduced in \cite{Kallosh:2018nrk}. The model \rf{7mG}, \rf{7m1} was shown in \cite{Cribiori:2019bfx} to produce many dS minima, without significant tuning of its parameters. There, this result came as an outcome of the numerical calculations.

In this paper we studied the seven-moduli model  with the same \K\, potential  \rf{7mG} and with uplifting anti-D6 branes as in  \cite{Cribiori:2019bfx},  but with a superpotential of the racetrack type \cite{Kallosh:2004yh}, where the constant part $f_6 $  comes from the six-flux, similarly like in  \cite{Cribiori:2019bfx}. The superpotential has the form:
\be
W = f_6 + \sum_i^7 \bigl( A_i e^{\rmi a_i \Phi^i}- B_i e^{\rmi b_i \Phi^i}\bigr)\,.
\label{7mW2}
\ee
The case of a two exponent KL type superpotential, where both exponents stem from non perturbative contributions, allows us to find a supersymmetric Minkowski vacuum without flat directions. This, in turn, makes it possible to satisfy the conditions 1-3 specified in the Introduction of this paper. According to \cite{Kallosh:2019zgd}, it is now \emph{predicted} that we will be able to construct a stable dS minimum, granted that the downshift and uplift are parametrically small. Stability translates to all of the 14 mass eigenvalues being positive and, indeed, this is what we find in our numerical investigation, as depicted in table \ref{tab:7modev}.

We also studied the isotropic version of this model, in which the fields were identified as $T_1=T_2=T_3  \equiv T$ and $U_1=U_2=U_3 \equiv U$. This is in fact a STU model for which, in addition to masses in table \ref{tab:3modev}, we have also presented various 2D and 3D plots.
We also produced plots describing the seven-moduli model, but we do not present them here as they do no contain any substantial novelty with respect to their simpler version in the STU model. Indeed, we believe that the latter  represents the situation in the IIA case quite well.

In our investigation of   dS vacua in type IIB theory in section \ref{IIB} we noticed that, in models associated with typical Calabi-Yau compactifications, some features of the mass matrix, and hence the analysis of the stability, become more complicated than in type IIA examples.
For example, the mass matrix \rf{nond1}  is still block diagonal, but the matrix $V_{\phi^i \phi^j}$ in \rf{nond} has 
 the off diagonal entries which are not small even in the 
supersymmetric Minkowski vacuum, see for example eq. \rf{nondiag}. Nevertheless, the prediction made in \cite{Kallosh:2019zgd} concerning  the existence of dS minima, under the conditions 1-3 specified here in the Introduction, remains valid. 

We have studied models with \K\, potentials of the form $K  =   -2 \log\left(\mathcal{V}_6(\tau_i) \right)$ where for the internal volume $\mathcal{V}_6(\tau_i)$ we had the following 4 examples in section 5:
 \begin{align}
 &
\mathcal{V}_6(\tau_i)= \frac{1}{2} \sqrt{\tau_1} \left( \tau_2 - \frac{2}{3}\tau_1\right)\, , \nonumber\\ 
 & \mathcal{V}_6(\tau_i)= \alpha \left( \sqrt{\tau_1} \left(\tau_2 - \beta \tau_1\right) - \gamma \tau_3^{3/2} \right) , \nonumber\\
 & \mathcal{V}_6(\tau_i)= \Big( {\frac{1}{108} \tau_1  [6 \tau_2-\tau_1 ] [2\tau_3 -\tau_1 ]}\Big )^{1/2} ,  \nonumber\\ \
 &\mathcal{V}_6(\tau_i) =\frac{1}{3 \sqrt{2}} \left( 2 [\tau_1 + \tau_2 + 2 \tau_3 ]^{3/2} -  [ \tau_2 + 2 \tau_3 ]^{3/2} - \tau_2^{3/2} \right).
\end{align}
Details on dS vacua in these models  with the corresponding superpotentials and with an uplifting anti-D3 brane, in accordance with the procedure outlined in the conditions 1-3 in the Introduction, 
 are presented in section \ref{IIB}. We have found a complete agreement of our numerical examples with the predictions in \cite{Kallosh:2019zgd} and in sections \ref{KL} and \ref{stability} of this paper.
 
We would like to stress the following features of all of the examples studied numerically in sections \ref{sec:IIAexamples} and \ref{IIB}:

1) The properties of the dS vacua are in a complete agreement with the predictions outlined in  \cite{Kallosh:2019zgd} and in sections \ref{KL} and \ref{stability} in this paper, including
the detailed properties of the mass matrix both in type the IIA and type IIB models.

2) In this work, we have presented just one set of parameters for each example. However, we have in fact performed additional investigations of the models with alternative parameters. In particular, it was easy to change the parameters within the required constraints and get additional dS minima. Thus, while not irrelevant, the choice of the parameters does not require fine-tuning in our constructions.

3) All of our examples are based on two  exponents in the superpotential for each of the moduli. Indeed, this is the simplest case where Minkowski progenitor models are available. However, as explained in section \ref{manymoduli}, one can easily generalize these models to the theories with any number of exponents  in the superpotential for each of the moduli.

An important advantage of the models  constructed from a supersymmetric Minkowski vacuum by its downshift and uplift is related to the requirement of vacuum stability in the early universe. In these models, the values of the moduli masses, as well as the height of the potential barrier protecting the   supersymmetric Minkowski vacuum  and the   dS minimum, can be made very large and independent on the scale of supersymmetry breaking \cite{Kallosh:2004yh,BlancoPillado:2005fn,Kallosh:2011qk,Kallosh:2019zgd}.

General theorems presented in \cite{BlancoPillado:2005fn}, in section 5 of \cite{Kallosh:2019zgd} and in the Appendix \ref{appA} of this paper, describe the conditions which guarantee the stability of dS states originating from a small downshift and uplift of  a  supersymmetric Minkowski vacuum. In section \ref{big}, we have shown that in some cases even very large downshift and uplift of  the   supersymmetric Minkowski vacuum   do not lead to dS instability, and may in fact significantly strengthen the dS vacuum stabilization.

To conclude, our $d=4$ supergravity models, inspired by string theory, represent a new way of constructing models with dS minima where the non-perturbative string theory contributions play an important role. The actual embedding of these models into a full string theory is a subject of a separate investigation. 
 One of the principal observations here is that de Sitter minima are relatively easy to find if the non-perturbative exponents in superpotential are present for all of the moduli.  Related ideas based on U-duality of string theory are believed to underline the unification of string theory into M-theory. 
%%%%%%%%%
\section*{Acknowledgement}
We are grateful to J. Bl\r{a}b\"{a}ck, G. Dibitetto and T. Wrase for important discussions. RK and AL  are supported by SITP and by the US National Science Foundation Grant  PHY-1720397, and by the  Simons Foundation Origins of the Universe program (Modern Inflationary Cosmology collaboration),  and by the Simons Fellowship in Theoretical Physics.
The work of NC and CR  is supported by an FWF grant with the number P 30265.
 CR is grateful to  SITP for the hospitality while this work was performed. CR is furthermore grateful to the Austrian Marshall Plan Foundation for making his stay at SITP possible.

\appendix

\section{More on dS stability}
\label{appA}

The purpose of this Appendix is to bring up some details about the scalar mass matrix in a dS vacuum and in particular to present  the important  steps in the derivation of  our equation 
\rf{dSmassN}.

As explained in \cite{Kallosh:2019zgd}, the scalar mass matrix in a dS vacuum derived in \cite{Denef:2004ze} is given in the concise notations with covariantly holomorphic objects like $m\equiv e^{K\over 2} W$, $m_I\equiv   D_I m$, $m_{IJ} \equiv   D_I D_J m$ etc.,  as follows:
\begin{align}
\label{VabbdS}
V_{i\bar j}^{dS} &= m_{iI} g^{I\bar J}\bar m_{\bar J\bar j}-2g_{i\bar j}m \bar m+g_{i \bar j}m_I \bar m^I - R_{i\bar j I \bar J}\bar m^I  m^{\bar J} - m_i \bar m_{\bar j},\\
\label{VaadS}
\cr
V_{ij}^{dS} &= -m_{ij}\bar m + m_{ij I}\bar m^I.
\end{align}
In order to show that  the  dS vacuum is stable, it was assumed in  \cite{Kallosh:2019zgd} that in the progenitor  Minkowski model the mass matrix has no flat directions,  and that there is a hierarchy
\be
\label{hier}
m_\chi^2 \gg |m_I|^2 \approx 0, \qquad m_\chi^2 \gg m_{3/2}^2 = m \bar m \approx 0.
\ee
In particular, this assumption means that the masses of the matter fermions $m_\chi$ which are  approximately equal to the scalar masses of stabilized moduli, are much heavier than the mass of  the  gravitino and the scale of the total supersymmetry breaking.
In this regime, \eqref{VabbdS} is dominated by the first term, which is positive definite, while \eqref{VaadS} is negligible
\begin{align}
\label{VabbdS2}
V_{i\bar j}^{dS} \approx m_{iI} g^{I\bar J}\bar m_{\bar J\bar j},  \qquad 
V_{ij}^{dS} \approx 0.
\end{align}
Furthermore, recalling that
\be
m_{iX} = \partial_i m_X + \frac12 \partial_i K m_X-\Gamma^I_{iX}m_I
\ee
and assuming the  dependence on $X$ in $W$ as in \eqref{genmodel}, as a consequence of $m\approx 0\approx m_i$ we have that
\be
m_{iX} \sim m_X, 
\ee 
which implies that the sum in \eqref{VabbdS2} is actually dominated by the first $I\neq X$ directions and thus
\be
V_{i\bar j}^{dS} \approx m_{ik} g^{k\bar k}\bar m_{\bar c\bar j}>0.
\ee
This is the result (implicitly) stated in \eqref{dSmassN}.

The prediction of the mass production of dS vacua for a general class of models can be explained by the properties of the matrix elements \rf{VabbdS} and \rf{VaadS}. In general, specific models have different values of the three geometric quantities appearing in these formulas, namely the metric $g_{IJ}$, the curvature $R_{i\bar j I \bar J}$ and the third derivative of the covariantly holomorphic superpotential $m_{ij I}$. However, the smallness of $m\bar m$ and $|m_I|^2$  is a property which is expected to cover all of the cases in which these three geometric terms are not extremely large.
 
On the other hand, in some models it is possible that these three geometric factors are actually  small, or some cancellation of negative contributions might take place, such that the requirement that $m_\chi^2 \gg |m_I|^2$ and $m_\chi^2 \gg  m \bar m$ may be relaxed, while dS minima are preserved.

\bibliographystyle{JHEP}
\bibliography{lindekalloshrefs}
%\bibliography{refs}
\end{document}